\documentclass[a4paper,11pt]{article}
 \pdfoutput=1
 \usepackage{jheppub} 
 \usepackage{graphicx}
 \usepackage{amssymb}
\usepackage{dcolumn}% Align table columns on decimal point
\usepackage{bm}% bold math 
\usepackage{color}
\usepackage{multirow}
 
\newcommand{\met}{{/\!\!\! E_T}} 
\newcommand{\mpt}{{\;/\!\!\!\! \vec{P}_T}} 
\newcommand{\mptx}{{\;/\!\!\!\! P_x}} 
\newcommand{\mpty}{{\;/\!\!\!\! P_y}} 
 
\newcommand{\mptsq}{{\;/\!\!\!\! P_T^2}} 
\usepackage{relsize}
\usepackage{cancel}
\usepackage[normalem]{ulem}

 \newcommand{\lsim}{{\;\raise0.3ex\hbox{$<$\kern-0.75em\raise-1.1ex\hbox{$\sim$}}\;}}
\newcommand{\gsim}{{\;\raise0.3ex\hbox{$>$\kern-0.75em\raise-1.1ex\hbox{$\sim$}}\;}}
\newcommand{\beq}{\begin{equation}}
\newcommand{\eeq}{\end{equation}}
\newcommand{\bea}{\begin{eqnarray}}
\newcommand{\eea}{\end{eqnarray}}

\def\baa{\begin{array}}
\def\eaa{\end{array}}

\mathchardef\minus="002D

\def\met{E_T\hspace{-0.45cm}/\hspace{0.25cm}}

\title{\boldmath Singularity Variables for Missing Energy Event Kinematics}
 
\author[a]{Konstantin T.~Matchev,} 
\author[a]{Prasanth Shyamsundar}

\affiliation[a]{Institute for Fundamental Theory, Physics Department, University of Florida, Gainesville, FL 32611, USA}

\emailAdd{matchev@ufl.edu}
\emailAdd{prasanths@ufl.edu}

\abstract{We discuss singularity variables which are properly suited for analyzing the kinematics of events with missing transverse energy at the LHC. 
We consider six of the simplest event topologies encountered in studies of leptonic $W$-bosons and top quarks, 
as well as in SUSY-like searches for new physics with dark matter particles.
In each case, we illustrate the general prescription for finding the relevant singularity variable, 
which in turn helps delineate the visible parameter subspace on which the singularities are located.
Our results can be used in two different ways --- first, as a guide for targeting the signal-rich regions of parameter space during the stage of discovery,
and second, as a sensitive focus point method for measuring the particle mass spectrum after the initial discovery.
}
 
\date{November 5 2019}

\begin{document} 
\maketitle
\flushbottom

\section{Introduction}
\label{sec:introduction}

Events with missing transverse energy (MET) at the Large Hadron Collider (LHC) are of great interest to both
theory and experiment. On the experimental side, the MET is a very challenging object, and a great amount of effort
has gone into the proper calibration of the detector and in the evaluation of its missing energy performance in 
both ATLAS \cite{TheATLAScollaboration:2013oia} and CMS \cite{CMS:2016ljj}.
On the theoretical side, events with MET are likely to hold the key to understanding some of the 
great unsolved puzzles of the Standard Model (SM). For example, the dark matter problem 
greatly motivates searches for new physics beyond the Standard Model (BSM) with dark matter candidates \cite{Feng:2010gw},
which would escape the detector leaving a MET signature \cite{Hubisz:2008gg}.
Similarly, the flavor problem provides impetus for focusing on the third generation in the SM, 
where the top quark, the bottom quark and the tau lepton all have decay channels with invisible
particles (neutrinos) in the final state. Finally, the $W$-bosons, whose leptonic decays exhibit a classic MET signature, 
have long been considered promising probes of the electroweak symmetry breaking sector \cite{Lee:1977eg,Duncan:1985vj},
and more recently have rounded out the suite of Higgs discovery channels \cite{Chatrchyan:2013iaa,ATLAS:2014aga}.
Therefore, the sound theoretical understanding of the event kinematics in MET events should be a high priority.

The fundamental problem with MET events is the incomplete information about the final state, since the 
energies $\varepsilon_i$ and momenta $\vec{q}_i$ of the invisible particles (neutrinos or dark matter particles) are not measured.
At the same time, there is partial information available in the form of the energies $e_j$ and momenta $\vec{p}_j$ 
of the visible final state particles, which typically are the (approximately massless) leptons, photons and/or QCD jets, 
but may also be massive\footnote{For this reason, in our analysis below we shall try to retain the visible particle masses 
$m_j\equiv \sqrt{e_j^2-\vec{p}_j^{\,2}}$ as arbitrary whenever possible.}
reconstructed visible particles like a $W$-boson, a $Z$-boson or a Higgs boson. 
In the spirit of the simplified models approach \cite{Alves:2011wf}, in what follows we shall remain agnostic about the underlying
physics which might give rise to a particular event topology, and instead shall focus on the salient features of its kinematics
as represented by phase space singularities.
Correspondingly, we shall also ignore any secondary dynamical effects such as spin correlations, helicity suppressions, etc.,
since they do not affect the singular phase space features which we are interested in.

With this backdrop, we are ready to introduce the three relevant (and related) questions 
regarding MET events which will be addressed in this paper.
\begin{itemize}
\item Are there any singular features in the {\em visible} phase space of a given event topology?
\item What are the relevant kinematic variables which best describe such singular features?
\item What do measurements of such features tell us about the underlying mass spectrum?
\end{itemize}
Let us briefly motivate and comment on each question before proceeding to the main analysis in the following sections.

\subsection{Visible phase space singularities}

In this paper, we define a singularity in the visible parameter space $\{ \vec{p}_j\}$ as a point where the event number density formally becomes 
infinite.\footnote{In the literature, kinematic endpoints, cusps and kinks are sometimes also referred to as singularities \cite{Kim:2009si}, 
even though they are defined in terms of suitable {\em derivatives} of the event number density. 
In what follows, we shall adopt the more narrow definition of a phase space singularity as stated in the text above.}
The origin of such singularities is very well understood \cite{Kim:2009si,Rujula:2011qn,DeRujula:2012ns,Kim:2019prx}:
they arise in the process of projecting the allowed region in the full phase space $\{ \vec{q}_i, \vec{p}_j\}$ (which does not exhibit any singularities)
onto the visible subspace $\{ \vec{p}_j\}$. Similar to the phenomenon of caustics in optics, astrophysics \cite{Sikivie:1997ng} or 
accelerator physics \cite{Charles:2016xph}, singularities are formed at points where the visible projection onto $\{ \vec{p}_j\}$ 
of the allowed phase space in $\{ \vec{q}_i, \vec{p}_j\}$ gets folded.
Mathematically this is expressed as the reduction in the rank of the Jacobian matrix of the coordinate transformation
from the relevant set of kinematic constraints to $\{\vec{q}_i\}$
(alternatively, from the generator-level event parameters to the visible space $\{ \vec{p}_j\}$), 
which is why such singularities are sometimes known as Jacobian peaks.

\subsection{Singularity variables}
\label{sec:singvar}

The visible parameter space $\{ \vec{p}_j\}$ may be parametrized simply in terms of the visible momentum components, as indicated,
but it also allows infinitely many alternative re-parametrizations, involving, e.g., angles, rapidities, invariant masses, etc.
An interesting question then is which of those re-parametrizations is ``the best". 
In his pioneering paper  \cite{Kim:2009si}, Kim proposed to construct an optimized {\em one-dimensional} kinematic variable,
called a singularity coordinate, which is defined in terms of the measured visible momenta $\{\vec{p}_j\}$, plus possibly some 
mass parameters $\{M_k\}$. The latter set includes the masses $\sqrt{\varepsilon_i^2-\vec{q}_i^{\,2}}$
of the invisible particles in the final state, as well as the masses of any intermediate resonances in the event topology.
As the name suggests, the singularity coordinate is designed to capture the singular behavior and so must satisfy the following 
criteria put forth in \cite{Kim:2009si}:
(i) it must vanish at the singularity locations; (ii) its direction must be perpendicular to the singularity hypersurface in the observable 
phase space $\{ \vec{p}_j\}$; (iii) events which are equally far away from the singularity should produce the same value.

Unfortunately, Kim's paper went largely unnoticed --- in the past ten years, there have been very few explicitly worked out examples of practical significance,
with the exception of two follow-up investigations by De Rujula and Galindo, who introduced and studied several different versions of 
a singularity coordinate for the case of single $W$ production  \cite{Rujula:2011qn} and $h\to WW$  \cite{DeRujula:2012ns}.
Our goal in this paper is to expand the set of worked out examples, on occasion taking the opportunity to point out connections
to other results in the literature which have been obtained by different means. For completeness, we shall also review 
and further expand on the two case studies in \cite{Rujula:2011qn,DeRujula:2012ns}.

In parallel with the one-dimensional approach of a singularity variable, we shall also try to analyze and 
visualize the singularity hypersurface from a multi-dimensional perspective.\footnote{At the same time, 
we shall strive to describe the phase space singularities with the minimal possible number 
of visible degrees of freedom while fully retaining the singular behavior in the observed (multivariate) distributions.
The benefit from this approach will become clear once we consider some concrete examples.} For example, consider
the classic supersymmetry (SUSY) example of a single decay chain proceeding through three successive two body decays. 
The relevant visible parameter space is three-dimensional and can be parametrized by the pairwise invariant masses of the 
final state visible particles. The singularity is then found at the two-dimensional {\em surface} boundary of the allowed region
\cite{Costanzo:2009mq,Lester:2013aaa,Agrawal:2013uka,Kim:2015bnd}. In order to maximize the sensitivity, 
the experimental analysis should target this two-dimensional surface; this can be done either directly
\cite{Debnath:2016mwb,Debnath:2016gwz,Altunkaynak:2016bqe}
(e.g., using Voronoi tessellations of the data \cite{Debnath:2015wra,Debnath:2015hva}) 
or by means of a suitable one-dimensional singularity coordinate \cite{Debnath:2018azt}.

\subsection{Using singular features for particle mass measurements}

As a result of the previous step, ideally we would end up with some parametric equation of the singularity hypersurface 
within the visible phase space 
\beq
g(\{\vec{p}_j\}; \{m_j\}; \{M_k\})=0,
\label{eq:generic}
\eeq
which contains three types of variables: the set $\{\vec{p}_j\}$ of measured 3-momenta of the visible final state particles,
the corresponding set $\{m_j\}$ of their masses, which are known SM parameters, and finally, a set $\{M_k\}$ of mass parameters, 
which, depending on the circumstances,  may or may not be known {\em a priori}. As already mentioned, $\{M_k\}$ 
would typically include the masses $m_i=\sqrt{{\cal \varepsilon}_i^2-\vec{q}_i^{\,2}}$ of the invisible particles in the final state.
Now, if one is willing to assume that the invisible particles are all neutrinos, as is usually\footnote{The only exception being an 
invisibly decaying $Z$-boson, or a leptonically decaying $W$-boson where the lepton is lost \cite{Bai:2012gs}.} 
the case with studies of SM signatures, those masses can be safely set to zero. On the other hand,
in studies of BSM signatures the invisible particles are some new dark matter particles whose masses are {\em a priori} unknown 
and should be explicitly retained in (\ref{eq:generic}). Also included in the set $\{M_k\}$ are the masses of intermediate resonances, 
which are known if the resonance is a SM particle and unknown otherwise.

The constraint (\ref{eq:generic}) can be viewed in different ways.
First, the function $g$ can be regarded as a bona fide singularity coordinate in the sense of Ref.~\cite{Kim:2009si}.
Second, given a choice of masses $\{M_k\}$,
the constraint~(\ref{eq:generic}) defines the locus of points where the signal event density becomes singular (by construction), 
while the behavior of the background event density is typically unremarkable. Therefore, this locus of points is precisely
the region of phase space which should be targeted (with suitable selection cuts) in an analysis aimed at a discovery \cite{Debnath:2018azt}.
Finally, we can turn the last argument around and instead of studying the phase space $\{\vec{p}_j\}$ for a given choice of mass parameters $\{M_k\}$,
we can study the mass parameter space $\{M_k\}$ for given points in phase space as sampled by the events in the data. 
Correspondingly, for each event, eq.~(\ref{eq:generic}) can be viewed as a constraint on the allowed values 
of the mass parameters for which the current event would be located on a singularity hypersurface.
In a companion paper \cite{Kim:2019prx}, this idea was formulated as a new mass measurement method\footnote{For
reviews of the large variety of mass measurement methods proposed for SUSY-like events with MET, see
\cite{Barr:2010zj,Matchev:2019sqa} and references therein.}, called the ``focus point method", which was
inspired by previous related work in \cite{Nojiri:2003tu,Kawagoe:2004rz,Cheng:2008hk,Gripaios:2011jm,Anagnostou:2011aa}.
The analysis in \cite{Kim:2019prx} was illustrated with one specific event topology, dilepton $t\bar{t}$ events,
and we shall now show that it is also applicable for the remaining 5 event topologies considered here.

The paper is organized as follows. In Section~\ref{subsec:notation} we first introduce the six event topologies to be studied in this paper, 
along with our notation and conventions, as well as some details on our simulations. Then in Section~\ref{subsec:idea}
we flesh out the general method for deciding whether a singularity feature exists, and if so, for deriving the corresponding
constraint (\ref{eq:generic}). In the next five Sections~\ref{sec:11}-\ref{sec:22}, we illustrate the general method for each
individual event topology. In the process, we shall sometimes rederive some existing results in the literature, albeit in a
different, universal and perhaps more intuitive way. We hope that the expert reader will enjoy seeing the underlying commonality between 
those familiar results, as well as appreciate the novelty of the others. The novice reader is perhaps best advised to first read 
Refs.~\cite{Kim:2009si,Rujula:2011qn,DeRujula:2012ns,Kim:2019prx} and be ready to consult a standard reference like
\cite{Byckling:1971vca,Barger:1987nn} when necessary.
Finally, in Section~\ref{sec:conclusions} we present our conclusions and outlook for future studies.

\section{Preliminaries}
\label{sec:prelim}

\subsection{Notations and Setup}
\label{subsec:notation}

\begin{figure}[t]
 \centering
 \includegraphics[width=.9\textwidth]{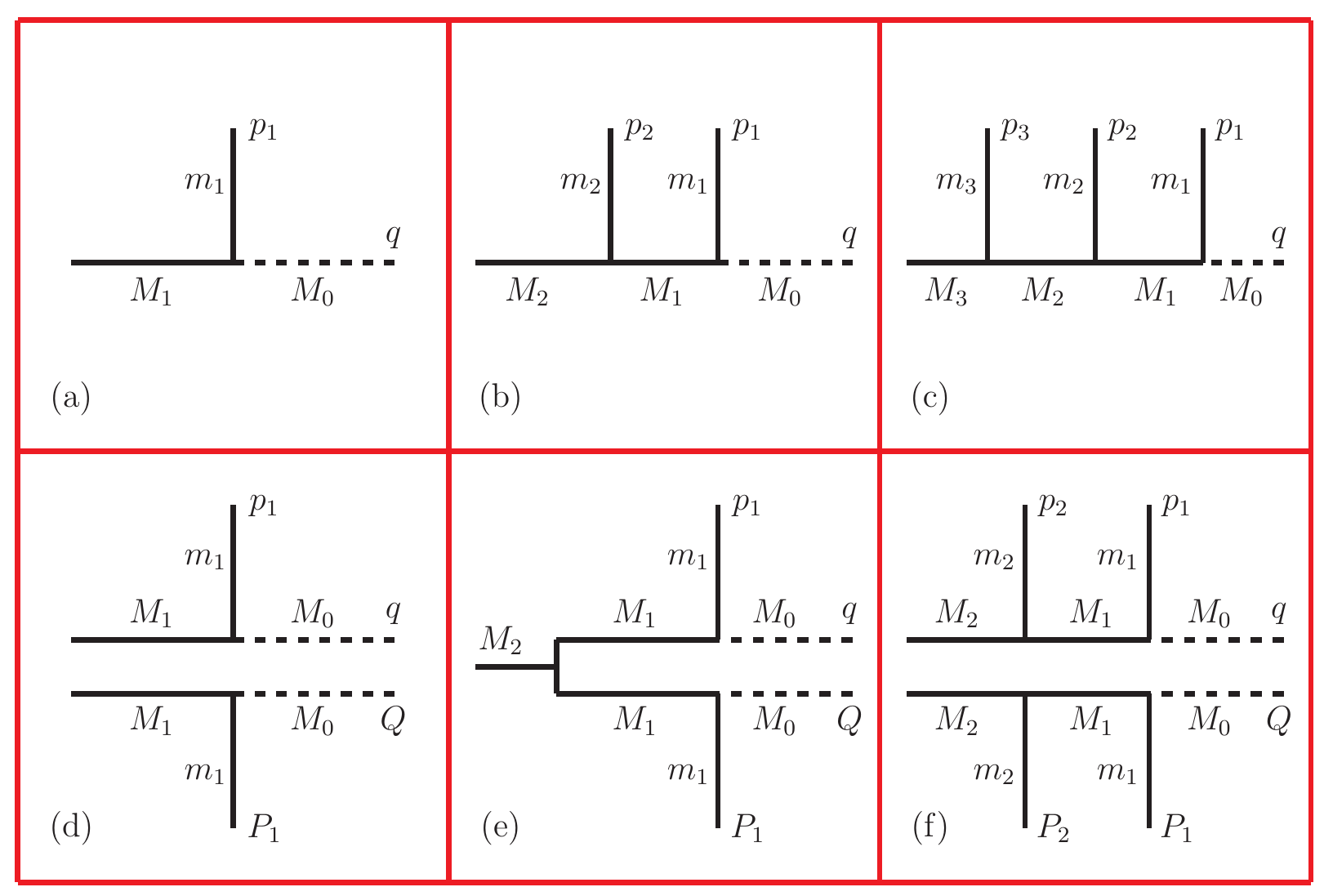}
 \caption{\label{fig:feynmandiag} The event topologies studied in this paper. Diagrams (a-c) in the top row
 represent single decay chains, each terminating in a single invisible particle with mass $M_0$ and 4-momentum $q=(\varepsilon,\vec{q})$, 
 while diagrams (d-f) in the bottom row represent pair-production processes leading to two decay chains 
 and two invisible particles with 4-momenta $q=(\varepsilon,\vec{q})$ and $Q=({\cal E},\vec{Q})$, respectively.
 $p_i=(e_i,\vec{p}_i)$ and $P_i=(E_i,\vec{P}_i)$ label the 4-momenta of visible final state particles of mass $m_i$,
 while $M_k$ with $k=1,2,3$ label the masses of on-shell intermediate resonances.}
\end{figure}

The six event topologies to be considered in this paper are shown in Fig.~\ref{fig:feynmandiag}, where,
in accordance with the notation of \cite{Barr:2011xt}, the letters p and q are reserved for denoting momenta of 
visible and invisible particles, respectively.
Diagrams (a-c) in the top row represent single decay chains, each terminating in a single invisible particle 
with mass $M_0$ and 4-momentum $q=(\varepsilon,\vec{q})$, which is taken to be 
on-shell\footnote{This is certainly true in the case of stable dark matter particles and neutrinos. 
However, in principle it is also possible that the invisible particle itself decays invisibly --- this case was treated in \cite{Kim:2017qdi}.}: 
$\varepsilon^2-\vec{q}^{\,2}=M_0^2$.
The decay chains in Fig.~\ref{fig:feynmandiag}(a-c) differ by the number of two-body decays --- 
here we have limited ourselves to at most three (in Fig.~\ref{fig:feynmandiag}(c)), but the generalization 
to four or more is straightforward \cite{Altunkaynak:2016bqe}.
Similarly,  $p_i=(e_i,\vec{p}_i)$, with $i=1,2,3$, label the 4-momenta of visible final state particles of mass $m_i$,
with $e_i^2-\vec{p}_i^{\, 2}=m_i^2$. The masses $M_k$ with $k=1,2,3$ label the masses of on-shell intermediate resonances.
The diagrams (d-f) in the bottom row of Fig.~\ref{fig:feynmandiag} represent pair-production processes leading to two decay chains 
 and thus to two invisible particles with 4-momenta $q=(\varepsilon,\vec{q})$ and $Q=({\cal E},\vec{Q})$, respectively.
 The 4-momenta of the visible particles in the second decay chain are similarly capitalized:
 $P_i=(E_i,\vec{P}_i)$, where $E_i^2-\vec{P}_i^{\, 2}=m_i^2$ and $i=1,2$.
  
The six diagrams in Fig.~\ref{fig:feynmandiag} cover a host of interesting physics processes at the LHC, both within and outside the SM.
For any given such process, there will be a corresponding choice for the masses $m_i$, $M_0$ and $M_k$. Consider, for example, 
single top decay 
\beq
t \to bW^+\to b \ell^+\nu
\label{topdecay}
\eeq
as an illustration for Fig.~\ref{fig:feynmandiag}(b). In that case, $m_1$ is the mass of a lepton (electron or muon),
$m_2$ is the mass of the reconstructed $b$-jet, $M_0$ is the neutrino mass, $M_1$ is the $W$-boson mass and $M_2$ is the top quark mass.
Any other process will dictate a similar choice of mass parameters; and of course, BSM processes may involve 
dark matter particles and intermediate resonances with {\em a priori} unknown masses $M_0$ and $M_k$. 
In order to keep everything on the same footing, in our numerical examples we shall use the same mass spectrum 
for all six processes as shown in Table~\ref{tab:mass}.
\begin{table}[t]
\centering
\begin{tabular}{||c||c|c|c|c||}
\hline
Parameter & $M_0$ & $M_1$& $M_2$ & $M_3$ \\
\hline
Value (in GeV) & 700 & 800 & 1000 & 1300 \\
\hline
\end{tabular}
\caption{\label{tab:mass}
The common mass spectrum used in the numerical examples, except for Fig.~\ref{fig:feynmandiag}(e) where $M_2=3000$ GeV instead. }
\end{table}

Events will be generated at parton level with {\sc MadGraph} \cite{Alwall:2011uj}. Intermediate resonances will be decayed by phase space,
i.e., ignoring any spin correlations. This assumption will not impact our results, since they are based on purely kinematics arguments 
and are thus independent of the model dynamics.  For clarity of the presentation, and to better see the singularity structures,
we will not employ any detector simulation and will keep the intermediate resonances strictly on-shell. Of course, 
in a real experiment there will be some smearing due to both the detector resolution and the finite particle widths; 
the extent of those effects depends on the particular realization of the topology in question and is beyond our scope here. 
The visible final state particles, which are typically leptons and QCD jets, will be taken to be massless. As we shall see below,
in the case of Fig.~\ref{fig:feynmandiag}(a), the presence of initial state radiation (ISR) is important and actually quite beneficial for the analysis.
This is why ISR will be simulated for that case, while in the remaining event topologies the ISR will not be modelled.
Finally, we shall ignore the combinatorial problem arising in Fig.~\ref{fig:feynmandiag}(f), since
it has already been addressed in the literature \cite{Baringer:2011nh,Choi:2011ys,Debnath:2017ktz}, 
and for simplicity we will assume that the visible particles have been properly assigned to the two decay branches. 

\subsection{The basic idea}
\label{subsec:idea}

We are now ready to outline the general method for deriving a singularity variable \cite{Kim:2009si,Rujula:2011qn,DeRujula:2012ns,Kim:2019prx}.
The first step is to collect all kinematic constraints on the invisible momenta $q$ (and $Q$, if applicable).
In general, the kinematic constraints fall into several categories:
\begin{itemize}
\item {\em On-shell relations.} These exist for the final state invisible particles
\begin{subequations}
\label{eq:M0}
\bea
M_0^2 &=& q^2, \\ [2mm]
M_0^2 &=& Q^2, \label{eq:M0Q} 
\eea
\end{subequations}
as well as for the intermediate resonances:
\begin{subequations}
\label{eq:Mk}
\bea
M_k^2 &=& \left(q+\sum_{i=1}^k p_i\right)^2, \\ [2mm]
M_k^2 &=& \left(Q+\sum_{i=1}^k P_i\right)^2. \label{eq:MkQ}
\eea
\end{subequations}
Notice that the equations (\ref{eq:M0Q}) and (\ref{eq:MkQ}) are only applicable in the case of the bottom row diagrams in Fig.~\ref{fig:feynmandiag}.
\item {\em Missing transverse momentum constraint.} Momentum conservation in the transverse plane 
constrains the sum of the transverse momenta of the invisible particles:
\beq
\mpt = \left\{ 
\begin{array}{ll}
\vec{q}_T & \text{\rm for\ Fig.~\ref{fig:feynmandiag}(a-c),}  \\ [2mm]
\vec{q}_T+\vec{Q}_T & \text{\rm for\ Fig.~\ref{fig:feynmandiag}(d-f).}
\end{array}
\right.
\label{eq:mpt}
\eeq
\item {\em Additional constraints due to the specific experimental setup.} Additional constraints may arise at specific experimental facilities, e.g., at a lepton collider,
where the center-of-mass (CM) energy and momentum of the initial state are known.
\end{itemize}

Combining all of these constraints, we obtain a set of $N_C$ equations\footnote{To simplify the notation, we suppress the index $i$ labelling the individual 
visible final state particles, and we do not include their masses $m_i$ among the mass parameters in (\ref{constraints}).} 
\beq
f_\alpha\left(\left\{p,P\right\};\left\{q,Q\right\};\left\{M_k\right\}\right)=0, \qquad \alpha=1,2,....,N_C,
\label{constraints}
\eeq
involving the set $\left\{p,P\right\}$ of measured 4-momenta of the visible final state particles, 
the unknown 4-momenta $\left\{q,Q\right\}$ of the invisible final state particles,
and the set of mass parameters $\left\{M_k\right\}$ introduced in Section~\ref{sec:singvar}.
For the single decay chain diagrams in the top row of Fig.~\ref{fig:feynmandiag}, the momenta $P_i$ and $Q$ are absent, and 
(\ref{constraints}) simplifies to
\beq
f_\alpha\left(\left\{p\right\};\left\{q\right\};\left\{M_k\right\}\right)=0, \qquad \alpha=1,2,....,N_C.
\label{constraintstoprow}
\eeq
Let $N_q$ be the total number of invisible 4-momentum components, i.e.
\beq
N_q = \left\{ 
\begin{array}{ll}
4, ~~& {\rm for\ the\ case\ of\ Figs.~\ref{fig:feynmandiag}(a-c)}, \\
8, ~~& {\rm for\ the\ case\ of\ Figs.~\ref{fig:feynmandiag}(d-f)}.
\end{array}
\right.
\label{Nqdefinition}
\eeq
In what follows, we shall focus on situations where the number of constraints $N_C$ is just enough so that one can solve
(\ref{constraints}) and (\ref{constraintstoprow}) for the invisible momenta in terms of the mass parameters $\left\{M_k\right\}$. 
In other words we shall always have\footnote{The case of $N_C\ne N_q$ will be treated in a future publication \cite{MS}.}
\beq
N_C= N_q.
\label{NceqNq}
\eeq
Note that this does not imply that the final state kinematics is fully solved --- we are just trading one set of unknowns,
the components of the invisible momenta, for another: the masses $\left\{M_k\right\}$ of the intermediate resonances 
and of the invisible final state particles. In other words, we are still dealing with an undetermined problem, 
in which we are not able to compute the exact momenta of the invisible particles in the event. 

In order to illustrate the basic idea, it is sufficient to consider the simpler version (\ref{constraintstoprow})
of the kinematic constraints --- the same argument goes through for any number of invisible particles in the final state,
as long as (\ref{NceqNq}) is in effect. Consider some particular solution $\tilde q^\mu$ of (\ref{constraintstoprow}):
\beq
f_\alpha\left(\left\{p\right\};\left\{\tilde q\right\};\left\{M_k\right\}\right)=0, \qquad \alpha=1,2,....,N_C.
\label{particularsol}
\eeq
A singularity at $\tilde q$ is obtained when at least one of the directions of the local tangent plane to the full phase space 
is aligned with an invisible momentum direction \cite{Kim:2009si}.
This means that we can make an infinitesimal change $\delta \tilde q$ in the unmeasured invisible 4-momentum components  
while continuing to satisfy the original system, i.e.
\beq
f_\alpha\left(\left\{p\right\};\left\{\tilde q + \delta\tilde q\right\};\left\{M_k\right\}\right)=0, \qquad \alpha=1,2,....,N_C.
\eeq
Upon expanding the last equation and taking into account (\ref{particularsol}), one finds
\beq
\frac{\partial f_\alpha}{\partial q^\mu} \ \delta \tilde q^\mu = 0, \qquad \alpha=1,2,....,N_C.
\label{expandedconstraints}
\eeq
The condition (\ref{NceqNq}) ensures that the Jacobian matrix
\beq
D_{\alpha \mu} \equiv \frac{\partial f_\alpha}{\partial q^\mu}
\label{defJac}
\eeq
is a square matrix. The existence of non-trivial solutions $\delta\tilde q^\mu$ for the system (\ref{expandedconstraints}) is guaranteed if
the determinant of $D_{\alpha\mu}$ vanishes:
\beq
{\rm Det}\, D_{\alpha\mu} 
(\left\{p\right\}, \left\{M_k\right\})
\equiv \left|\frac{\partial f_\alpha}{\partial q^\mu}\right|_{q=\tilde q}
=0.
\label{det0toprow}
\eeq
Note that the left-hand side of this equation is a function of only visible momenta $\left\{p\right\}$ and mass parameters $\left\{M_k\right\}$,
since the invisible momenta can be eliminated via eqs.~(\ref{particularsol}).
Comparing to (\ref{eq:generic}), we see that the left-hand side of (\ref{det0toprow}) can be taken to be the desired singularity 
coordinate\footnote{Of course, any function {\em proportional to} the left-hand side of (\ref{det0toprow}) is also a singularity coordinate~\cite{Rujula:2011qn,DeRujula:2012ns}.}.
Repeating the same argument for the case with two invisible particles, i.e., starting from eq.~(\ref{constraints}), 
leads us to a similar condition
\beq
\left| \frac{\partial f_\alpha}{\partial (q, Q)} \right|
=0.
\label{det0}
\eeq
From a mathematical point of view, eqs.~(\ref{det0toprow}) and (\ref{det0}) are simply the reduced rank conditions leading to a critical point.
In our case their importance lies in the fact that the distribution of the corresponding kinematic variable is singular at the critical point,
which is why Refs.~\cite{Rujula:2011qn,DeRujula:2012ns} also referred to these reduced rank conditions as ``singularity conditions".

In the subsequent sections we shall explore the implication of (\ref{det0toprow}) or (\ref{det0}) for the event topologies of Fig.~\ref{fig:feynmandiag}.
As outlined in the introduction, for each event topology we shall focus on the following three issues:
\begin{itemize}
\item Derivation of the relevant singularity coordinate.
\item Delineation of the signal-rich regions of the visible phase space, i.e., where the signal density becomes singular. In doing so,
we shall be careful to use the symmetries of the problem in order to maximally reduce the dimensionality of the observable phase space 
{\em without} washing out any singular kinematic features. 
\item Demonstration of the focus point method for mass measurements proposed in \cite{Kim:2019prx}.
\end{itemize}

\section{Single decay chain, one two-body decay} 
\label{sec:11}

In this section we shall consider the single two-body decay diagram from Fig.~\ref{fig:feynmandiag}(a).
We shall revisit and expand the discussion in Ref.~\cite{Rujula:2011qn}, which showed that
the singularity coordinate in this case is nothing but the usual transverse mass variable $m_T$ \cite{Barger:1983wf,Smith:1983aa}.
By now, the transverse mass is one of the standard kinematic variables, which has been widely used in precision 
measurements of the $W$-boson mass \cite{Abazov:2012bv,Aaltonen:2013vwa,Aaboud:2017svj} as well as in new physics searches 
for $W'$ resonances \cite{Abazov:2007ah,Aaltonen:2010jj,Sirunyan:2018mpc,Aad:2019wvl}.
The new element in our discussion will be the role of the focus point method of~\cite{Kim:2019prx}
and its connection to the kink method for mass measurements \cite{Cho:2007qv,Gripaios:2007is,Barr:2007hy,Cho:2007dh,Matchev:2009fh}. 

\subsection{Derivation of a singularity coordinate}

Let us begin by listing the four kinematic constraints for Fig.~\ref{fig:feynmandiag}(a):
\begin{subequations}
\begin{eqnarray}
q^2 &=& M_0^2, \\
(q+p_1)^2 &=& M_1^2, \\
\vec{q}_T &=& \mpt ,
\end{eqnarray}
\end{subequations}
or after a simple rearrangement,
\begin{subequations}
\begin{eqnarray}
q^2 &=& M_0^2, \label{aqq}\\
2p_1\cdot q &=& M_1^2-M_0^2-m_1^2, \label{apq}\\
q_x &=& \mptx, \label{aqx}\\
q_y &=& \mpty. \label{aqy}
\end{eqnarray}
\label{systemFiga}%
\end{subequations}
The Jacobian matrix (\ref{defJac}) in this case is \cite{Rujula:2011qn}
\beq
D = 
\left(
\begin{array}{cccc}
2\varepsilon & -2 q_{x} & -2 q_{y} & -2 q_{z} \\
2e_1 & -2 p_{1x} & -2 p_{1y} & -2 p_{1z} \\
0 & 1 & 0 & 0 \\
0 & 0 & 1 & 0 
\end{array}
\right)
\eeq
and the corresponding singularity condition (\ref{det0toprow}) reads \cite{Rujula:2011qn}
\beq
{\rm Det}\, D= -4(\varepsilon p_{1z}-e_1 q_z)=0.
\label{singularityFiga}
\eeq
The latter is nothing but the equal rapidity condition
\beq
\frac{\varepsilon}{q_z}=\frac{e_1}{p_{1z}},
\label{equalrapiditycondition}
\eeq
which is known \cite{Barr:2011xt} to provide the link between transverse invariant mass variables (like the transverse mass $m_T$,
the Cambridge $m_{T2}$ and others) and their respective 3+1 dimensional analogues 
\cite{Ross:2007rm,Barr:2008ba,Konar:2008ei,Konar:2010ma,Mahbubani:2012kx,Cho:2014naa,Cho:2014yma,Kim:2014ana,Swain:2014dha,Cho:2015laa,Konar:2016wbh,Goncalves:2018agy}.
In order to cast the singularity condition in the desired form (\ref{eq:generic}), we must eliminate the invisible 4-momentum $q$
by using four out of the five equations appearing in eqs.~(\ref{systemFiga}) and (\ref{equalrapiditycondition}).
For example, using (\ref{aqq}), (\ref{aqx}), (\ref{aqy}) and (\ref{equalrapiditycondition}), one can find the components of 
the 4-vector $q=(\varepsilon,q_x,q_y,q_z)$ as
\begin{subequations}
\bea
\varepsilon &=& e_1\, \sqrt{\frac{M_0^2+\mptsq}{m_1^2+p_{1T}^2}}, \\ [2mm]
q_x &=& \mptx, \\ [2mm]
q_y &=& \mpty, \\ [2mm]
q_z &=& p_{1z}\, \sqrt{\frac{M_0^2+\mptsq}{m_1^2+p_{1T}^2}}.
\eea
\label{aqsol}%
\end{subequations}
Substituting the result (\ref{aqsol}) into the remaining fifth equation (\ref{apq}),
we obtain the final singularity condition explicitly in the form
\beq
m_1^2+M_0^2 + 2
\left(
\sqrt{m_1^2+p_{1T}^2}\sqrt{M_0^2+\mptsq}-\vec{p}_{1T}\cdot\mpt
\right)
- M_1^2=0,
\label{aMTcond}
\eeq
where in the left-hand side we recognize the transverse mass variable $m_T$ written as
\beq
m^2_T(M_0)\equiv m_1^2+M_0^2 + 2
\left(
\sqrt{m_1^2+p_{1T}^2}\sqrt{M_0^2+\mptsq}-\vec{p}_{1T}\cdot\mpt
\right).
\label{amtdef1}
\eeq
If we introduce the transverse energies
\bea
e_{1T}&\equiv& \sqrt{m_1^2+p_{1T}^2}, \\
\varepsilon_T &\equiv& \sqrt{M_0^2+\mptsq},
\eea
the previous result (\ref{amtdef1}) can be rewritten more compactly in the familiar form \cite{Barger:1987nn}
\beq
m^2_T(M_0) = \left[e_{1T}+\varepsilon_T(M_0)\right]^2-\left[\vec{p}_{1T}+\mpt\right]^2.
\label{amtdef2}
\eeq
Note that while we have succeeded in eliminating the unknown components of the invisible 4-momentum $q$,
there still remains one {\em a priori} unknown parameter, namely, the mass $M_0$ of the invisible particle, which
enters through the transverse energy $\varepsilon_T$. The singularity condition (\ref{aMTcond}) can 
then be rewritten in the very compact form
\beq
m_T(M_0) = M_1.
\label{mTeqM1}
\eeq
This confirms that, up to the additive constant $M_1$, the relevant singularity variable for the event topology of Fig.~\ref{fig:feynmandiag}(a)
is indeed the transverse mass $m_T$. Furthermore, eq.~(\ref{mTeqM1}) shows that the singularity 
occurs at the mass $M_1$ of the parent particle, i.e.
\beq
\lim_{m_T\to M_1} \left(\frac{dN}{dm_T}\right)= \infty.
\label{asingularitylocation}
\eeq
However, there is an important subtlety --- the latter statement is true only if we have made the correct choice 
for the invisible mass parameter $M_0$ entering the definition (\ref{amtdef2}).
In general, the true value of $M_0$ is {\em a priori} unknown, and we have to adopt a certain ansatz for it
(denoted in the following by $\tilde M_0$) in order to compute the transverse mass $m_T$ from (\ref{amtdef2}).\footnote{In some sense, 
the situation here is analogous to the well-known behavior of the kinematic endpoint 
\beq
m_T^{max}(\tilde M_0)\equiv \max_{\vec{p}_1}\left\{m_T(\vec{p}_1,\tilde M_0)\right\},
\label{eq:mtmax}
\eeq 
which is obtained by considering the whole sample of events and finding the maximum value of $m_T$.
Since $m_T(M_0)\le M_1$ by construction, the measured endpoint value $m_T^{max}(\tilde M_0)$ can be 
interpreted as the corresponding mass $\tilde M_1$ of the parent particle for this choice of $\tilde M_0$:
\beq
\tilde M_1 = m_T^{max}(\tilde M_0).
\label{eq:M1fromM0}
\eeq
}
Once the ansatz $\tilde M_0$ differs from the true value $M_0$, the existence of a singularity 
in the $m_T(\tilde M_0)$ distribution is generally not guaranteed, and the singular feature in the 
$dN/dm_T$ distribution predicted by eqs.~(\ref{mTeqM1}) and (\ref{asingularitylocation}) will be washed out. 
This is the key observation behind the focus point method for mass measurements proposed in \cite{Kim:2019prx}.
There the method was illustrated for the case of the dilepton $t\bar{t}$ event topology of Fig.~\ref{fig:feynmandiag}(f).
In Sec.~\ref{sec:aFP} below we shall show that it is also applicable to the simple event topology of Fig.~\ref{fig:feynmandiag}(a) as well. 
But first let us understand the phase space geometry behind the singularity condition (\ref{mTeqM1}).

\subsection{The phase space geometry of the singularity condition} 
\label{sec:mtsingularity}

In general, the parent particle is produced {\em inclusively}, and the event depicted in Fig.~\ref{fig:feynmandiag}(a)
would also contain additional visible particles due to initial state radiation (ISR), or decays upstream {\em to} the 
parent particle. Let us denote the total transverse momentum of these additional visible particles as $\vec{P}_T^{ISR}$,
which can be measured in the detector. Then, by definition, the missing transverse momentum $\mpt$ 
is due to the invisible recoil against both $\vec{p}_{1T}$ and $\vec{P}_T^{ISR}$:
\beq
\mpt \equiv - \vec{p}_{1T} - \vec{P}_T^{ISR}.
\label{mptdefgen}
\eeq

\subsubsection{The case of no upstream visible momentum: $P_T^{ISR}=0$}
\label{sec:noPTISR}

In order to gain some intuition, it is instructive to first consider the simpler case of no upstream visible momentum, $P_T^{ISR}=0$.
In that case, (\ref{mptdefgen}) reduces to $\mpt = - \vec{p}_{1T}$ and the transverse mass (\ref{amtdef2}) can be written simply as
\beq
m_T(\tilde M_0) =  e_{1T}+\varepsilon_T(\tilde M_0)= \sqrt{m_1^2+p_{1T}^2} + \sqrt{\tilde M_0^2+p_{1T}^2}.
\label{amTnoISR}
\eeq
Note that in this case $m_T$ is a function of only one degree of freedom, namely, the magnitude of $\vec{p}_{1T}$.
This can be easily understood in terms of the symmetries of the problem --- $p_{1z}$ does not enter 
due to the invariance under longitudinal boosts, while the direction of $\vec{p}_{1T}$ is irrelevant due to the azimuthal symmetry.

\begin{figure}[t]
 \centering
 \includegraphics[width=.45\textwidth]{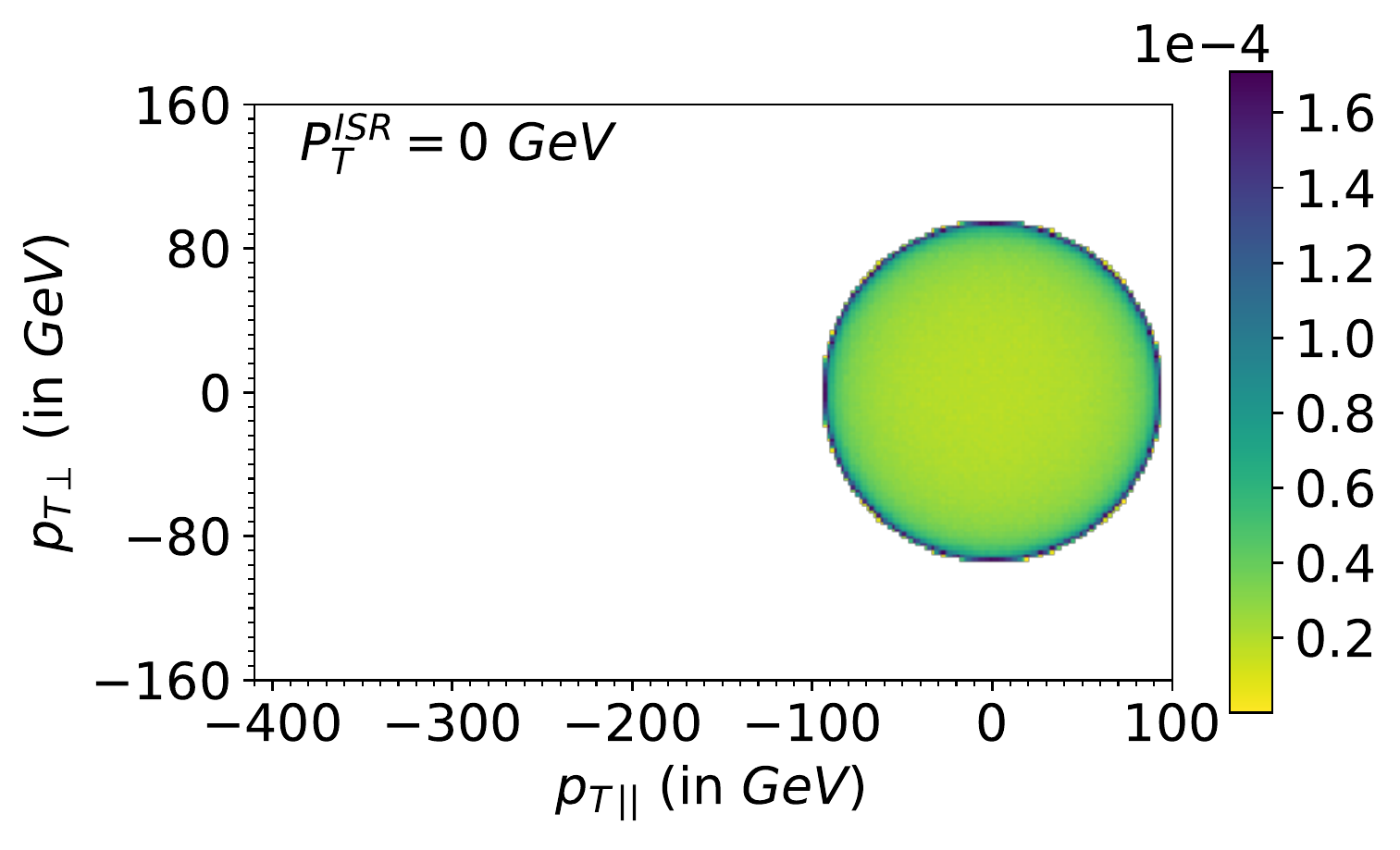}
 \hskip 5mm
 \includegraphics[width=.45\textwidth]{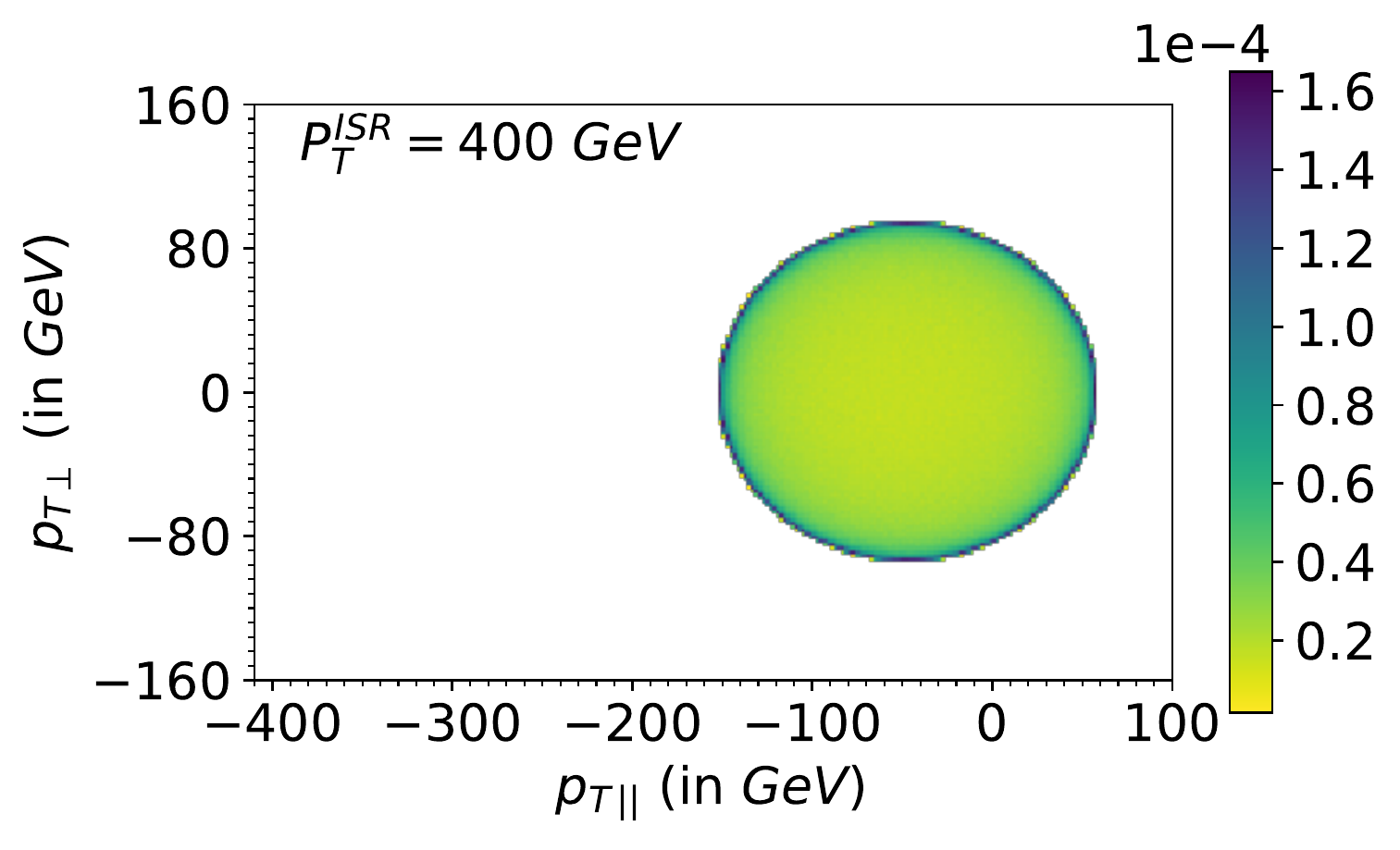}
 \\
 \includegraphics[width=.45\textwidth]{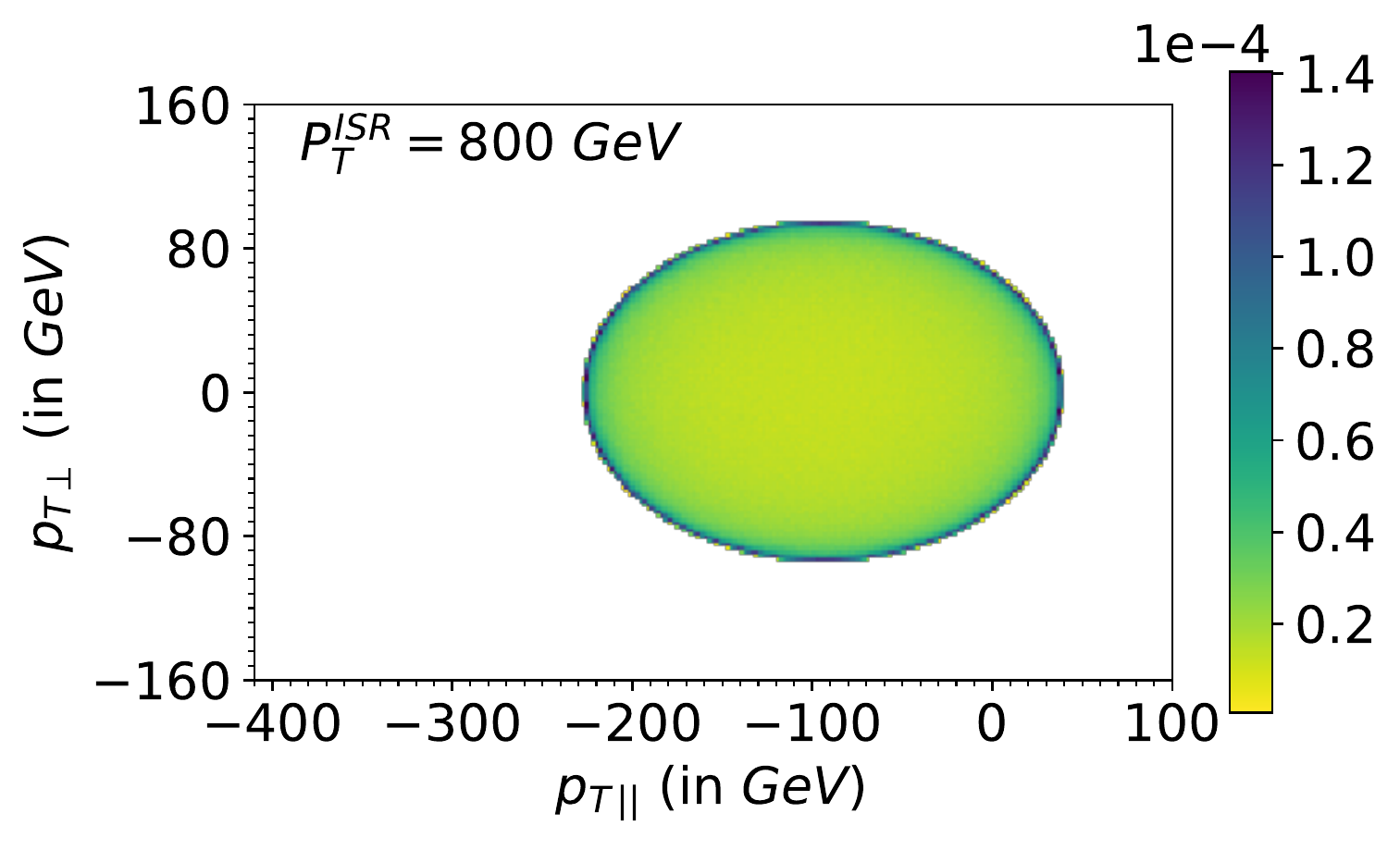}
 \hskip 5mm
 \includegraphics[width=.45\textwidth]{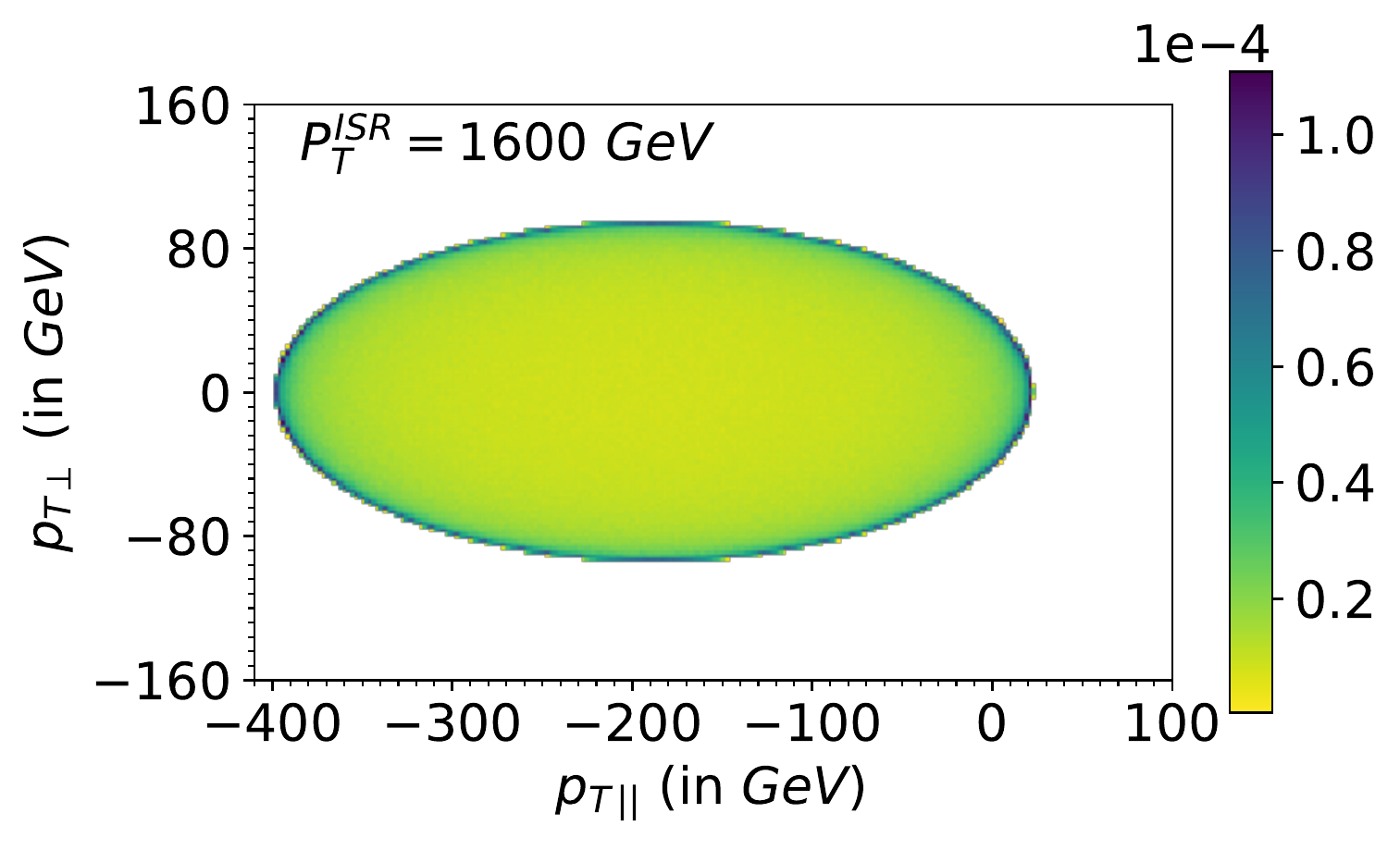}
\caption{\label{fig:1stepISR} Allowed values for the transverse momentum of the visible particle (assumed massless for simplicity)
for different values of $P_T^{ISR}$:
$P_T^{ISR}=0$ (top left), $P_T^{ISR}=400$ GeV (top right),
$P_T^{ISR}=800$ GeV (bottom left) and $P_T^{ISR}=1600$ GeV (bottom right). 
Following Refs.~\cite{Matchev:2009ad,Konar:2009wn}, the component in the direction of $\vec{P}_T^{ISR}$ 
is denoted by $p_{T\parallel}$ and defined in (\ref{eq:ptpardef}), while 
the component in a direction orthogonal to $\vec{P}_T^{ISR}$ is denoted by $p_{T\perp}$ and defined in (\ref{eq:ptperpdef}). 
The color scale is indicative of the event number density. 
In particular, points colored in deep purple mark the locations in phase space where the number density becomes singular.
}
\end{figure}

\begin{figure}[t]
 \centering
 \includegraphics[width=.3\textwidth]{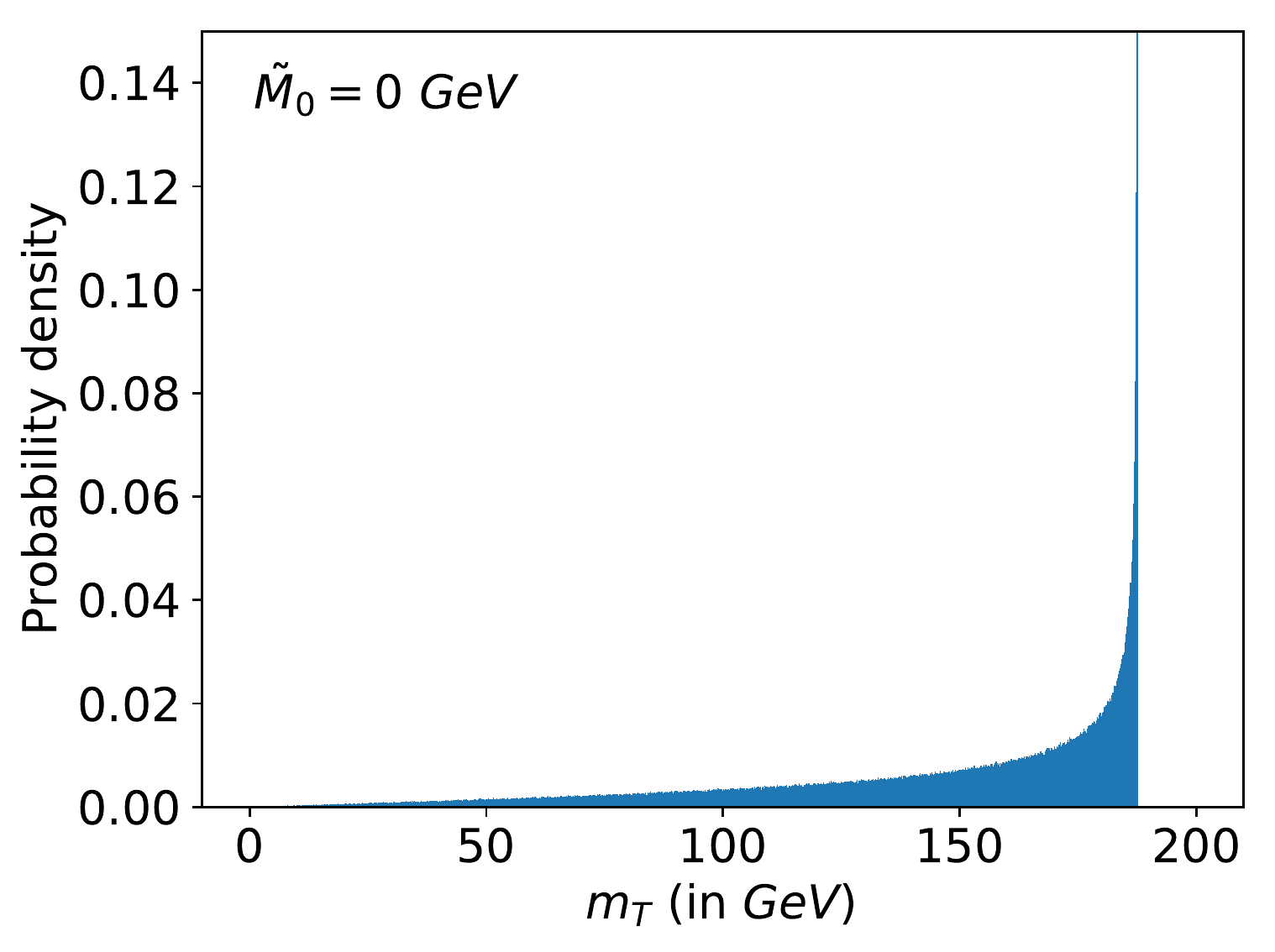}
 \hskip 5mm
 \includegraphics[width=.3\textwidth]{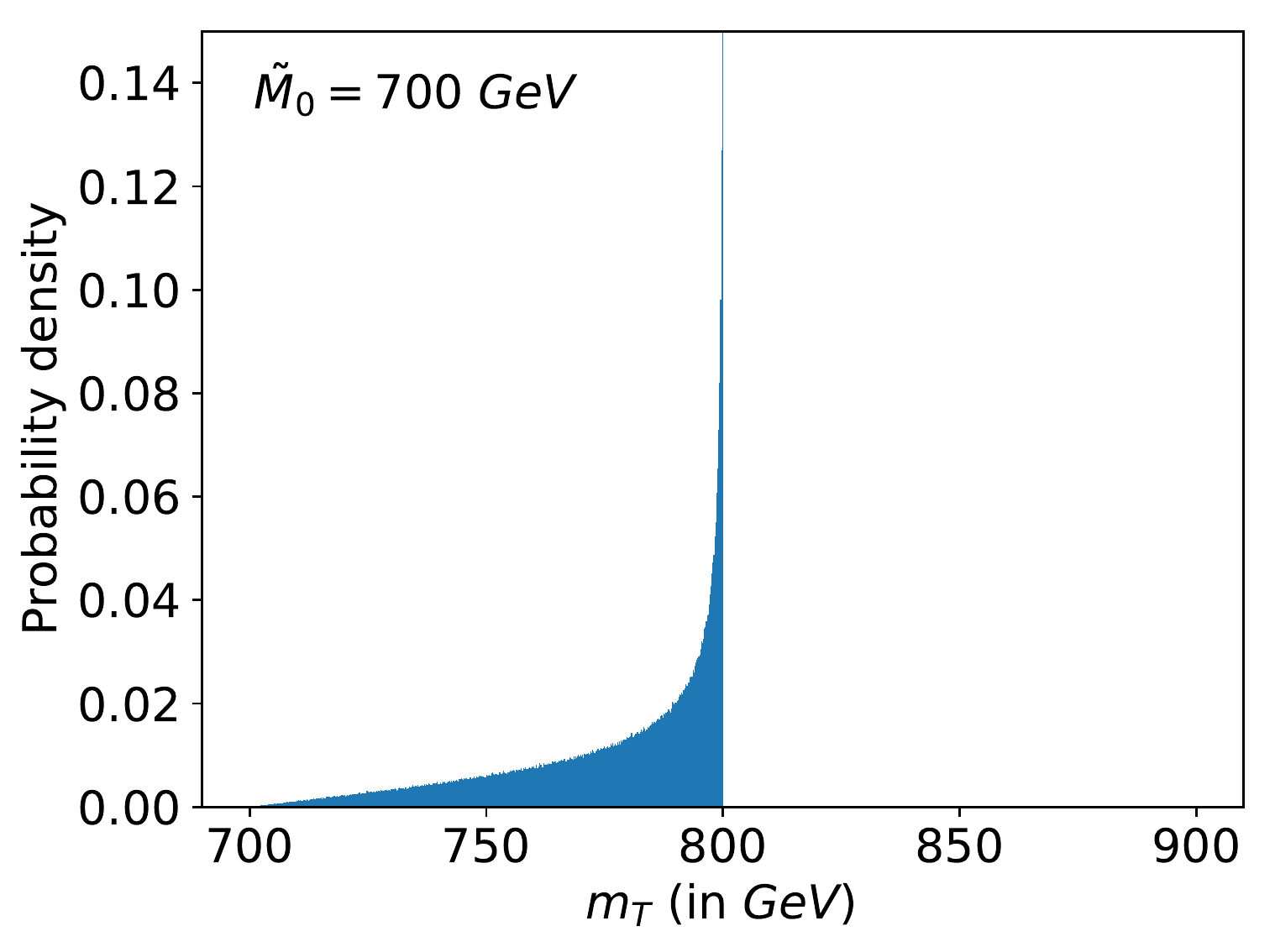}
 \hskip 5mm
 \includegraphics[width=.3\textwidth]{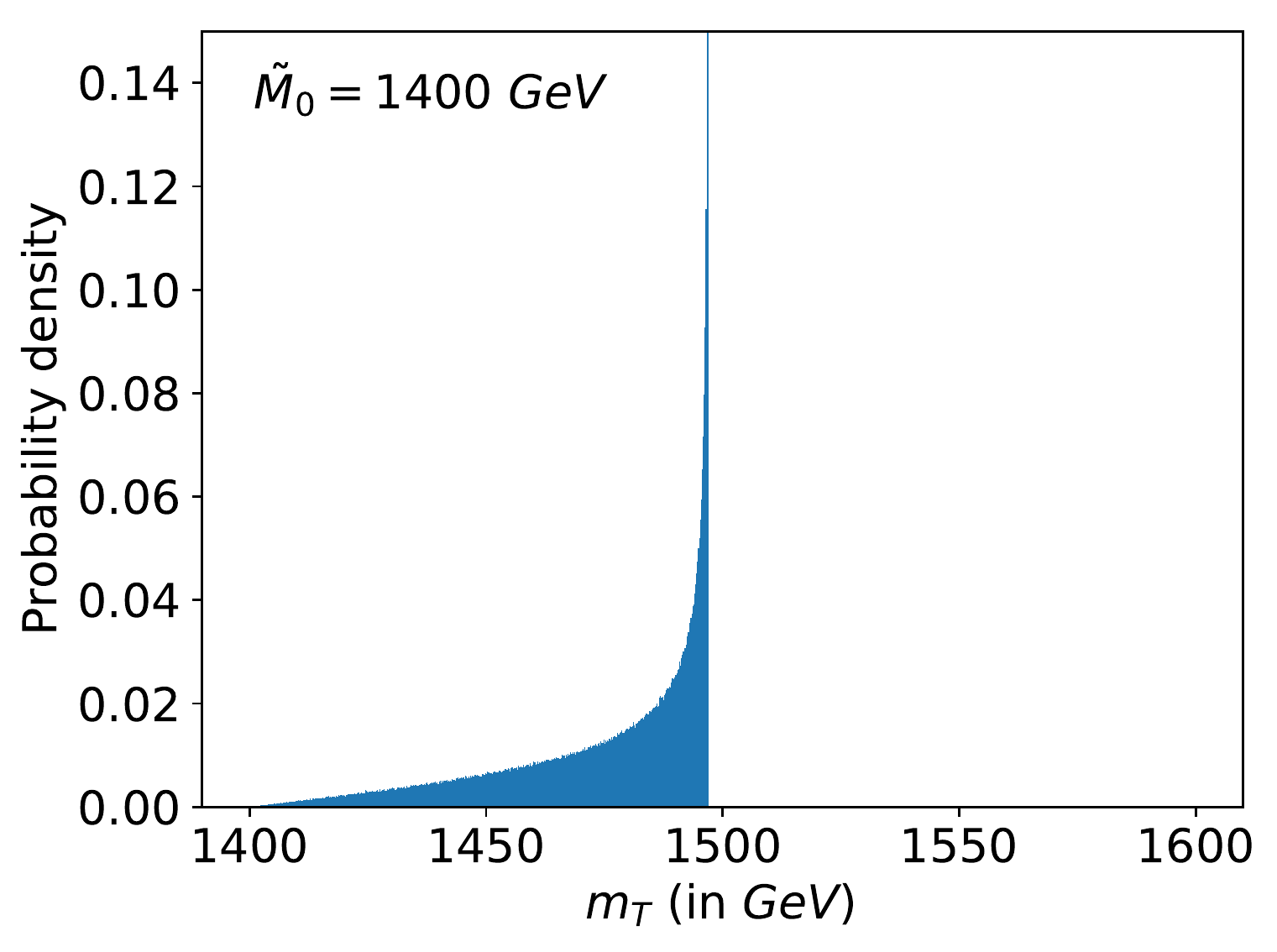}
\caption{\label{fig:1stepMTdistributions} Unit-normalized distributions of the relevant transverse mass (\ref{amTnoISR}) for the case of $P_T^{ISR}=0$ and with the
mass spectrum from Table~\ref{tab:mass}: $M_1=800$ GeV and $M_0=700$ GeV.
The test mass is chosen as $\tilde M_0=0$ GeV (left panel), $\tilde M_0=700$ GeV (middle panel)  and $\tilde M_0=1400$ GeV (right panel).  
In each case, the upper kinematic endpoint of the $m_T$ distribution is given by (\ref{eq:singlocptisreq0}).
}
\end{figure}

The singularity condition (\ref{mTeqM1}) can then be written as
\beq
\sqrt{m_1^2+p_{1T}^2} + \sqrt{M_0^2+p_{1T}^2} = M_1.
\label{asingPTISR0}
\eeq
The set of points satisfying this relation belong to a circle in the $\vec{p}_{1T}$ plane which is centered at the origin and has radius  equal to
\beq
p_{1T}^{max} = \frac{\sqrt{\lambda(M_1^2,M_0^2,m_1^2)}}{2M_1},
\label{eq:p1Tradius}
\eeq
where $\lambda(x,y,z)\equiv x^2+y^2+z^2-2xy-2xz-2yz$.
Such points were categorized as ``extreme" in \cite{Kim:2019prx}, since they delineate the 
boundary of the allowed phase space, where the solutions for the invisible momenta become degenerate. 
This is pictorially illustrated\footnote{For simplicity, in Fig.~\ref{fig:1stepISR} the visible particle is assumed massless ($m_1=0$)
in which case (\ref{eq:p1Tradius}) reduces to
\beq
p_{1T}^{max} = \frac{M_1^2-M_0^2}{2M_1}.
\label{p1Tmaxm1eq0}
\eeq
} in Fig.~\ref{fig:1stepISR}, where
the top left panel corresponds to the current case of $P_T^{ISR}=0$.
We plot the event number density (as indicated by the color bar) in the $\vec{p}_{1T}$ plane, which we choose to
parametrize as $(p_{T\parallel},p_{T\perp})$, where 
$p_{T\parallel}$ ($p_{T\perp}$) is the component in the direction along (orthogonal to) $\vec{P}_T^{ISR}$ 
\cite{Matchev:2009ad,Konar:2009wn}.
We see that the allowed phase space in the $\vec{p}_{1T}$ plane is indeed a circle, and furthermore, 
that the maximal event density is found along the circumference of the circle, in agreement with (\ref{asingPTISR0}).
With the mass spectrum from Table~\ref{tab:mass}, eq.~(\ref{eq:p1Tradius}) predicts the radius of the circle to be $p_{1T}^{max}=93.75$ GeV, 
which is confirmed in the top left panel of Fig.~\ref{fig:1stepISR}.
Since the extreme events along the circumference of the circle have the same value of $p_{1T}=p_{1T}^{max}$, 
they will also share the same value of $m_T$, regardless of the choice of test mass $\tilde M_0$, see eq.~(\ref{amTnoISR}).
This means that for {\em any} value of $\tilde M_0$, the $m_T(\tilde M_0)$ distribution will continue to exhibit a singularity at 
\beq
m_T^{max}(\tilde M_0) =  \sqrt{m_1^2+(p_{1T}^{max})^2} + \sqrt{\tilde M_0^2+(p_{1T}^{max})^2}.
\label{eq:singlocptisreq0}
\eeq
This is illustrated explicitly in Fig.~\ref{fig:1stepMTdistributions}, which shows
several unit-normalized distributions of the transverse mass (\ref{amTnoISR}) for the case of $P_T^{ISR}=0$ and with the
mass spectrum from Table~\ref{tab:mass} ($M_1=800$ GeV and $M_0=700$ GeV).
The test mass is chosen as $\tilde M_0=0$ GeV in the left panel, $\tilde M_0=700$ GeV in the middle panel  and $\tilde M_0=1400$ GeV in the right panel.  
We see that in all three cases, the $m_T$ distribution has a very sharp singularity at its upper kinematic endpoint (\ref{eq:singlocptisreq0}).
However, this situation is rather atypical --- we shall see below that, in general, when we use the wrong value for the 
invisible mass parameter $\tilde M_0$, the singularity will be washed out. The reason why it persists here
is that the singularity coordinate (\ref{amTnoISR}) is parametrized by a single degree of freedom, $p_{1T}$.

\subsubsection{The case with non-zero upstream visible momentum: $P_T^{ISR}\ne0$}
\label{sec:PTISR}

We are now in position to discuss the more general case of non-vanishing upstream visible momentum, $\vec{P}_T^{ISR}\ne 0$.
Using (\ref{mptdefgen}), the transverse mass formula (\ref{amtdef2}) becomes \cite{Barr:2011xt}
\beq
m^2_T(\tilde M_0) 
 = \left[\sqrt{m_1^2+p_{1T}^2} + \sqrt{\tilde M_0^2+(\vec{p}_{1T}+\vec{P}_T^{ISR})^2 }\right]^2 - (P_T^{ISR})^2.
 \label{amTwithISR}
\eeq 
Since $\vec{P}_T^{ISR}$ breaks the azimuthal symmetry, the transverse mass is now a function of two 
visible momentum degrees of freedom, namely both the magnitude and the direction of $\vec{p}_{1T}$ in the transverse plane.
We can parametrize the latter degree of freedom by the angle $\varphi$ measured with respect to the direction defined by $\vec{P}_T^{ISR}$,
in which case the doubly projected transverse components of $\vec{p}_{1T}$ used in Fig.~\ref{fig:1stepISR}
are given by 
\bea
p_{T\parallel} &\equiv & p_{1T} \cos\varphi ,  \label{eq:ptpardef}\\ [2mm]
p_{T\perp} &\equiv & p_{1T} \sin\varphi . \label{eq:ptperpdef}
\eea

\begin{figure}[t]
 \centering
 \includegraphics[width=.4\textwidth]{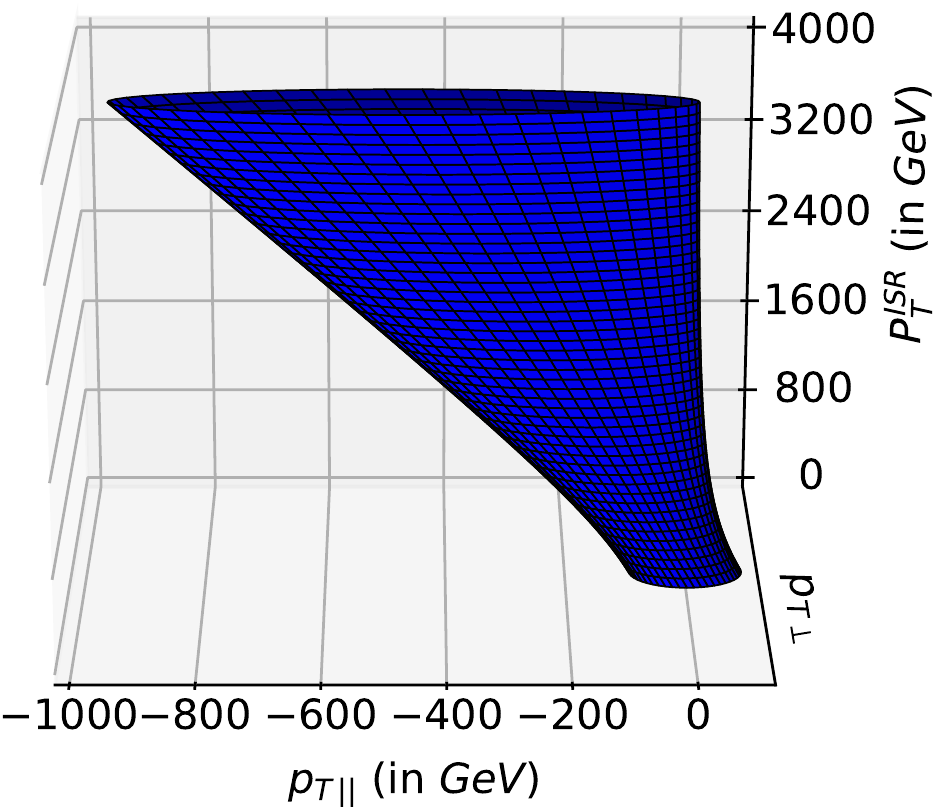}
  \hskip 5mm 
 \includegraphics[width=.4\textwidth]{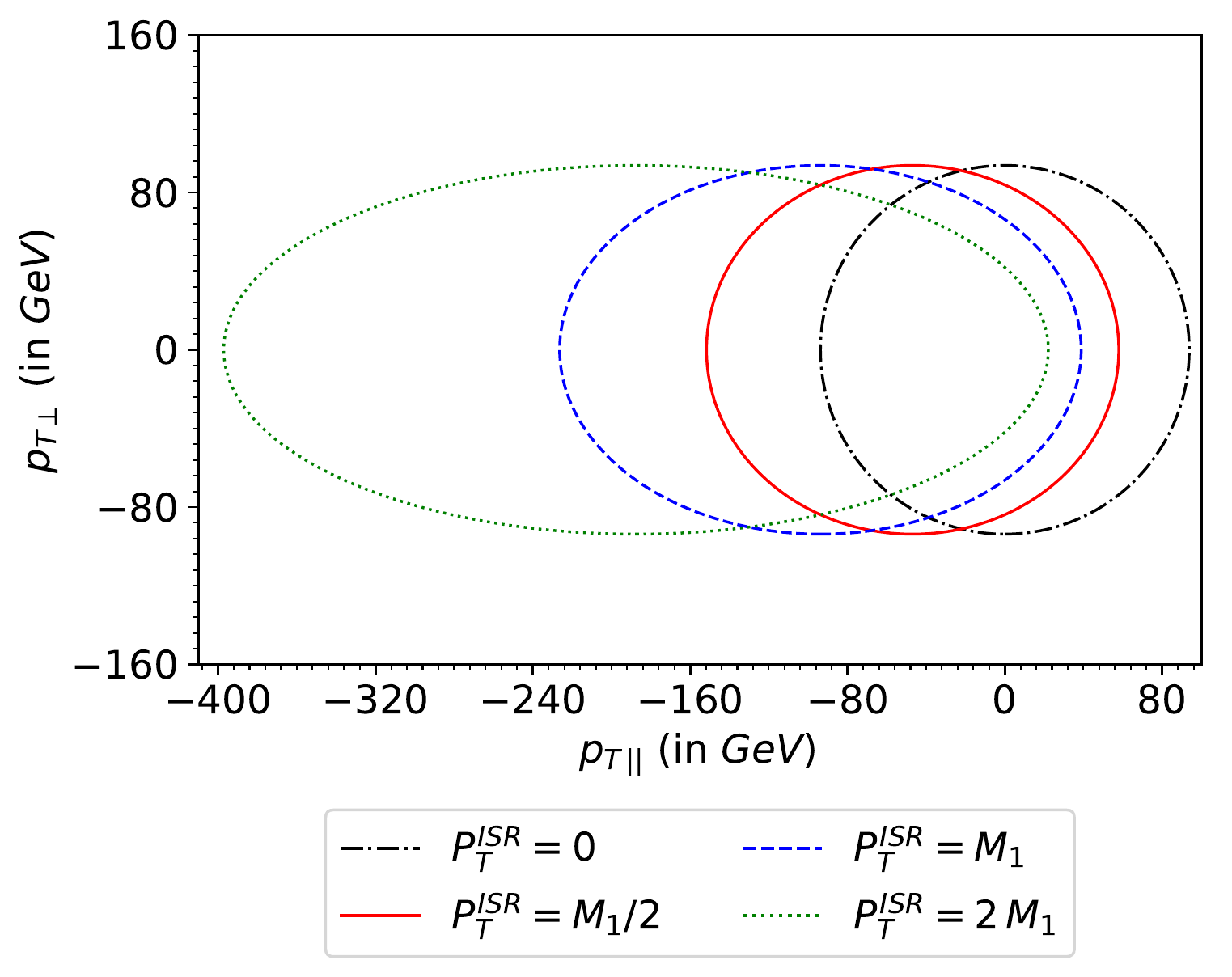}
 \caption{\label{fig:ellipses} 
 Illustration of the geometry of the singularity condition (\ref{eq:ellipse}) for the mass spectrum 
 $M_1=800$ GeV and $M_0=700$ GeV.
Left panel: a 3-dimensional plot in the $(p_{T\parallel},p_{T\perp},P_T^{ISR})$ space 
 of the two-dimensional singularity surface defined by eq.~(\ref{eq:ellipse}). 
 Right panel: the boundary ellipses from Fig.~\ref{fig:1stepISR} which were 
 obtained at fixed values of $P_T^{ISR}=\{0, M_1/2,M_1,2M_1\}$.  }
\end{figure}

As before, let us find the locus of points which satisfy the singularity condition (\ref{mTeqM1}),
now in the presence of non-zero $P_T^{ISR}$. For simplicity, let us only focus on massless visible particles, $m_1=0$,
which is an excellent approximation for leptons and jets. In that case, the singularity condition (\ref{mTeqM1}) reads
\beq
M_0^2 + 2p_{1T}
\left(
\sqrt{M_0^2 +p_{1T}^2+ (P_T^{ISR})^2 + 2p_{1T}P_T^{ISR}\cos\varphi}+p_{1T}+P_T^{ISR}\cos\varphi
\right)=M_1^2,
\label{eq:amtM0eqM1}
\eeq
which can be solved to give the location of the singularity surface in parametric form (see the left panel in Fig.~\ref{fig:ellipses})
\beq
p_{1T}^{max}(\varphi)=\frac{M_1^2-M_0^2}{2\left(\sqrt{M_1^2+(P_T^{ISR})^2}+P_T^{ISR}\cos\varphi\right)}.
\label{eq:ellipse}
\eeq
As a consistency check, we see that eq.~(\ref{eq:ellipse}) reduces to (\ref{p1Tmaxm1eq0}) in the limit of $P_T^{ISR}\to 0$.
For a fixed value of $P_T^{ISR}$, the result (\ref{eq:ellipse}) can be recognized as the equation of an ellipse \cite{Byckling:1971vca}
centered at $(-c,0)$, where $c$ is the linear eccentricity
\beq
c \equiv \sqrt{a^2-b^2} =  \frac{M_1^2-M_0^2}{2M_1^2}\, P_T^{ISR},
\label{eq:eccentricity}
\eeq 
while $a$ and $b$ are the semi-major and semi-minor axes, respectively:
\bea
a &=&  \frac{M_1^2-M_0^2}{2M_1}\, \sqrt{1+\left(\frac{P_T^{ISR}}{M_1}\right)^2}, \label{eq:a-axis}\\ [2mm]
b &=&  \frac{M_1^2-M_0^2}{2M_1}. 
\label{eq:b-axis}
\eea
Note that the eccentricity $c$ is linearly proportional to $P_T^{ISR}$, 
thus in the case of $P_T^{ISR}\to 0$ considered in the previous Section~\ref{sec:noPTISR},
the ellipse (\ref{eq:ellipse}) became the circle (\ref{p1Tmaxm1eq0}). 
The results (\ref{eq:ellipse}-\ref{eq:b-axis}) are visualized in the remaining three panels of Fig.~\ref{fig:1stepISR}, 
where we plot the allowed values for $\vec{p}_{1T}=(p_{T\parallel},p_{T\perp})$
for several fixed non-zero values of $P_T^{ISR}$: $P_T^{ISR}=M_1/2=400$ GeV (top right panel),
$P_T^{ISR}=M_1=800$ GeV (bottom left panel) and $P_T^{ISR}=2M_1=1600$ GeV (bottom right panel). 
Since the upstream visible momentum $\vec{P}_T^{ISR}$ is always oriented along the positive 
$p_{T\parallel}$ axis, the recoil of the mother particle is in the negative $p_{T\parallel}$ direction,
which explains the increasing preference for negative $p_{T\parallel}$ values as $P_T^{ISR}$ gets larger (see also the right panel in Fig.~\ref{fig:ellipses}).
At the same time, the (doubly) transverse components $p_{T\perp}$ are unaffected by the boost of the mother particle,
and the maximal $p_{T\perp}$ value in each panel stays the same.
This is also reflected in the fact that the semi-minor axis (\ref{eq:b-axis}) of the ellipse remains constant, independent of $P_T^{ISR}$.

In deriving the equation of the singularity surface (\ref{eq:ellipse}), we have achieved our main goal for this subsection.
From here on, how the result (\ref{eq:ellipse}) will be used in practice, depends on the specific purpose of the experimental analysis.
If the aim is a {\em discovery} of signal events with the event topology of  Fig.~\ref{fig:feynmandiag}(a), 
one should study the distribution of events in the three-dimensional space $(p_{T\parallel},p_{T\perp},P_T^{ISR})$
depicted in the left panel of Fig.~\ref{fig:ellipses}, where the signal-rich regions will be found at the locations singled out by eq.~(\ref{eq:ellipse}).
If, on the other hand, the goal is to {\em measure the mass spectrum}, the singularity surface contains 
all the kinematic information to do that as well, and the masses $M_0$ and $M_1$ can be extracted from a parameter fit
to eq.~(\ref{eq:ellipse}). 

The fitting procedure will have to account for the different available statistics at different values of $P_T^{ISR}$. 
Operationally this can be accomplished as follows (following on an idea from Ref.~\cite{Matchev:2009fh}).
One can select a subset of events with (approximately) the same value of $P_T^{ISR}$,
and plot them in the $\vec{p}_{1T}=(p_{T\parallel},p_{T\perp})$ plane as in Fig.~\ref{fig:1stepISR}.
The signal events will exhibit an overdensity along the singularity ellipse given by (\ref{eq:ellipse}), as also illustrated in the right panel in Fig.~\ref{fig:ellipses}. 
Then, by fitting to (\ref{eq:ellipse}), one can find the distances from the focus to the two vertices of the ellipse, i.e., the maximum value of $p_{1T}$ 
in the direction along $\vec{P}_T^{ISR}$ and in the direction opposite to $\vec{P}_T^{ISR}$:
\bea
p_{1T}^{max}(\varphi=0) &=& \frac{M_1^2-M_0^2}{2\left(\sqrt{M_1^2+(P_T^{ISR})^2}+P_T^{ISR}\right)}, \label{eq:apt0}\\
p_{1T}^{max}(\varphi=\pi) &=& \frac{M_1^2-M_0^2}{2\left(\sqrt{M_1^2+(P_T^{ISR})^2}-P_T^{ISR}\right)}. \label{eq:aptpi}
\eea
These two relations can be easily inverted and solved for $M_1$ and $M_0$:
\beq
M_1 = 2 P_T^{ISR}\, \frac{\sqrt{p_{1T}^{max}(\varphi=\pi)p_{1T}^{max}(\varphi=0)}}{p_{1T}^{max}(\varphi=\pi)-p_{1T}^{max}(\varphi=0)}, 
\eeq
\beq
M_0 = 2 P_T^{ISR}\, \frac{\sqrt{p_{1T}^{max}(\varphi=\pi)p_{1T}^{max}(\varphi=0)}}{p_{1T}^{max}(\varphi=\pi)-p_{1T}^{max}(\varphi=0)}
\, \sqrt{1-\frac{p_{1T}^{max}(\varphi=\pi)-p_{1T}^{max}(\varphi=0)}{P_T^{ISR}}}.
\eeq
This demonstrates that the two measurements (\ref{eq:apt0}-\ref{eq:aptpi}) are sufficient to determine the masses $M_1$ and $M_0$
\cite{Matchev:2009fh}. The procedure can be repeated for different ranges of $P_T^{ISR}$, as long as there is sufficient statistics 
to reconstruct the singularity ellipse and from there extract the values of $p_{1T}^{max}(\varphi=0)$ and $p_{1T}^{max}(\varphi=\pi)$.

\subsection{The focus point method} 
\label{sec:aFP}

The discussion in the previous Sec.~\ref{sec:mtsingularity} revealed that for the event topology of Fig.~\ref{fig:feynmandiag}(a),
the location of the singularity is nicely exhibited as a two-dimensional surface in a three-dimensional space of observables 
$(p_{T\parallel},p_{T\perp},P_T^{ISR})$, as shown in the left panel in Fig.~\ref{fig:ellipses}. But is it possible to simplify matters further, e.g.,
by projecting onto an observable space of even lower dimensionality, while retaining all singular features? As an extreme example, 
is it possible to define a {\em single} kinematic variable whose distribution will capture {\em all} of the singular behavior, 
over the whole surface parametrized by eq.~(\ref{eq:ellipse})? In Section~\ref{sec:noPTISR} we saw that for
the special case of $P_T^{ISR}=0$ this is possible, and the relevant kinematic variable was the transverse mass  $m_T(\tilde M_0)$
(regardless of the choice of test mass $\tilde M_0$, see Fig.~\ref{fig:1stepMTdistributions}),
or alternatively, the magnitude $p_{1T}$ of the transverse visible momentum. However, the subsequent discussion in Sec.~\ref{sec:PTISR}
makes it clear that if we wish to continue using $m_T(\tilde M_0)$ in the more general case of $P_T^{ISR}\ne 0$, we run into a problem --- 
the parametrization of the singularity surface (\ref{eq:ellipse}) involves the mass spectrum, and in particular the 
mass $M_0$ of the invisible particle. Therefore, as implied in the singularity condition (\ref{mTeqM1}), the transverse mass $m_T(\tilde M_0)$ will continue 
to be the relevant singularity coordinate, but only for the correct choice of the test mass $\tilde M_0=M_0$,
since only in that case the singularity surface (\ref{eq:ellipse}) is a surface of constant $m_T$. 
If we make a wrong choice for $\tilde M_0$, which is different from the true mass $M_0$, the singularity surface is {\em not} a surface of constant $m_T$,
and therefore, the singular behavior observed in the three-dimensional picture of the left panel in Fig.~\ref{fig:ellipses}
will tend to be washed out in the one-dimensional $m_T(\tilde M_0)$ distribution.

\begin{figure}[t]
 \centering
 \includegraphics[width=.45\textwidth]{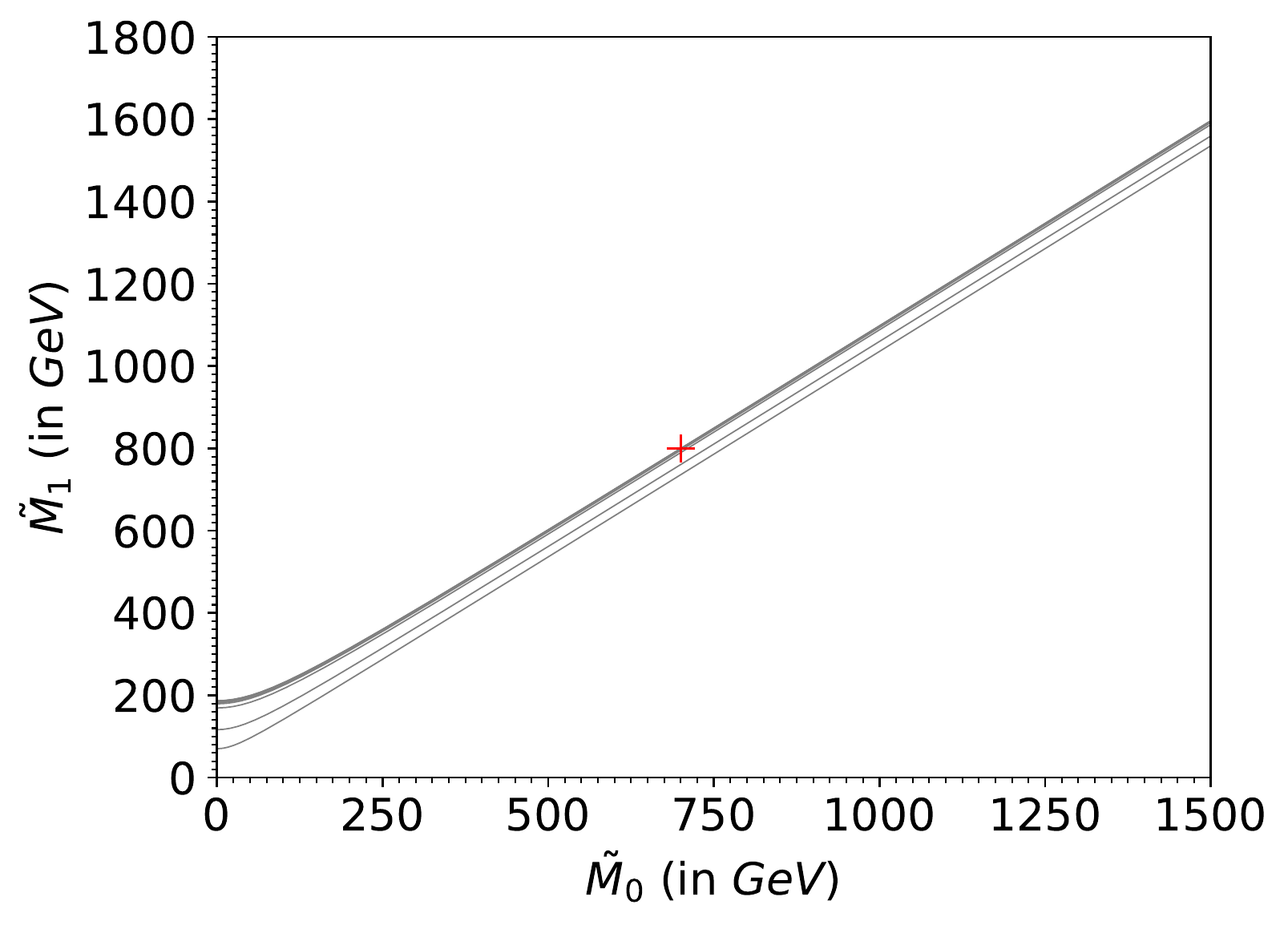}
 \hskip 5mm
 \includegraphics[width=.45\textwidth]{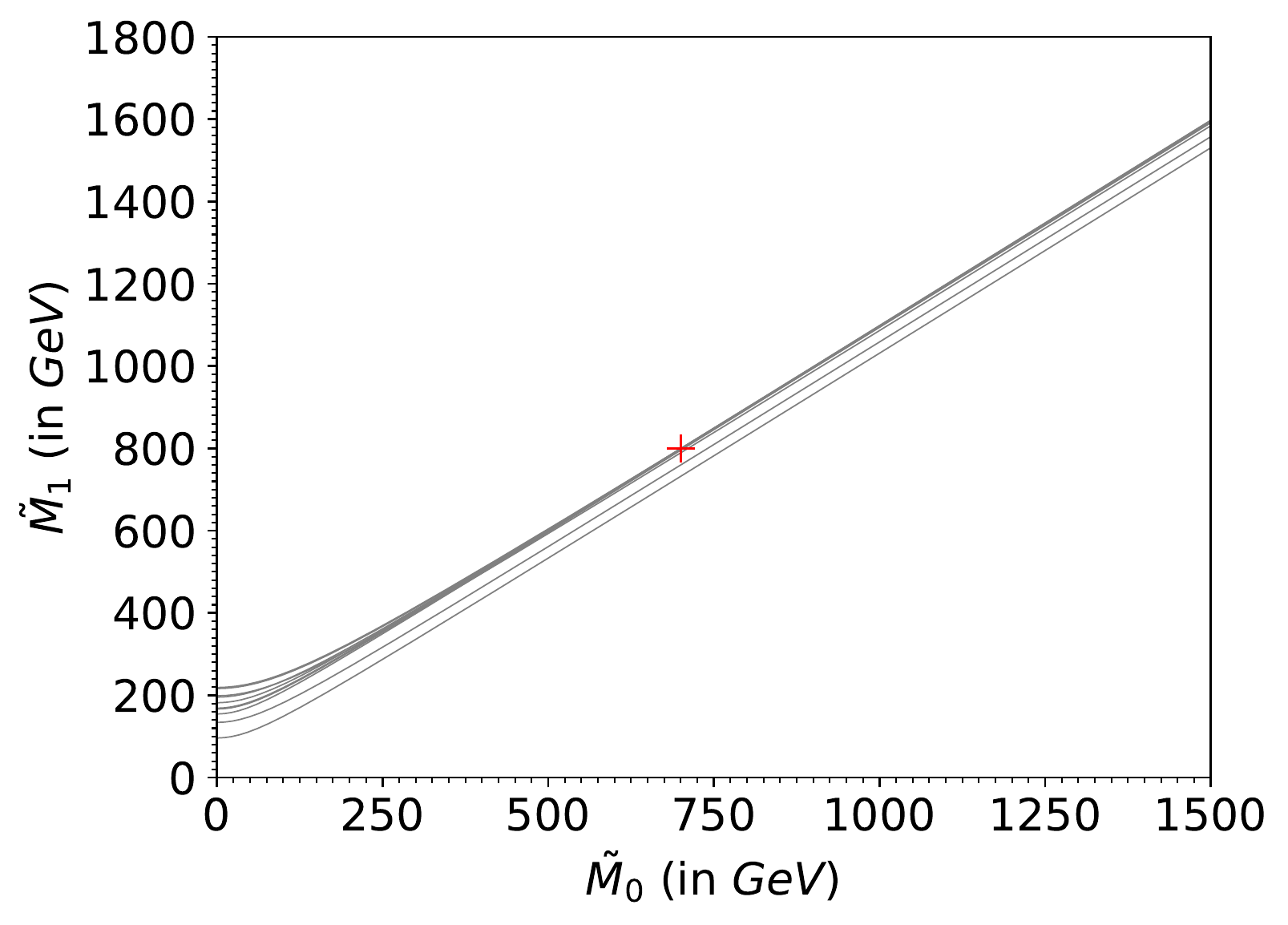}
 \\
 \includegraphics[width=.45\textwidth]{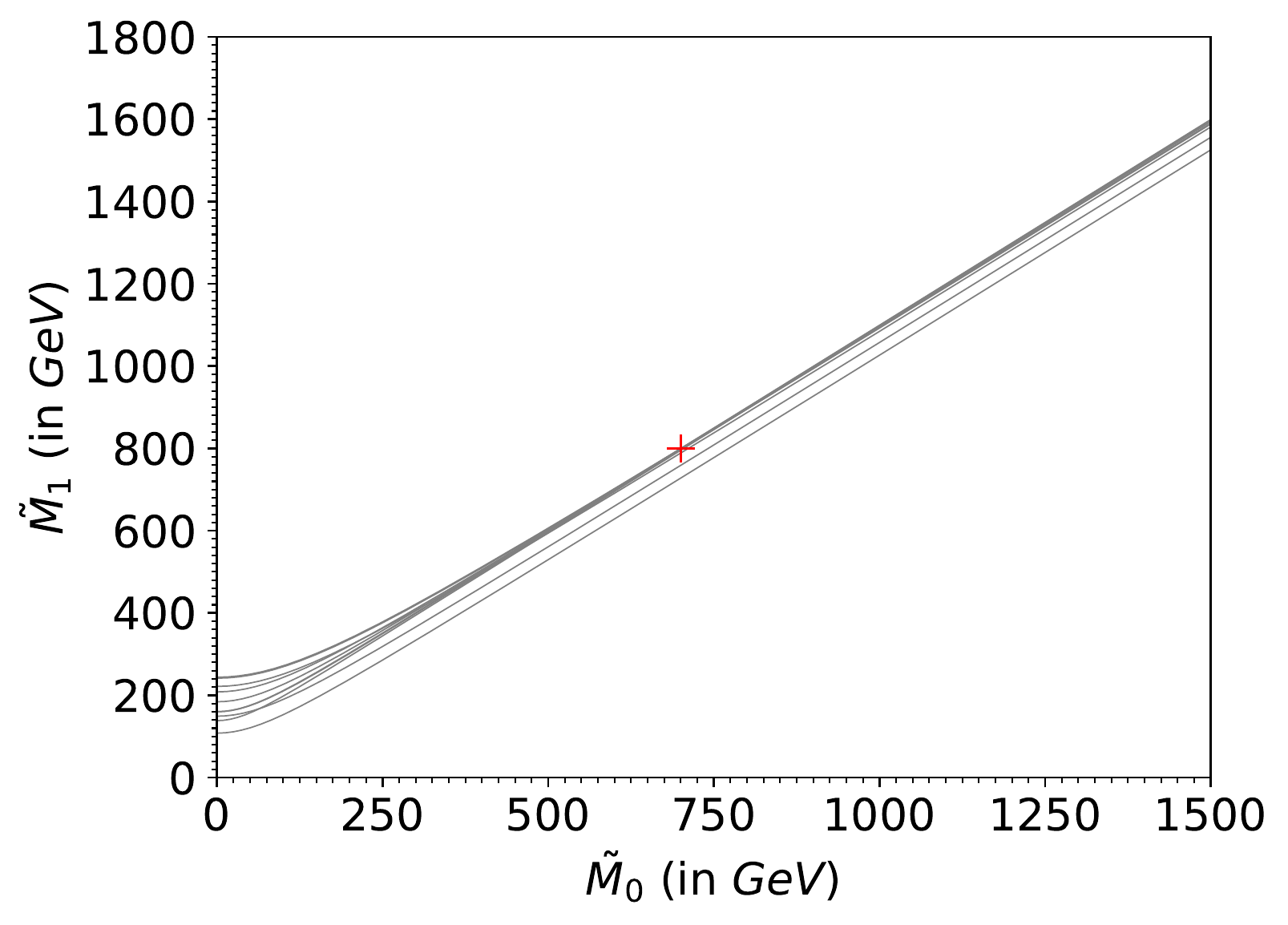}
 \hskip 5mm
 \includegraphics[width=.45\textwidth]{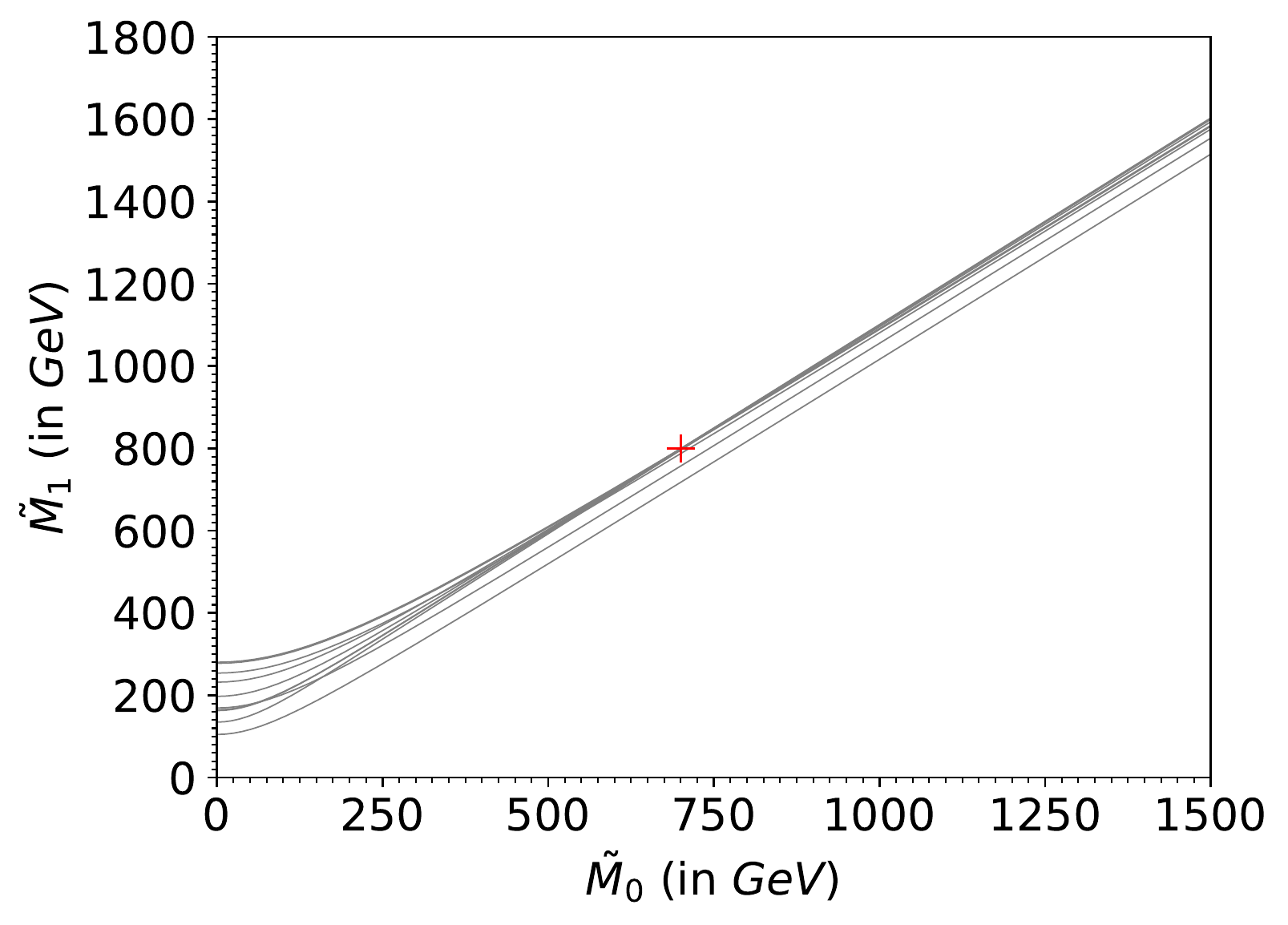}
 \\
 \includegraphics[width=.45\textwidth]{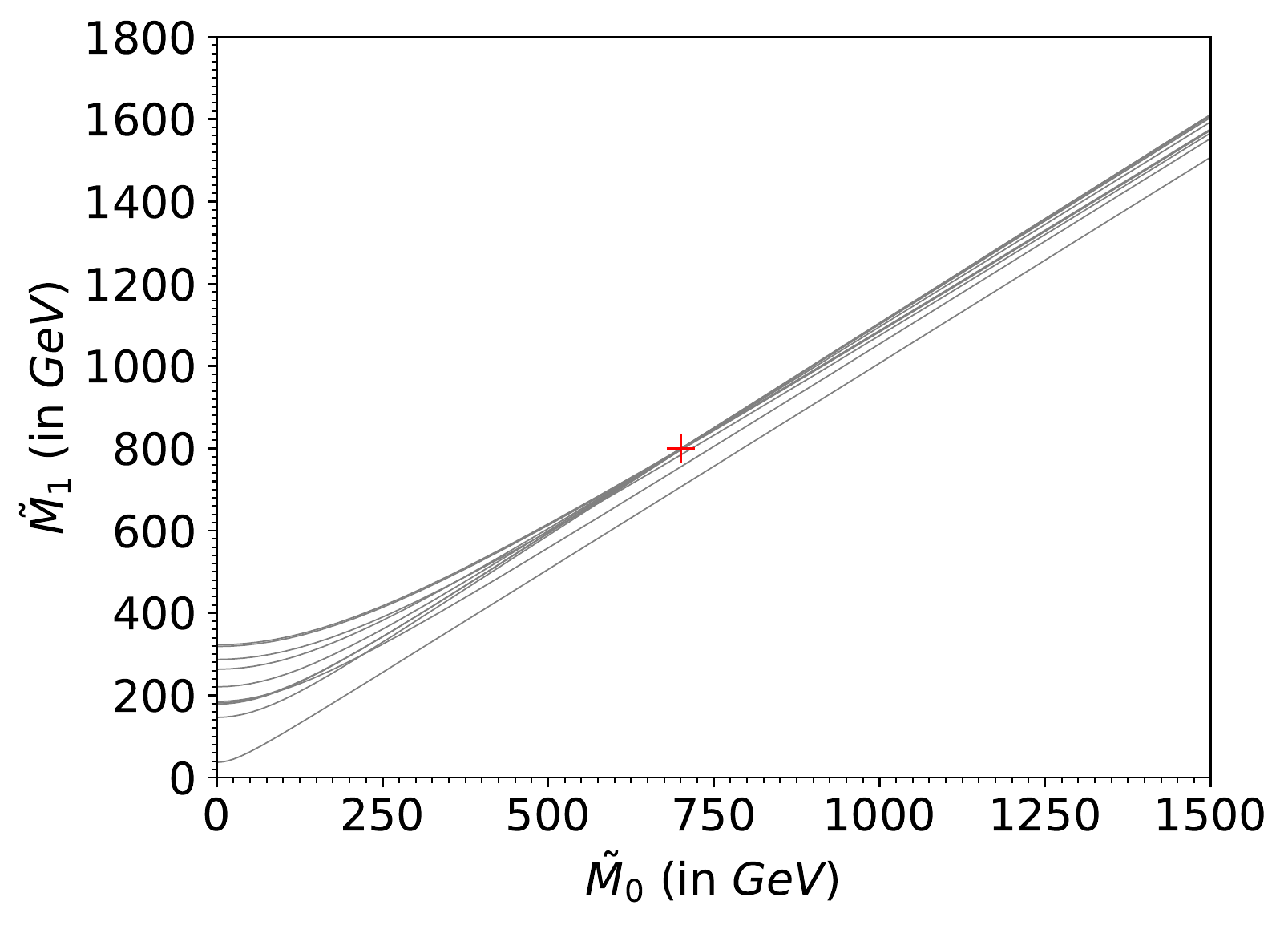}
 \hskip 5mm
 \includegraphics[width=.45\textwidth]{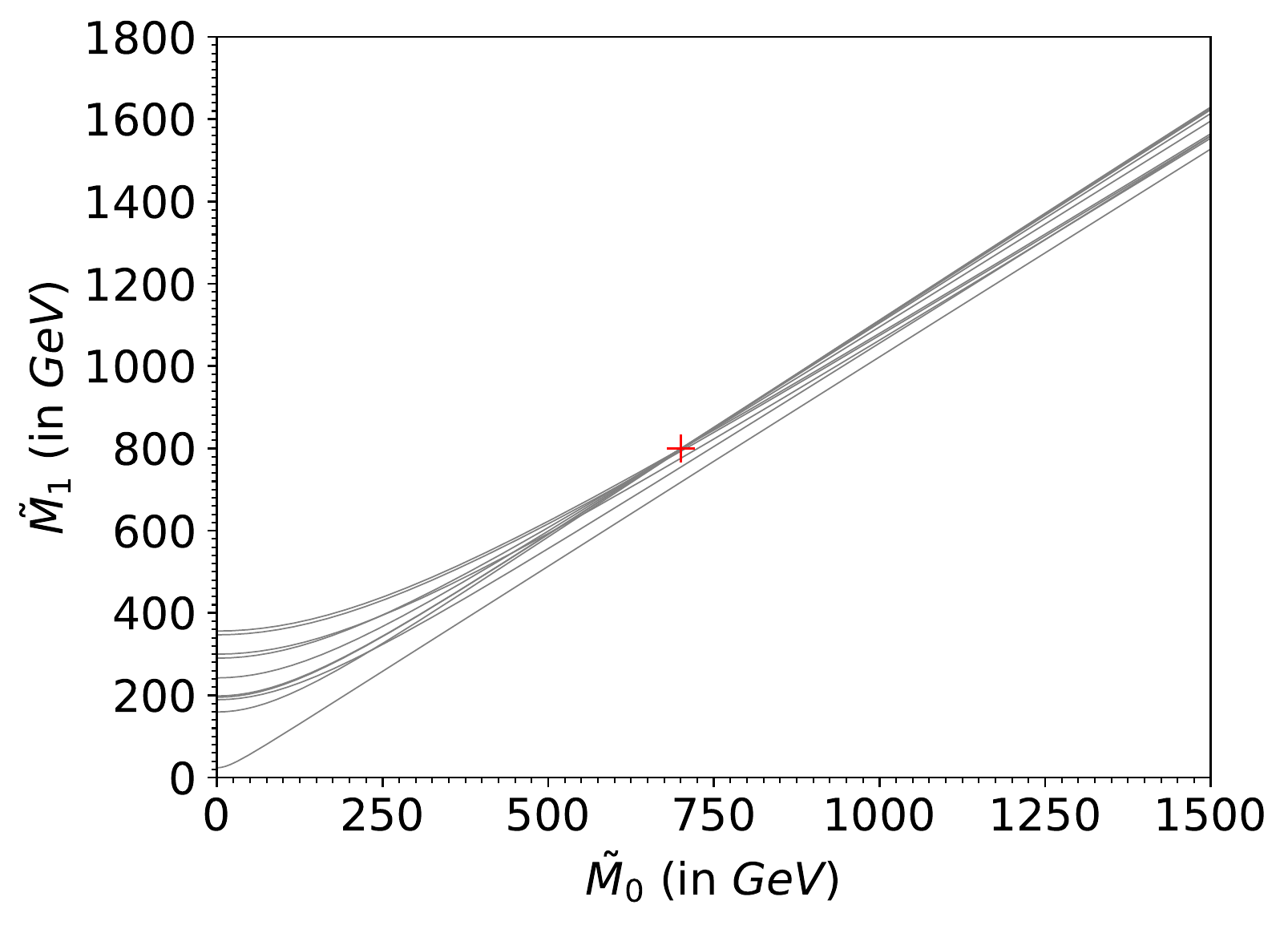}
\caption{\label{fig:1stepFP} Focus point plots in the $(\tilde M_0, \tilde M_1)$ mass parameter space 
for different values of $P_T^{ISR}$:
$P_T^{ISR}=0$ (upper left panel),
$P_T^{ISR}=50$ GeV (upper right panel),
$P_T^{ISR}=100$ GeV (middle left panel),
$P_T^{ISR}=200$ GeV (middle right panel),
$P_T^{ISR}=400$ GeV (lower left panel) and
$P_T^{ISR}=800$ GeV (lower right panel).
The plots show the solvability boundaries (\ref{eq:aM1tvsM0t}) for (the same) 10 randomly chosen events, 
boosted according to the corresponding $P_T^{ISR}$. The red ``$+$" symbol marks the location of the true 
masses $M_0=700$ GeV and $M_1=800$ GeV.}
\end{figure}

\begin{figure}[t]
 \centering
 \includegraphics[width=.45\textwidth]{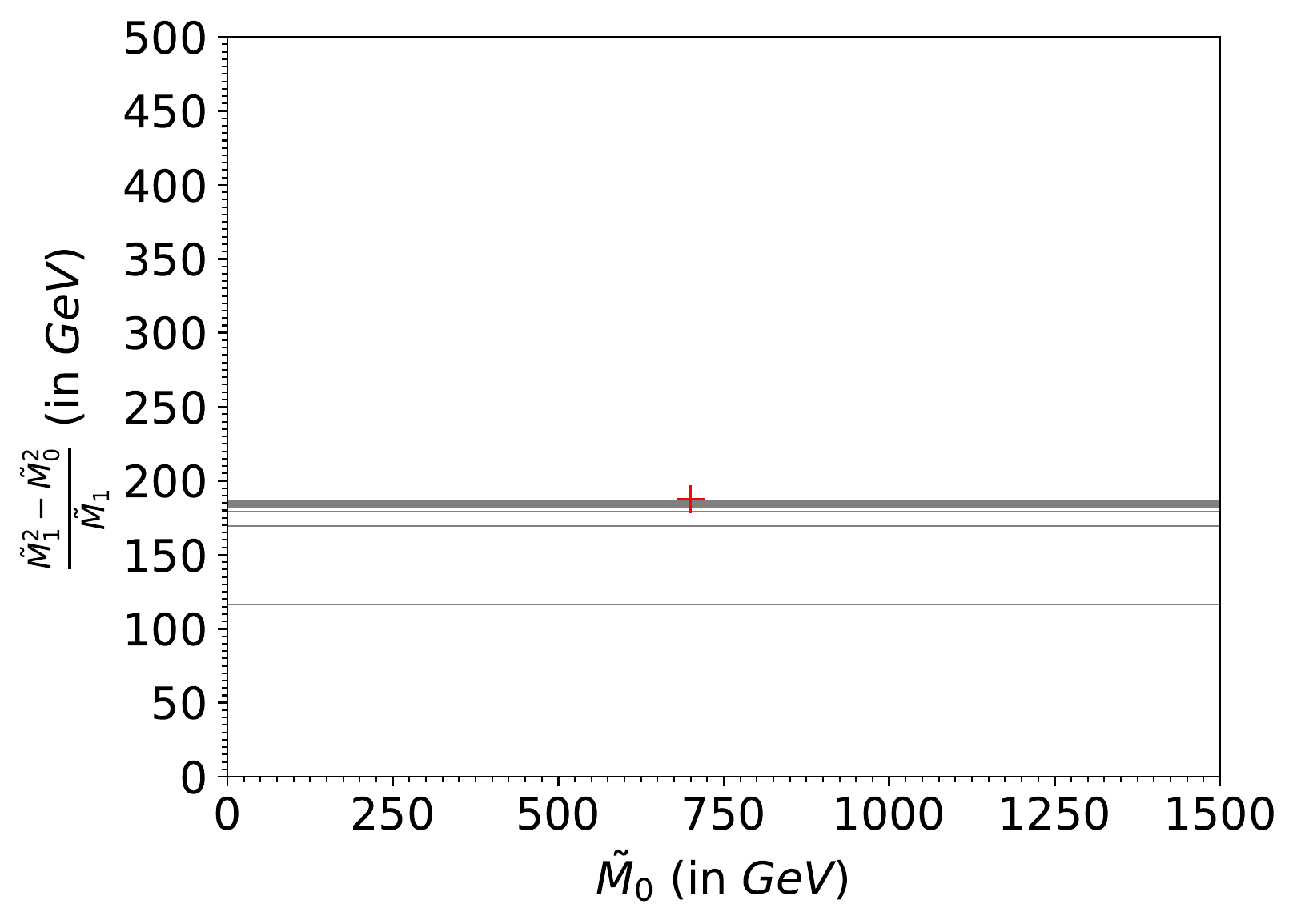}
 \hskip 5mm
 \includegraphics[width=.45\textwidth]{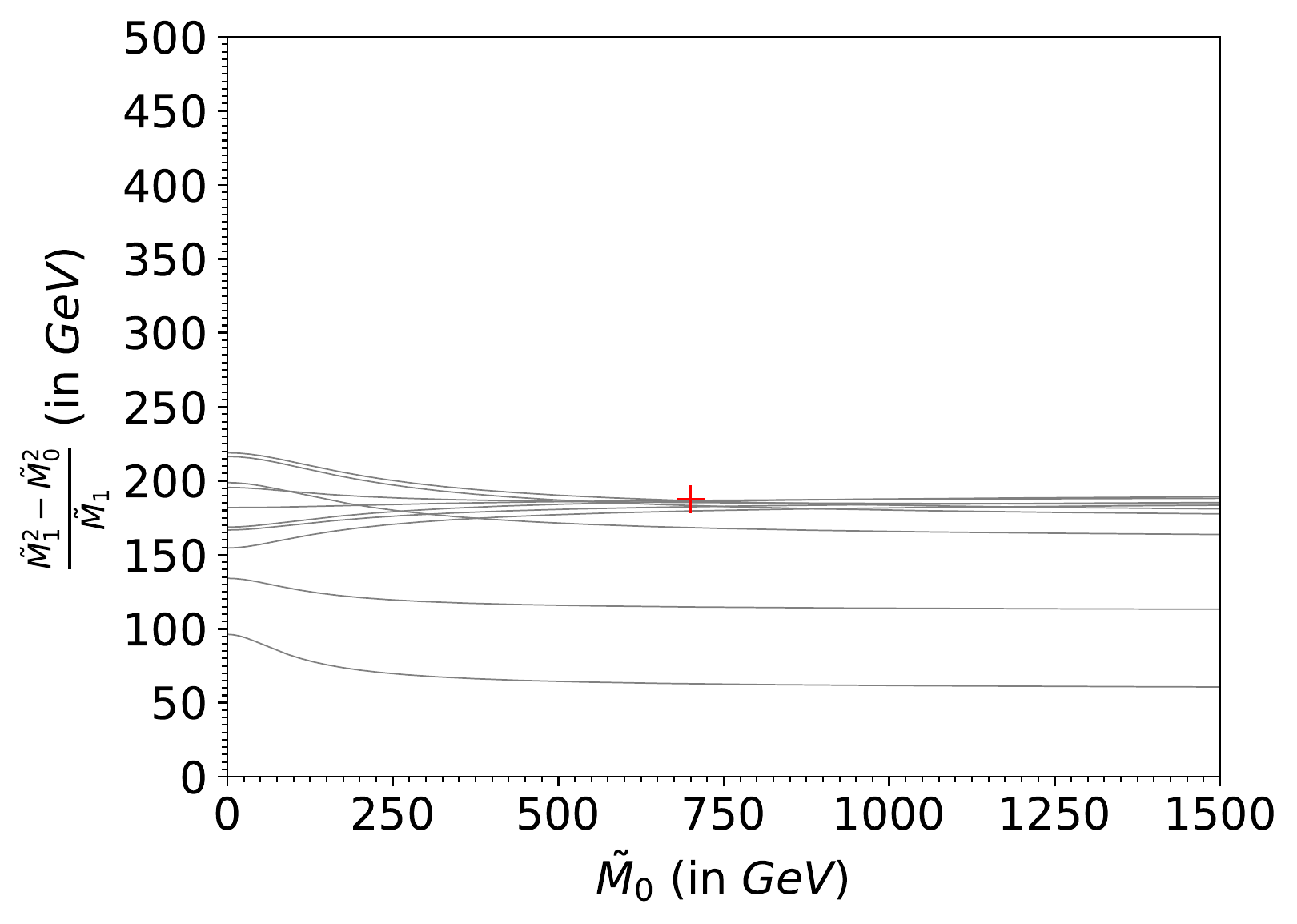}
 \\
 \includegraphics[width=.45\textwidth]{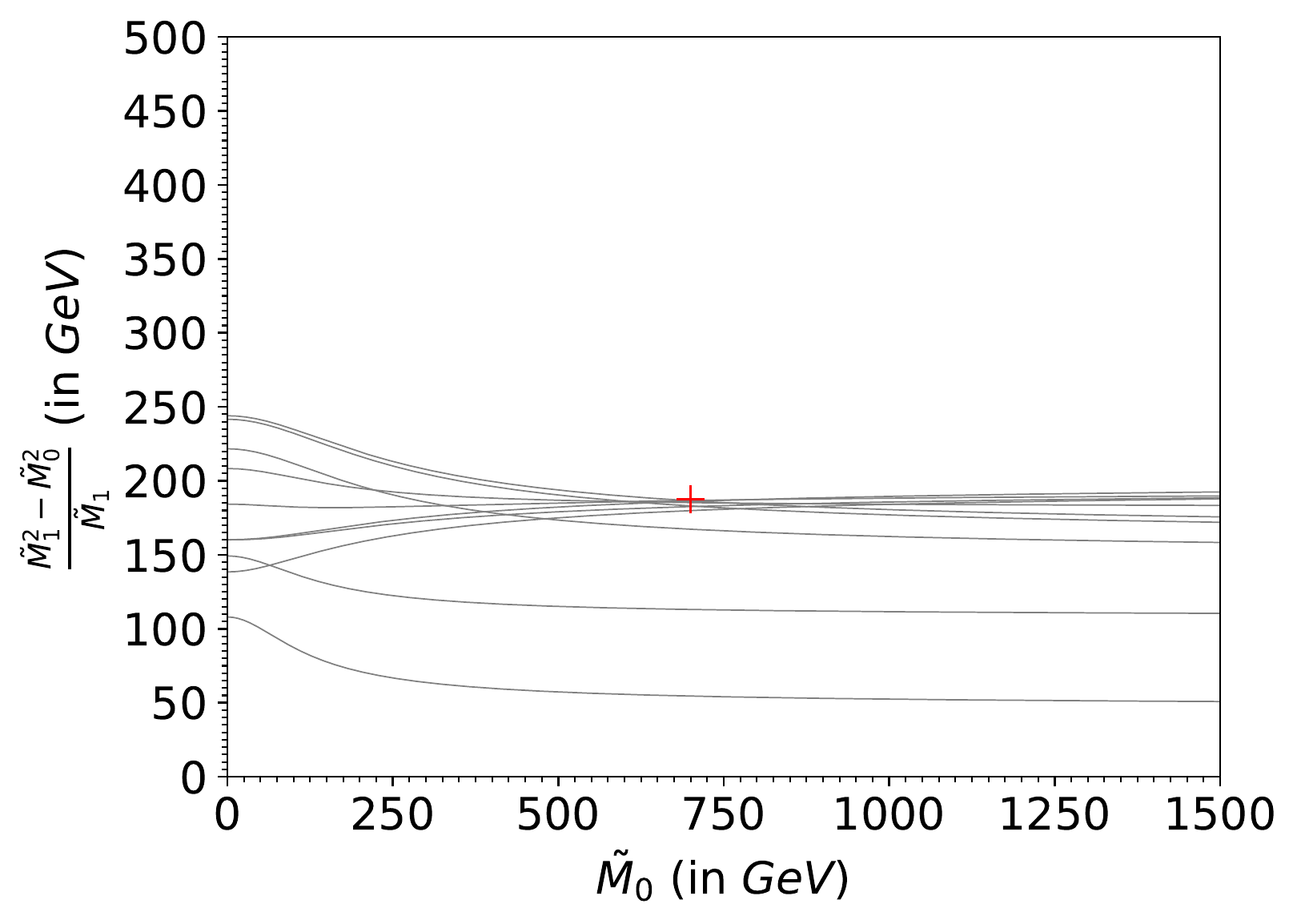}
 \hskip 5mm
 \includegraphics[width=.45\textwidth]{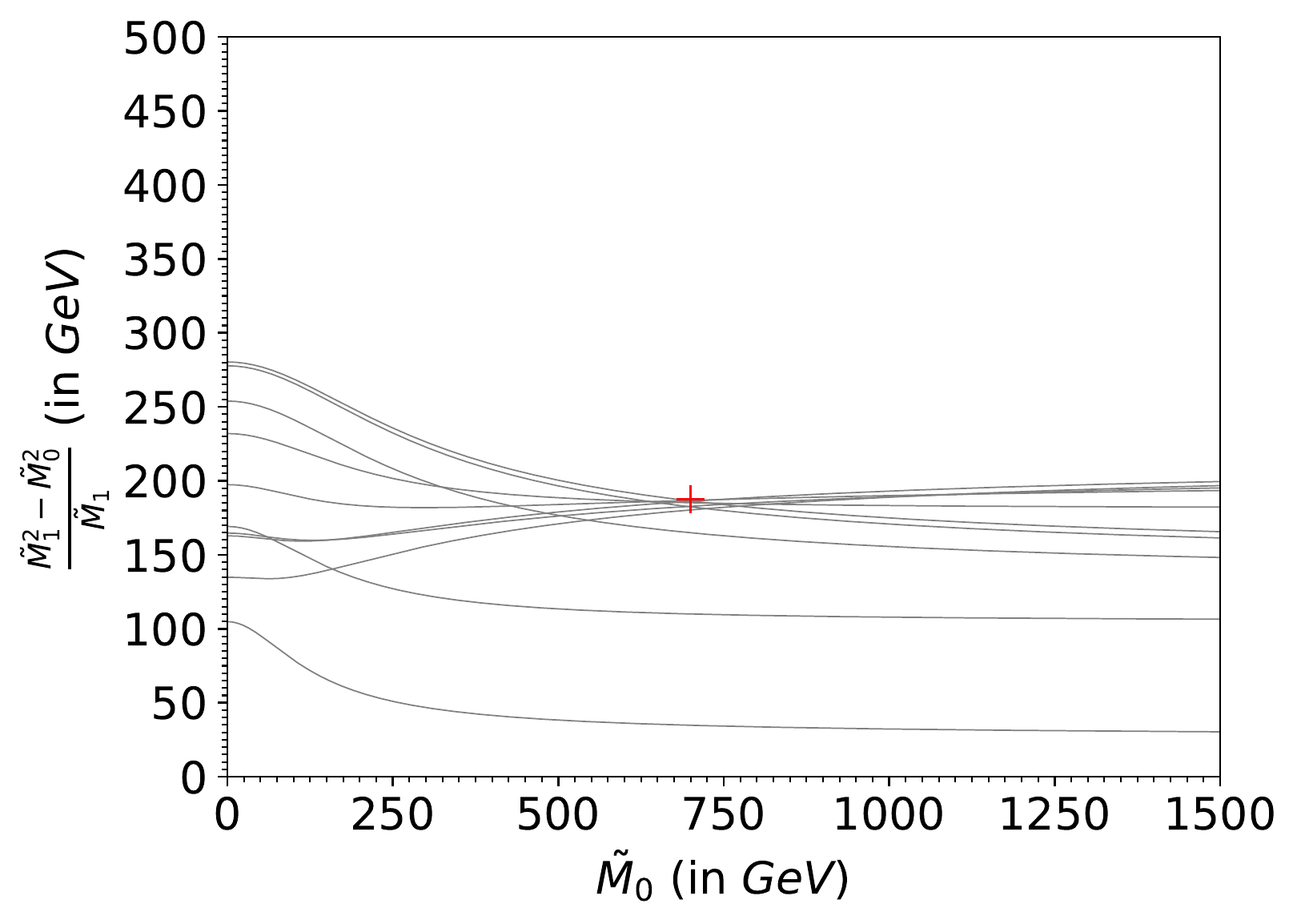}
 \\
  \includegraphics[width=.45\textwidth]{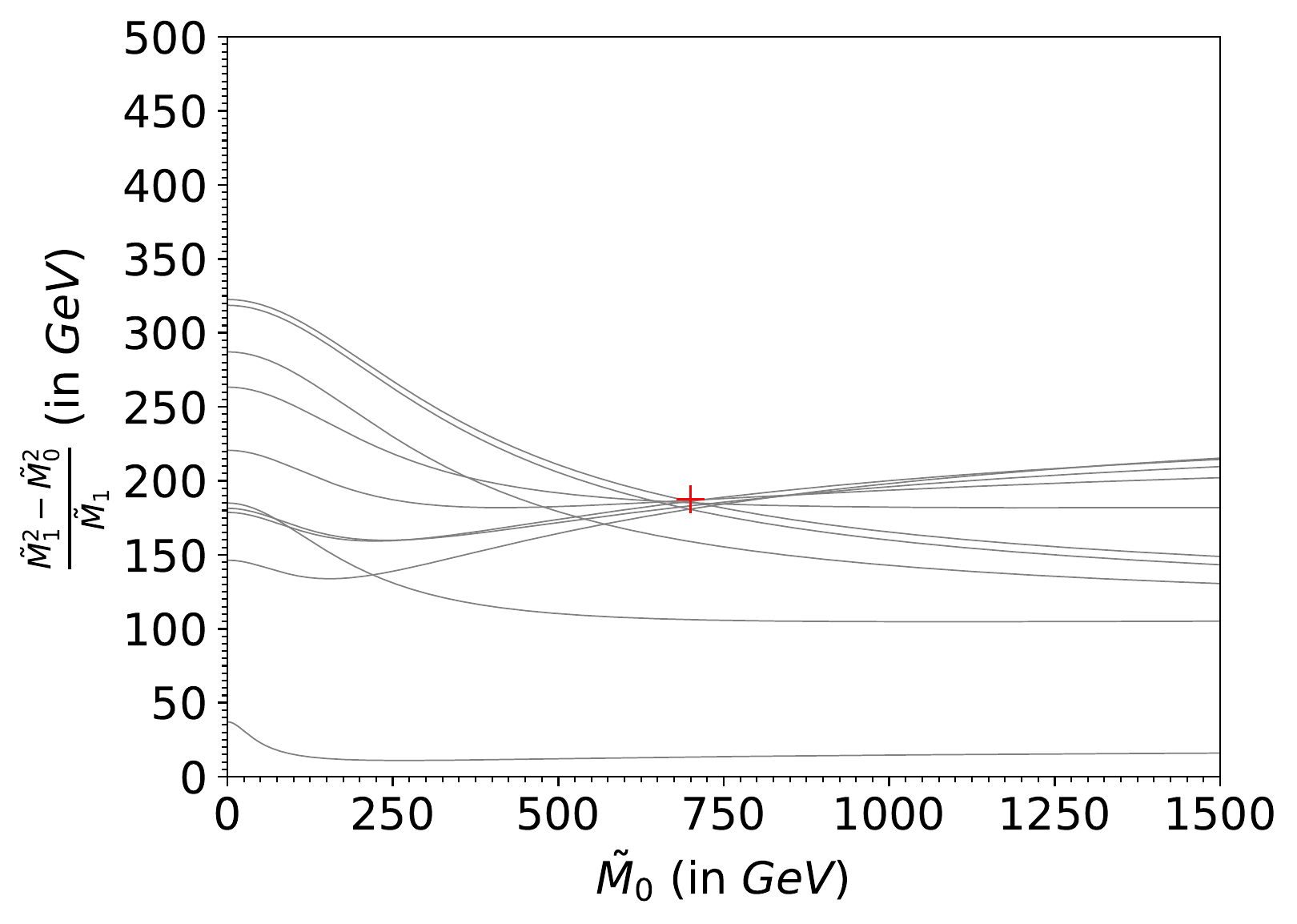}
 \hskip 5mm
 \includegraphics[width=.45\textwidth]{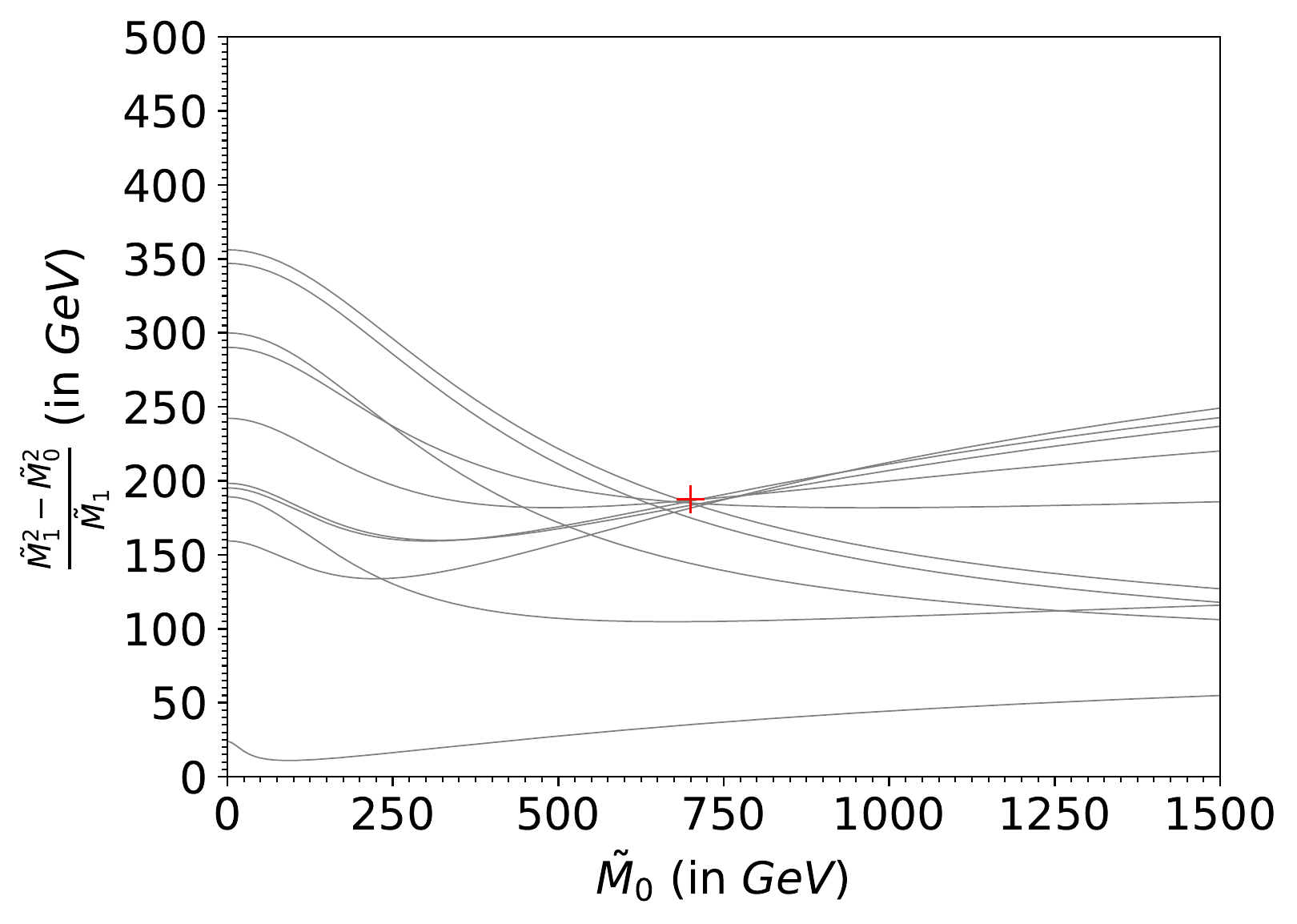}
\caption{\label{fig:1stepFP2} The same as Fig.~\ref{fig:1stepFP}, but for better visibility the quantity plotted on the $y$-xis 
is redefined to be $(\tilde M_1^2-\tilde M_0^2)/\tilde M_1$.}
\end{figure}

These observations are precisely the motivation behind the focus point method for mass measurement \cite{Kim:2019prx},
which can also be applied to the event topology of Fig.~\ref{fig:feynmandiag}(a) considered here,
with the intent of measuring  the two masses $M_1$ and $M_0$. The method is illustrated in 
Figs.~\ref{fig:1stepFP} and \ref{fig:1stepFP2}, where, in order to avoid overcrowding the plot, 
we use just a handful of events (in this case ten, chosen at random).
The main idea is for each event to delineate the allowed region in the hypothesized mass parameter space
$(\tilde M_0, \tilde M_1)$, which would lead to viable solutions for the invisible momentum $q$, given the
kinematic constraints (\ref{systemFiga}). It is well known that for a given test mass $\tilde M_0$, 
the transverse mass $m_T(\tilde M_0)$ provides the lowest kinematically allowed value for the parent mass $M_1$ 
\cite{Gripaios:2007is,Cheng:2008hk,Barr:2011xt}, therefore the boundary of the allowed region in our case will be given simply by the function
\beq
\tilde M_1 = m_T(\tilde M_0).
\label{eq:aM1tvsM0t}
\eeq
After superimposing these kinematic boundaries from many different events as in Figs.~\ref{fig:1stepFP} and \ref{fig:1stepFP2},
the true values of $M_0$ and $M_1$ are revealed by the location of the focus point of the kinematic boundaries \cite{Kim:2019prx}.
This is indeed what is seen in Figs.~\ref{fig:1stepFP} and \ref{fig:1stepFP2} --- even with the low statistics of just 10 events,
a focus point emerges near the red ``$+$" symbol marking the true mass point $(M_0,M_1)=(700\ {\rm GeV},800\ {\rm GeV})$.
In an actual experimental analysis, the available statistics is expected to be much larger, perhaps as much as several orders of magnitude, 
thus there should be no problem observing this focus point. Note that when plotted in the $(\tilde M_0, \tilde M_1)$ plane as in Fig.~\ref{fig:1stepFP},
the lines tend to be parallel to each other, and their crossing is difficult to trace with the naked eye without zooming in
on the relevant region near the red cross. This is why in Fig.~\ref{fig:1stepFP2} we have replotted the same data, only now replacing
$\tilde M_1$ on the $y$-axis with the more relevant combination of masses $(\tilde M_1^2-\tilde M_0^2)/\tilde M_1$ which enters the
analytical formulas (see, e.g., eqs.~(\ref{p1Tmaxm1eq0}) and (\ref{eq:eccentricity}-\ref{eq:b-axis})). As seen in Fig.~\ref{fig:1stepFP2},
this has the benefit of removing the dependence on the overall scale $\tilde M_0$, which allows us to better concentrate on the relative differences 
exhibited by lines corresponding to different events.

As observed in Figs.~\ref{fig:1stepFP} and \ref{fig:1stepFP2} (and supported by the previous discussion in Sec.~\ref{sec:PTISR}), 
the focusing effect in the event topology of Fig.~\ref{fig:feynmandiag}(a) 
relies on the presence of some non-zero $P_T^{ISR}$ in the event; and the larger the $P_T^{ISR}$, the more pronounced the effect.
In order to illustrate this, Figs.~\ref{fig:1stepFP} and \ref{fig:1stepFP2} 
depict results for several different fixed values of $P_T^{ISR}$:
$P_T^{ISR}=0$ (upper left panels), $P_T^{ISR}=50$ GeV (upper right panels),
$P_T^{ISR}=100$ (middle left panels), $P_T^{ISR}=200$ GeV (middle right panels),
$P_T^{ISR}=400$ GeV (lower left panels) and $P_T^{ISR}=800$ GeV (lower right panels).
Although higher values of $P_T^{ISR}$ would be even more beneficial to showcase our method, 
we limit ourselves here to these more typical values expected to come from initial state radiation 
(which has a steeply falling $P_T$ spectrum) or from decays upstream, 
where the typical $P_T$ is governed by the mass splitting of the new particles, which is likely to be around the $M_1$ mass scale.

Let us first discuss the upper left panels corresponding to the case of $P_T^{ISR}=0$ which was the subject of Sec.~\ref{sec:noPTISR}. 
These plots demonstrate that in the absence of any $P_T^{ISR}$, different values of $\tilde M_0$ are indistinguishable. 
In the plots, this is indicated by the fact that the lines stay roughly parallel to each other, 
and there are no line crossings at all. Notice that quite a few events in the figure are close to being extreme, 
i.e., their lines pass very close to the true mass point marked with the red cross.  
This is simply a consequence of the singularity condition (\ref{mTeqM1}) --- since the ten 
events entering the plot were picked at random, it is more likely that they belong to the region 
where the event density exhibits a (singular) peak. Correspondingly, their lines appear to be ``bunched up",
in the sense that their $m_T$ values for $\tilde M_0=M_0$ tend to be very similar, being so close to the true mass $M_1$
of the parent particle. Then, as we vary the test mass $\tilde M_0$ away from the true value $M_0$,
the lines in the upper left panels continue to stay bunched up, confirming the presence of a singularity in the $m_T(\tilde M_0)$ distribution
for all other values of $\tilde M_0$ as well  (recall Fig.~\ref{fig:1stepMTdistributions}); the exact location of the singularity as a function of $\tilde M_0$ is given by eq.~(\ref{eq:singlocptisreq0}).

\begin{figure}[t]
 \centering
 \includegraphics[width=.45\textwidth]{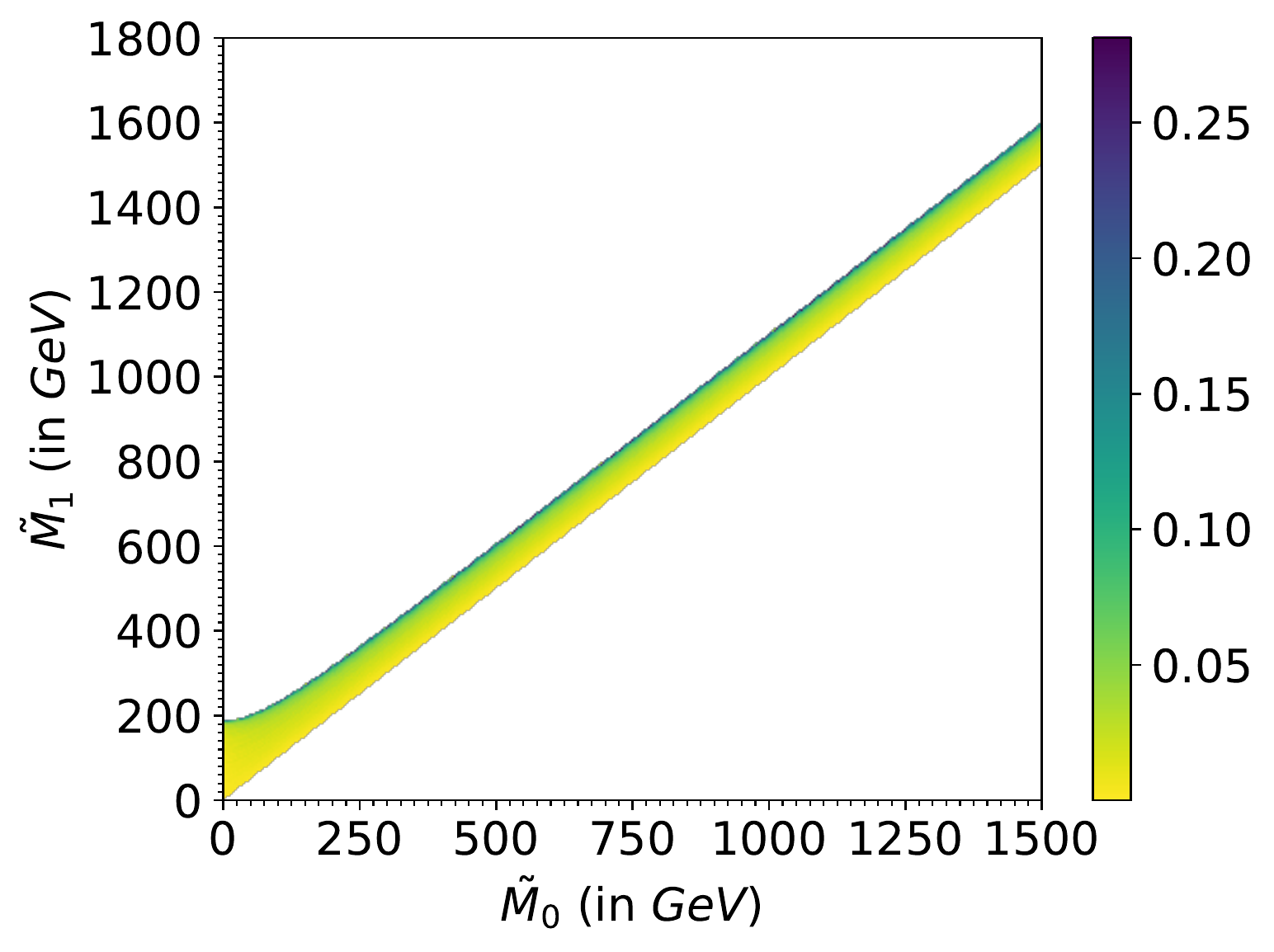}
 \hskip 5mm
 \includegraphics[width=.45\textwidth]{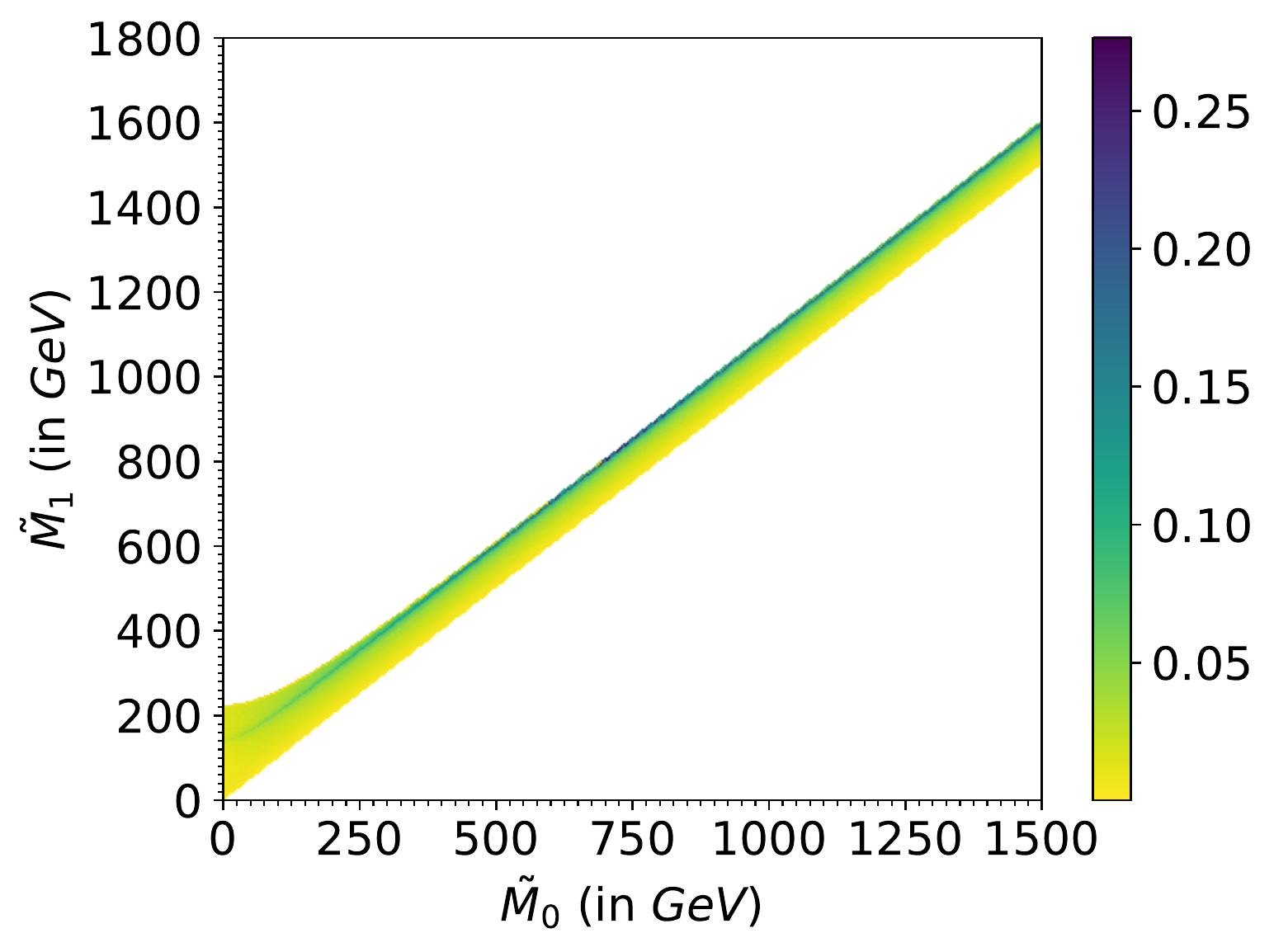}
 \\
 \includegraphics[width=.45\textwidth]{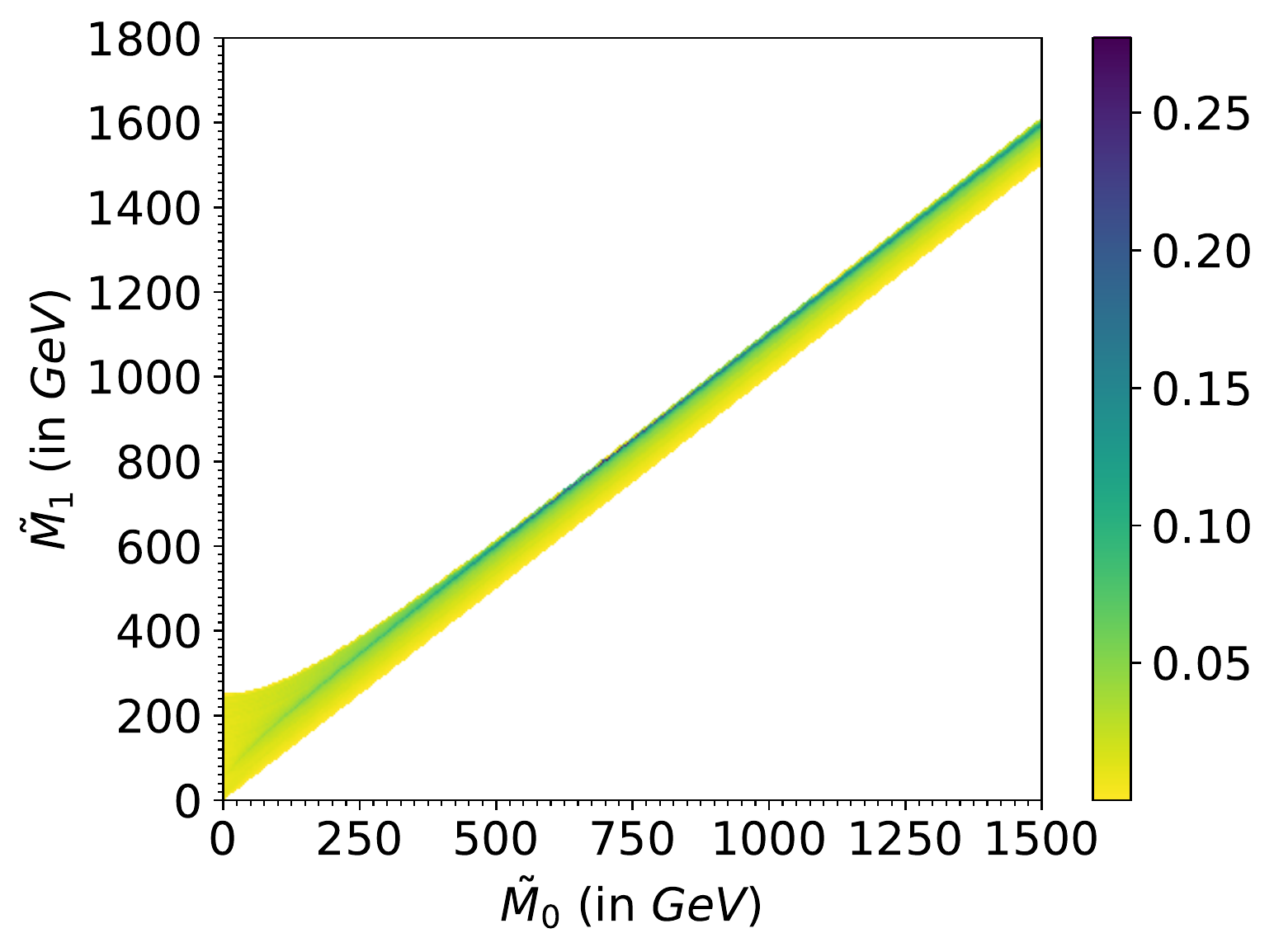}
 \hskip 5mm
 \includegraphics[width=.45\textwidth]{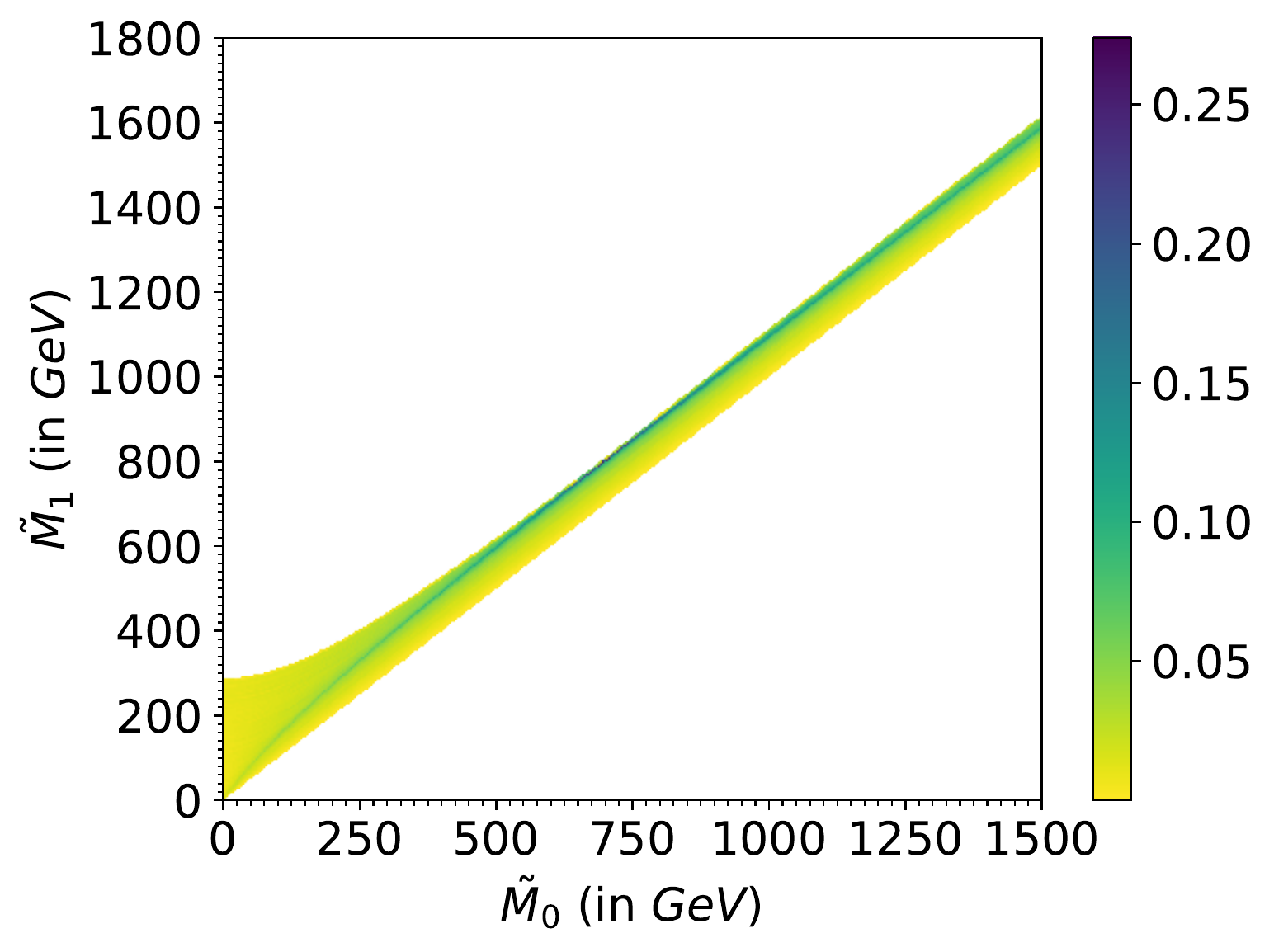}
 \\
 \includegraphics[width=.45\textwidth]{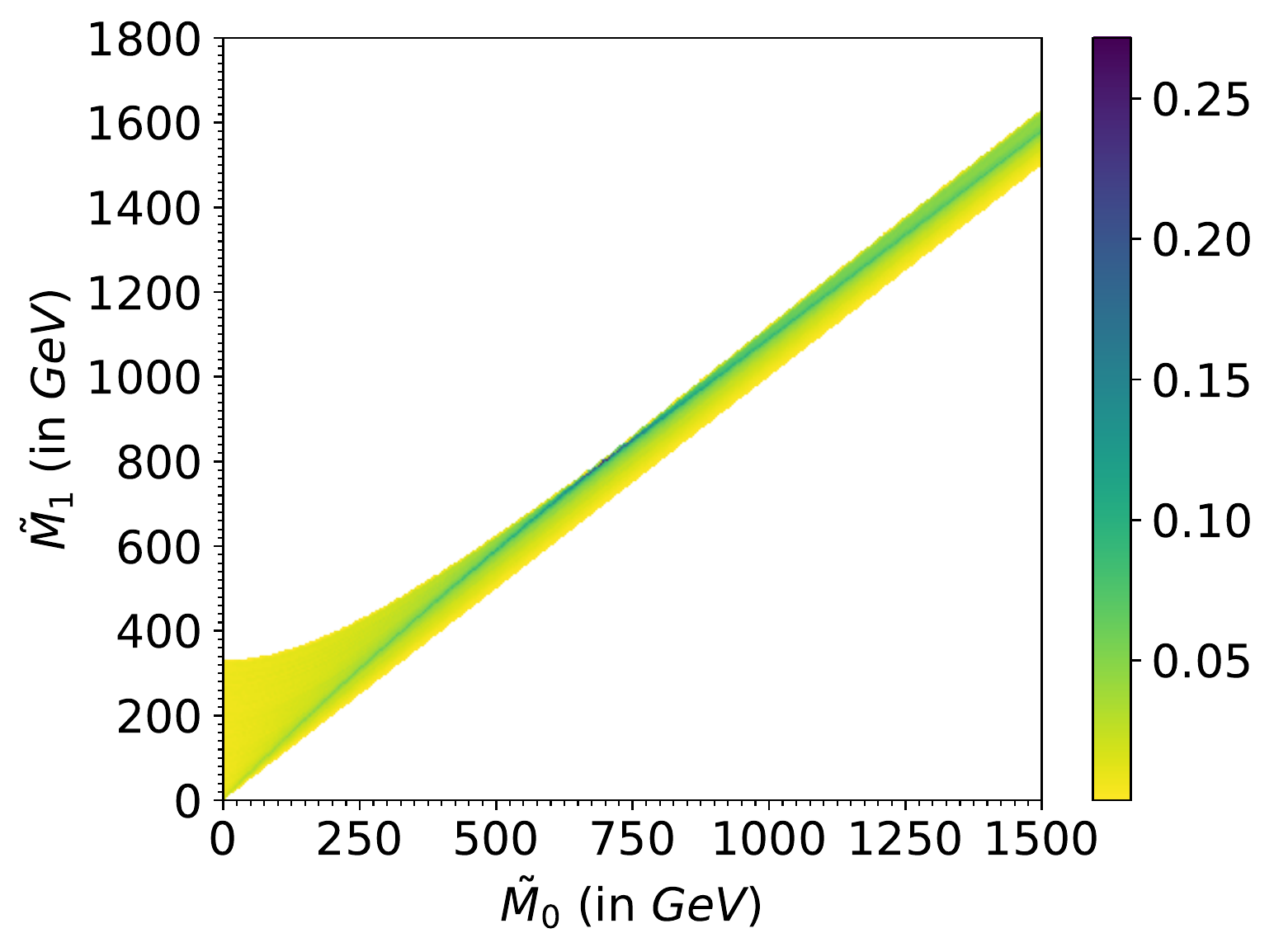}
 \hskip 5mm
 \includegraphics[width=.45\textwidth]{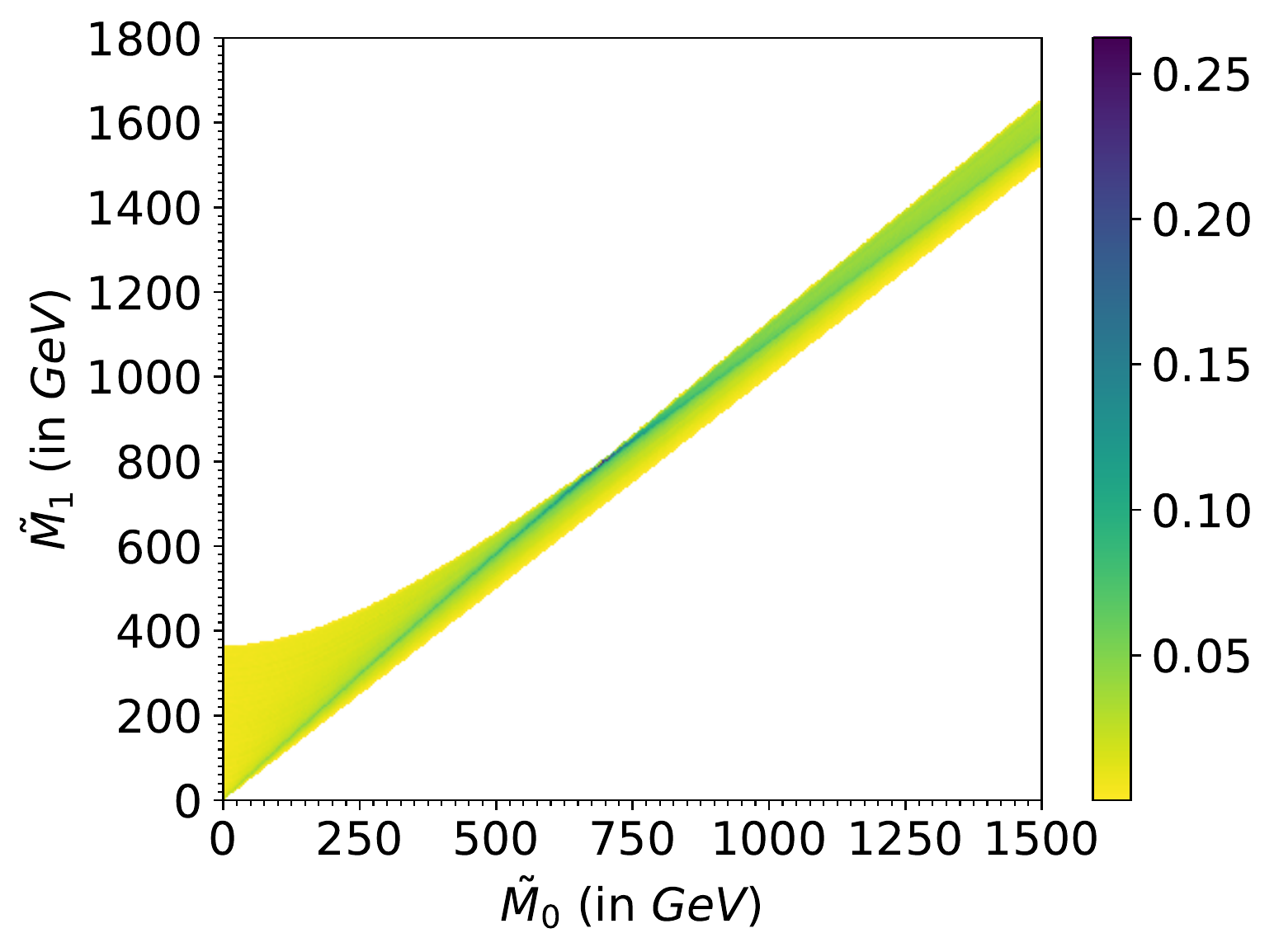}
\caption{\label{fig:1stepHM} The heatmaps corresponding to Fig.~\ref{fig:1stepFP}, produced with a full data sample.}
\end{figure}

Fortunately, the situation changes completely in the presence of non-zero $P_T^{ISR}$, as illustrated in the remaining 
five panels in each of Figs.~\ref{fig:1stepFP} and \ref{fig:1stepFP2}. We use the same 10 events as before, only now they have been 
boosted accordingly in order to generate the desired $P_T^{ISR}$. For illustration, let us only focus on the lower right panels with the largest $P_T^{ISR}=800$ GeV, where the effects are easiest to see. 
%which was the largest $P_T^{ISR}$ value considered in Figs.~\ref{fig:1stepISR} and \ref{fig:ellipses}.\footnote{The panels in the bottom rows of Figs.~\ref{fig:1stepFP} and \ref{fig:1stepFP2} are for demonstration purposes only, since such large values of $P_T^{ISR}$ are unlikely to be generated in a real data sample. Nevertheless, they are qualitatively very similar to the case of $P_T^{ISR}=1600$ GeV, implying that there is not much to gain from considering the asymptotic limit $P_T^{ISR}\to \infty$; in fact, the more realistic values of $P_T^{ISR}$ are already sufficient to see the focusing effect.}
We observe that the lines for the near-extreme events are still bunched up in the vicinity of the red $+$ symbol, 
but significantly diverge away from it in the region of either very low or very high values of $\tilde M_0$ 
(this is especially easy to see with the parametrization used in Fig.~\ref{fig:1stepFP2}). In other words, the crossing pattern of the lines 
has formed a focus point in the vicinity of the true mass point, and therefore, finding this focus point is a way to find the true masses \cite{Kim:2019prx}.
Note that the defocusing of the lines away from the true value $\tilde M_0=M_0$ implies that the singularity in the $m_T(\tilde M_0)$
distribution is getting washed out when the chosen value of $\tilde M_0$ is away from the true mass $M_0$. This offers an alternative method for finding the true value of $M_0$, namely,
by studying the sharpness of the singularity peak in the $m_T(\tilde M_0)$ distribution as a function of the test value $\tilde M_0$ \cite{Kim:2019prx}.

\begin{figure}[t]
 \centering
 \includegraphics[width=.45\textwidth]{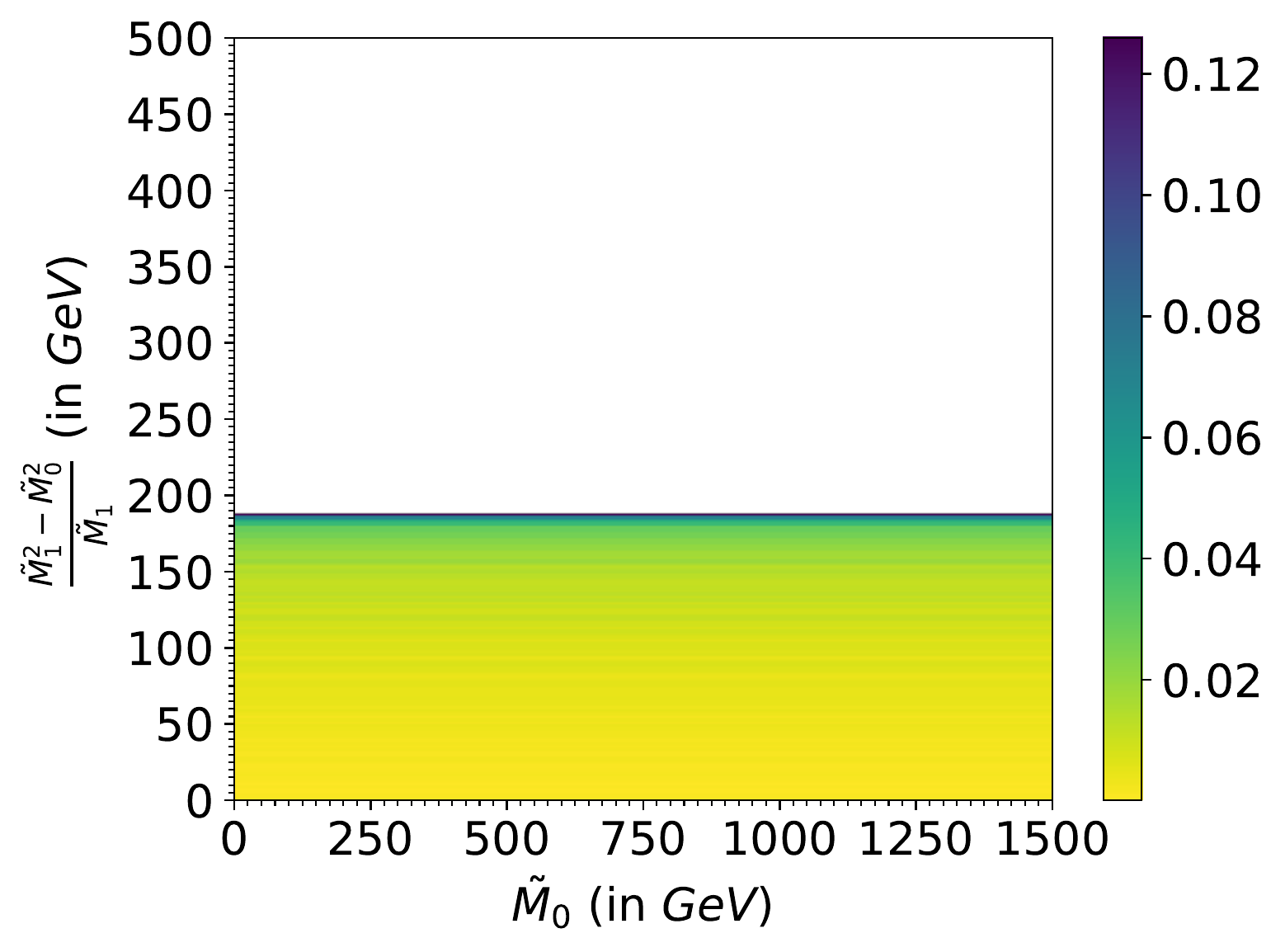}
 \hskip 5mm
 \includegraphics[width=.45\textwidth]{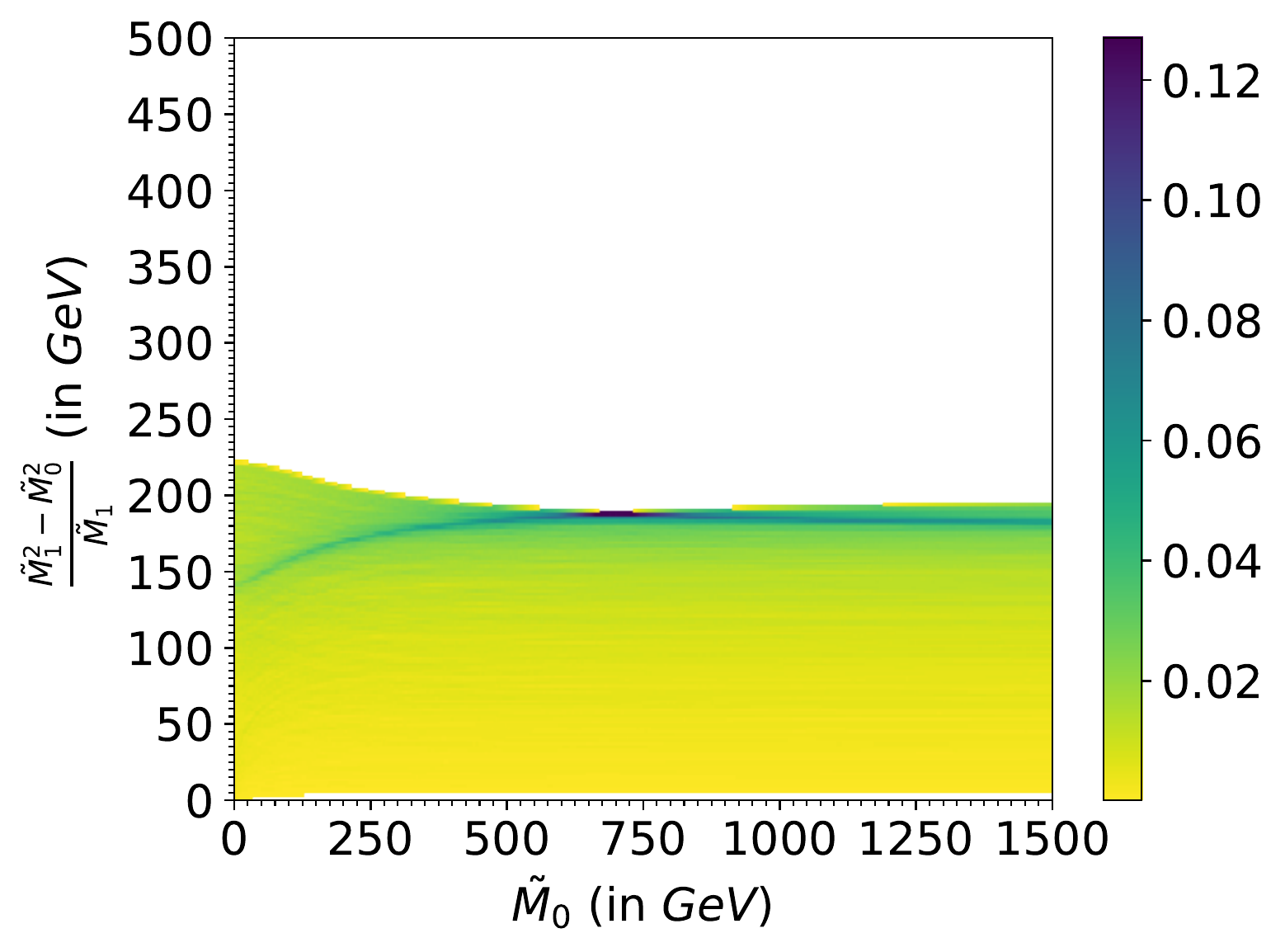}
 \\
 \includegraphics[width=.45\textwidth]{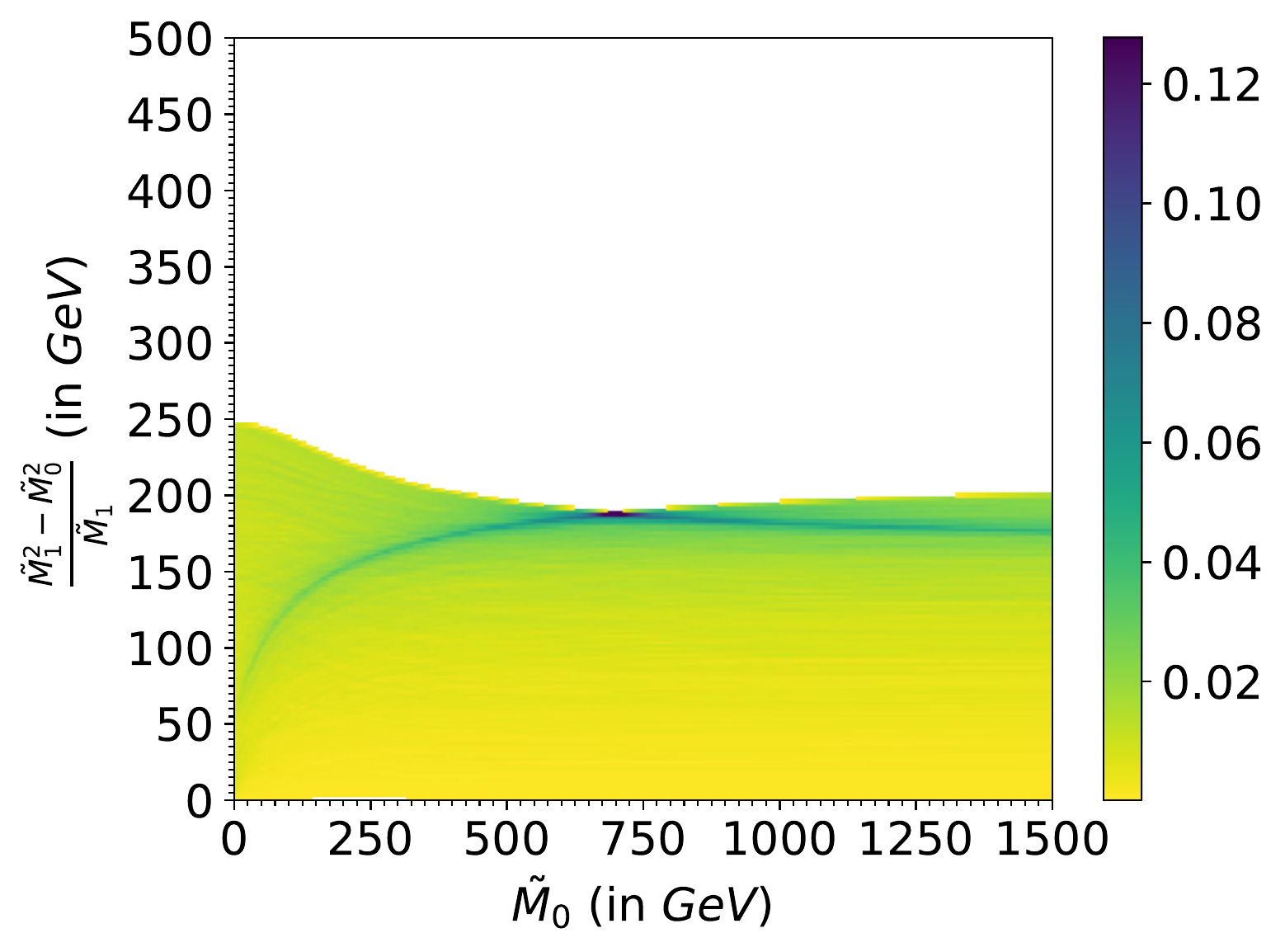}
 \hskip 5mm
 \includegraphics[width=.45\textwidth]{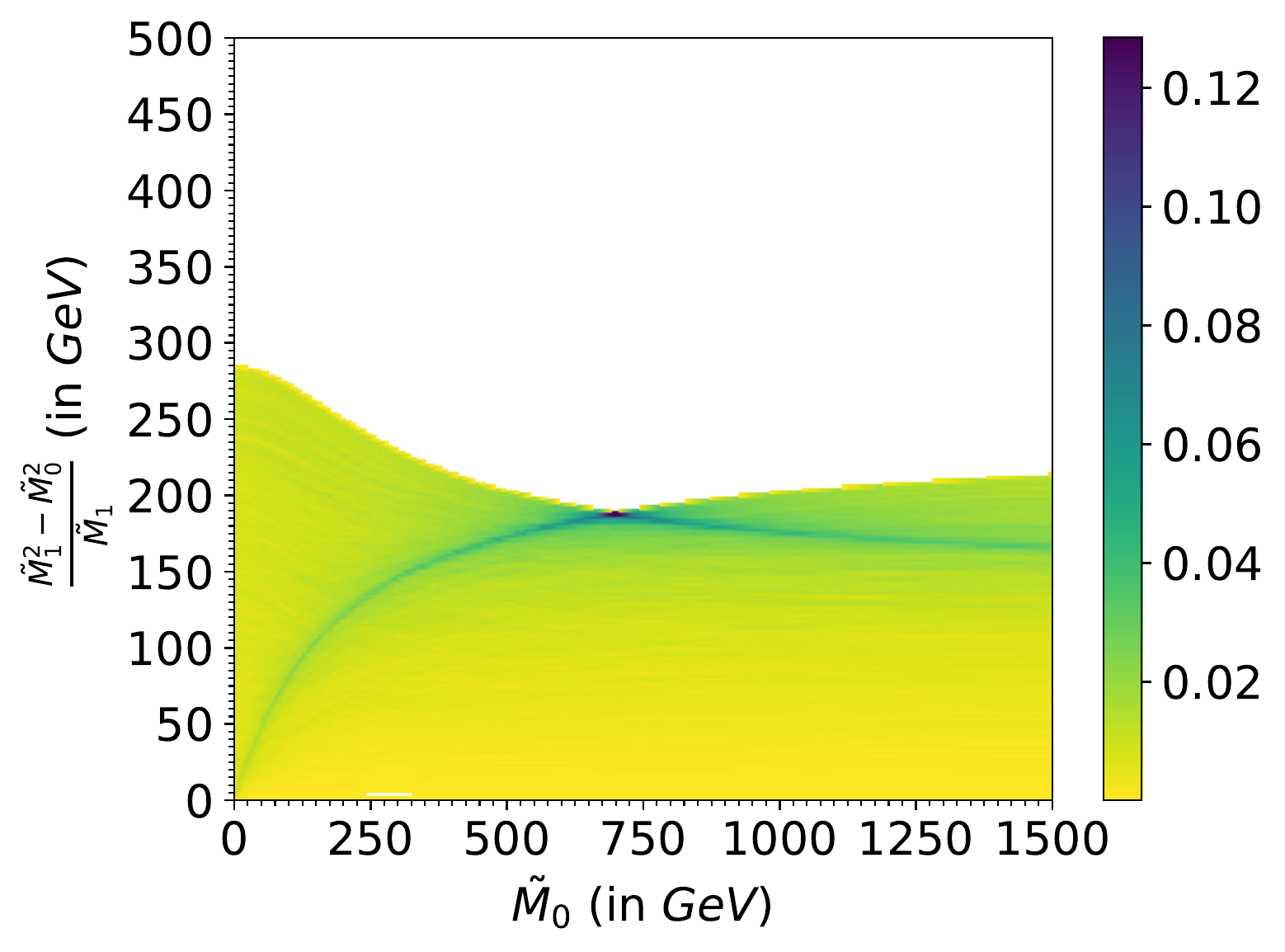}
 \\
 \includegraphics[width=.45\textwidth]{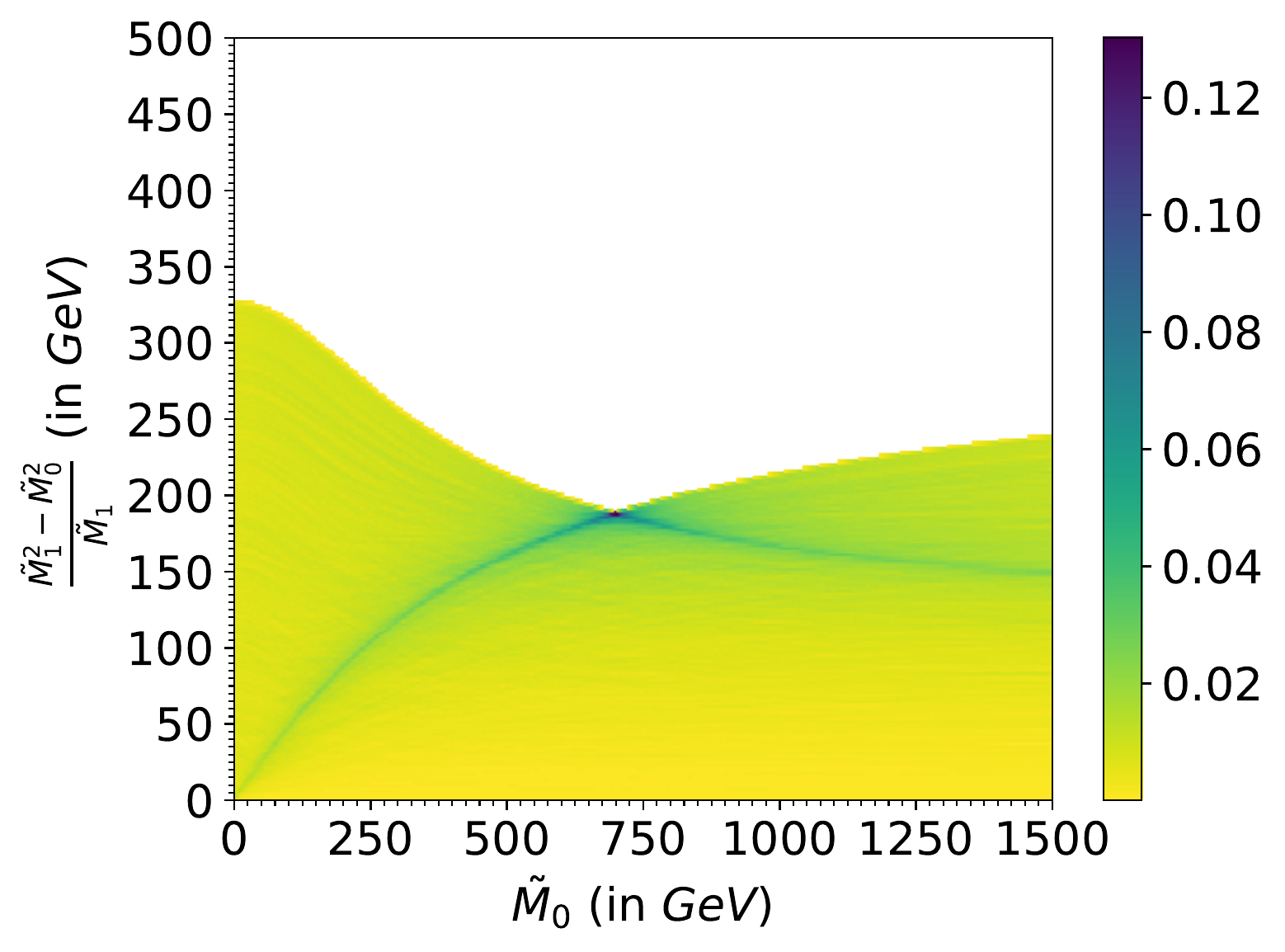}
 \hskip 5mm
 \includegraphics[width=.45\textwidth]{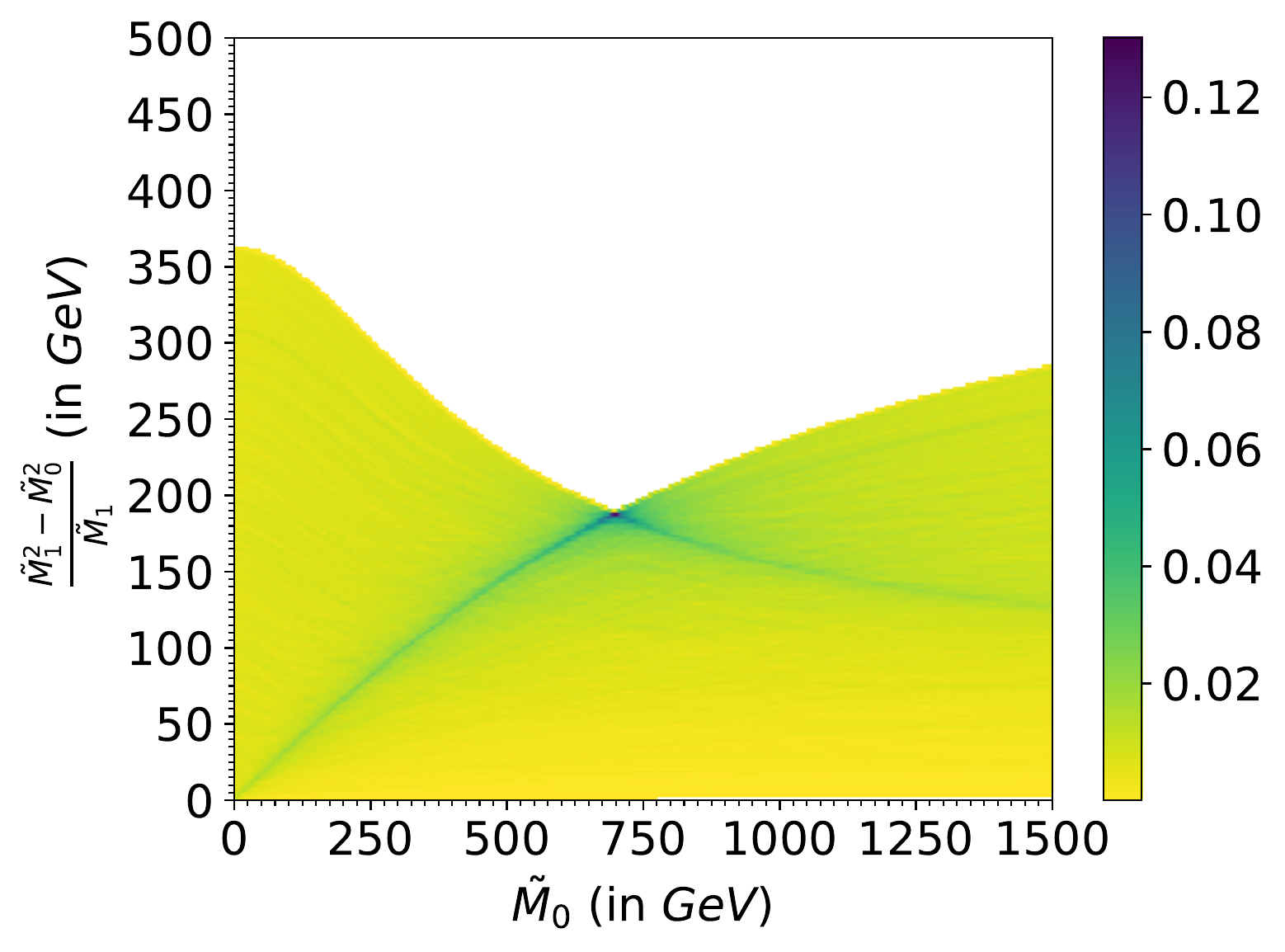}
\caption{\label{fig:1stepHM2} The same as Fig.~\ref{fig:1stepHM}, but with the $y$-axis reparametrization used in Fig.~\ref{fig:1stepFP2}.}
\end{figure}

Both of those mass measurement methods are illustrated in Figs.~\ref{fig:1stepHM} and \ref{fig:1stepHM2}, which are the analogues 
of Figs.~\ref{fig:1stepFP} and \ref{fig:1stepFP2}, only this time we are using the full data sample and, following \cite{Kim:2019prx},
we represent the density of curves as a heatmap where the color corresponds to the fraction of events\footnote{By normalizing to 
a fraction, our results become insensitive to the statistics used to make the plots; in this particular case, 
Figs.~\ref{fig:1stepHM} and \ref{fig:1stepHM2} were made using $10^4$ events.} whose lines pass through a square bin of width $2$ GeV. 
Fig.~\ref{fig:1stepHM} shows the heatmaps in the $(\tilde M_0, \tilde M_1)$ space as in Fig.~\ref{fig:1stepFP}, 
while Fig.~\ref{fig:1stepHM2} uses the alternative $y$-axis reparametrization of Fig.~\ref{fig:1stepFP2}.

The heat maps in Figs.~\ref{fig:1stepHM} and \ref{fig:1stepHM2} reveal that the highest density of lines is indeed found
in the vicinity of the true mass point. In the cases of non-zero $P_T^{ISR}$ (upper right, middle row and bottom row panels),
the location of the highest density bin is unique, and this completely fixes the values of $M_0$ and $M_1$. Furthermore, 
when we add the statistics from all the different non-zero $P_T^{ISR}$ samples, 
the contributions to the true mass bin will add coherently, further enhancing the sharpness of the focus point.
On the other hand, in the case of $P_T^{ISR}=0$ (upper left panels), there is a tie for the highest density bin all along the line parametrized
by eq.~(\ref{eq:singlocptisreq0}); as a result, we can only determine $M_1$ as a function of the test mass $\tilde M_0$, 
but the true value of $M_0$ remains unknown. This is why, such events with low $P_T^{ISR}$ should not be used in the analysis for measuring $M_0$ with the focus point method.

Before concluding this section, let us discuss the connection between the focus point method 
presented here and the kink method for mass measurements \cite{Cho:2007qv,Gripaios:2007is,Barr:2007hy,Cho:2007dh,Matchev:2009fh}. 
The two methods are closely related --- in fact, Figs.~\ref{fig:1stepFP}-\ref{fig:1stepHM2} also nicely illustrate the kink method itself, where one
instead tries to measure the {\em maximal} possible value of $m_T$ as a function of the test mass $\tilde M_0$.
In other words, the kink method is essentially targeting the upper boundaries of the colored regions in Fig.~\ref{fig:1stepHM},
which for $P_T^{ISR}\ne 0$ exhibit a kink\footnote{The kink is easier to see with the alternative parametrization of Fig.~\ref{fig:1stepHM2}.} 
at $\tilde M_0=M_0$. In contrast, the focus point method is targeting the highest line density bin on the plot,
and thus is designed to take full advantage of the available statistics --- note that the boundaries of the colored regions
in Figs.~\ref{fig:1stepHM} and \ref{fig:1stepHM2} are defined by just a handful of events and the extraction of the boundary 
(from kinematic endpoints or otherwise) is statistics-limited. On the other hand, as illustrated in Figs.~\ref{fig:1stepFP} and \ref{fig:1stepFP2},
events which do not contribute to the boundary may still pass close to the focus point and thus usefully contribute
to the focus point method. Of course, the kink occurs precisely at the location of the highest line density bin\footnote{This connection
was present, but overlooked in the existing literature, e.g., note the resemblance of Figs.~6(c) and 6(d) in Ref.~\cite{Barr:2007hy} 
to our Figs.~\ref{fig:1stepHM}(a) and \ref{fig:1stepHM}(b), respectively.}, 
thus the two methods in principle give the same results, as confirmed in Figs.~\ref{fig:1stepFP}-\ref{fig:1stepHM2}.

\section{Single decay chain, two successive two-body decays} 
\label{sec:12}

In this section we shall discuss the event topology of Fig.~\ref{fig:feynmandiag}(b) which is a cascade decay involving 
three new particles with masses $M_0$, $M_1$ and $M_2$. Correspondingly, there are three on-shell conditions
\begin{subequations}
\begin{eqnarray}
q^2 &=& M_0^2, \\
(q+p_1)^2 &=& M_1^2, \\
(q+p_1+p_2)^2 &=& M_2^2,
\end{eqnarray}
\end{subequations}
which can be rewritten as
\begin{subequations}
\begin{eqnarray}
q^2 &=& M_0^2, \label{bqq}\\
2 p_1\cdot q &=& M_1^2-M_0^2-m_1^2, \label{bp1q}\\
2 p_2\cdot q &=& M_2^2-M_1^2-m_2^2-2p_1\cdot p_2, \label{bp2q}
\end{eqnarray}
\label{systemFigb}%
\end{subequations}
At this point, we have $N_C=3$ constraints and $N_q=4$ unknowns which are the components of $q^\mu$.
Thus in order to obtain the necessary match (\ref{NceqNq}), we can do one of two things --- either add an additional 
constraint, e.g., in the form of a measurement of one of the $\mpt$ components (but not the other), 
or reduce the number of unknowns by going to $2+1$ dimensions, where $N_q=3$.
Both of these options are of mostly academic interest, so for concreteness we choose the latter, 
which has the added benefit of somewhat simpler math (involving $3\times 3$ instead of $4\times 4$ matrices). 
Consequently, for the remainder of this section, we shall be working in $2+1$ dimensions, where
the momenta have only transverse and no longitudinal components, i.e., $p_i=(e_i,p_{ix},p_{iy})$ and $q_i=(\varepsilon,q_x,q_y)$.

The next step is to construct the Jacobian matrix (\ref{defJac}) for the set of constraints (\ref{systemFigb}):
\beq
D =
\left(
\begin{array}{ccc}
2\varepsilon & -2 q_{x} & -2 q_{y}  \\
2e_1 & -2 p_{1x} & -2 p_{1y}  \\
2e_2 & -2 p_{2x} & -2 p_{2y}  
\end{array}
\right).
\eeq
The singularity condition (\ref{det0toprow}) becomes\footnote{Note that it can be equivalently written in a more compact form as 
\beq
\epsilon_{\alpha\beta\gamma}q^\alpha p_1^\beta p_2^\gamma = 0, \qquad \alpha,\beta,\gamma=\{0,1,2\}.
\label{epsilonabg}
\eeq}
\beq
{\rm Det}\, D
= 8 
\left|
\begin{array}{ccc}
\varepsilon & - q_{x} & - q_{y}  \\
e_1 & - p_{1x} & - p_{1y}  \\
e_2 & - p_{2x} & - p_{2y}  
\end{array}
\right|
= 8 
\left|
\begin{array}{ccc}
\varepsilon & q_{x} & q_{y}  \\
e_1 & p_{1x} & p_{1y}  \\
e_2 & p_{2x} & p_{2y}  
\end{array}
\right|
=0.
\label{detDFigb}
\eeq
Now, in order to rewrite the singularity variable appearing on the left-hand side in terms of observable momenta only, we just need to eliminate the invisible momentum components, 
for example, using eqs.~(\ref{systemFigb}). However, a more straightforward approach is to note that 
\beq
\mathrm{Det}~D = 0 \quad \Longleftrightarrow \quad
\mathrm{Det}\, D^T
= 8 
\left|
\begin{array}{ccc}
\varepsilon & e_1 & e_2   \\
- q_{x} & - p_{1x} & - p_{2x}  \\
- q_{y} & - p_{1y} & - p_{2y}  
\end{array}
\right|=0
\eeq
and combine the last two equations as
\bea
&&64
\left|
\left(
\begin{array}{ccc}
\varepsilon & q_{x} & q_{y}  \\
e_1 & p_{1x} & p_{1y}  \\
e_2 & p_{2x} & p_{2y}  
\end{array}
\right)
\left(
\begin{array}{ccc}
\varepsilon & e_1 & e_2   \\
- q_{x} & - p_{1x} & - p_{2x}  \\
- q_{y} & - p_{1y} & - p_{2y}  
\end{array}
\right)
\right|
= 64
\left|
\begin{array}{ccc}
q^2 & p_1\cdot q & p_2\cdot q  \\
p_1\cdot q & p_1^2 & p_1\cdot p_2 \\
p_2\cdot q & p_1\cdot p_2 & p_2^2   
\end{array}
\right|
\nonumber \\ [2mm]
&=& 64
\left|
\begin{array}{ccc}
M_0^2 & ~~~\frac{M_1^2-M_0^2-m_1^2}{2}~~~ & \frac{M_2^2-M_1^2-m_2^2}{2}-p_1\cdot p_2  \\ [2mm]
\frac{M_1^2-M_0^2-m_1^2}{2}  & m_1^2 & p_1\cdot p_2 \\ [2mm]
\frac{M_2^2-M_1^2-m_2^2}{2}-p_1\cdot p_2 & p_1\cdot p_2 & m_2^2   
\end{array}
\right|
=0.
\label{detDsquaredFigb}
\eea
By rewriting the singularity condition in this form, we have not only eliminated any reference to the invisible momentum components, 
but we also verified explicitly that the event-wise kinematic information only enters through the dot product $p_1\cdot p_2$, 
or equivalently, through the invariant mass $m_{12}$ of the two visible particles since the latter is related to $p_1\cdot p_2$ as
\beq
m_{12}^2\equiv (p_1+p_2)^2 = m_1^2+m_2^2+2p_1\cdot p_2.
\label{eq:m12def}
\eeq	
Eq.~(\ref{detDsquaredFigb}) suggests that instead of taking the whole determinant as the singularity coordinate for this event topology, 
a much simpler choice would be either $p_1\cdot p_2$ or $m_{12}$, both of which have the important advantage of avoiding the necessity of
an ansatz for the {\em a priori} unknown masses $M_0$, $M_1$ and $M_2$. 
For definiteness here we shall choose to work with $p_1\cdot p_2$,
but $m_{12}$ will be equally good\footnote{The reader should keep in mind that we are working in $2+1$ dimensions here, and it is only in that case that 
$m_{12}$ is a singularity coordinate and its distribution has singularities. In $3+1$ dimensions, the distribution of $m_{12}$ 
has no singularities and is simply proportional to $m_{12}$ (in other words, the distribution of $m_{12}^2$ is flat). }.

The locations of the singularities in the $p_1\cdot p_2$ distribution can be found from eq.~(\ref{detDsquaredFigb}), 
which leads to a quadratic equation and correspondingly two solutions. In order to simplify the formulas, let us again focus only on the case
of massless visible particles, i.e., $m_1=m_2=0$, when (\ref{detDsquaredFigb}) becomes
\beq
\left|
\begin{array}{ccc}
M_0^2 & ~~~\frac{M_1^2-M_0^2}{2}~~~ & \frac{M_2^2-M_1^2}{2}-p_1\cdot p_2  \\ [2mm]
\frac{M_1^2-M_0^2}{2}  & 0 & p_1\cdot p_2 \\ [2mm]
\frac{M_2^2-M_1^2}{2}-p_1\cdot p_2 & p_1\cdot p_2 & 0 
\end{array}
\right|
=0,
\eeq
leading to the quadratic equation
\beq
(p_1\cdot p_2)^2 - \frac{(M_2^2-M_1^2)(M_1^2-M_0^2)}{2M_1^2} (p_1\cdot p_2) =0,
\eeq
whose solutions for $(p_1\cdot p_2)$
\begin{subequations}
\bea
(p_1\cdot p_2)_{min} &=& 0, \label{p1p2singularitiesmin}
\\[1mm]
(p_1\cdot p_2)_{max} &=& \frac{(M_2^2-M_1^2)(M_1^2-M_0^2)}{2M_1^2},
\label{p1p2singularitiesmax}
\eea
\label{p1p2singularities}%
\end{subequations}
also happen to be the two kinematic endpoints of the $p_1\cdot p_2$ distribution. The actual shape of the distribution according to pure phase space is given by
\beq
\frac{dN}{d (p_1\cdot p_2)}= \frac{1}{\pi \sqrt{(p_1\cdot p_2)\left[(p_1\cdot p_2)_{max}-(p_1\cdot p_2)\right]}}.
\label{dNdp1p2}
\eeq

\begin{figure}[t]
 \centering
 \includegraphics[width=.45\textwidth]{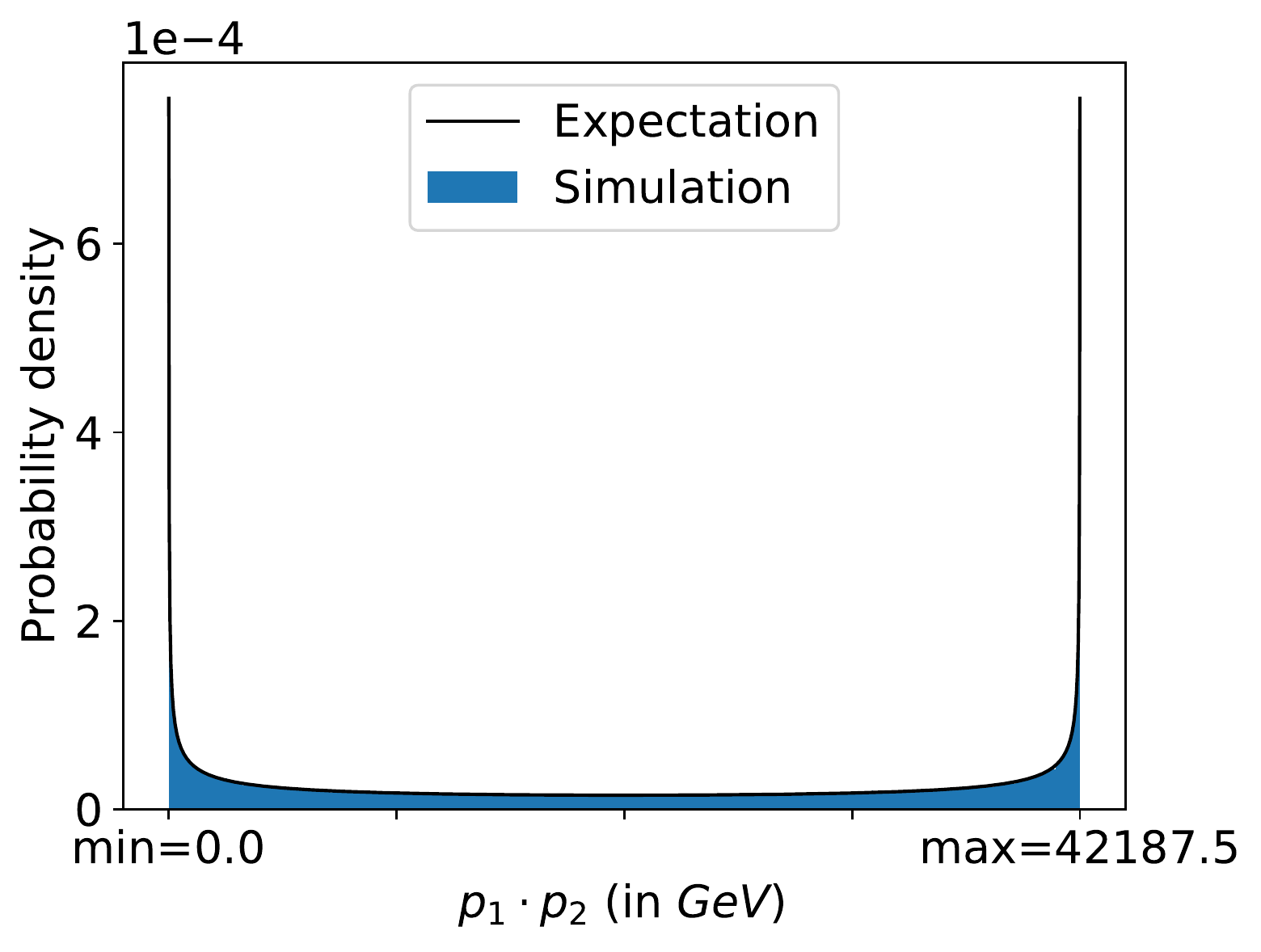}
 \hskip 5mm
 \includegraphics[width=.45\textwidth]{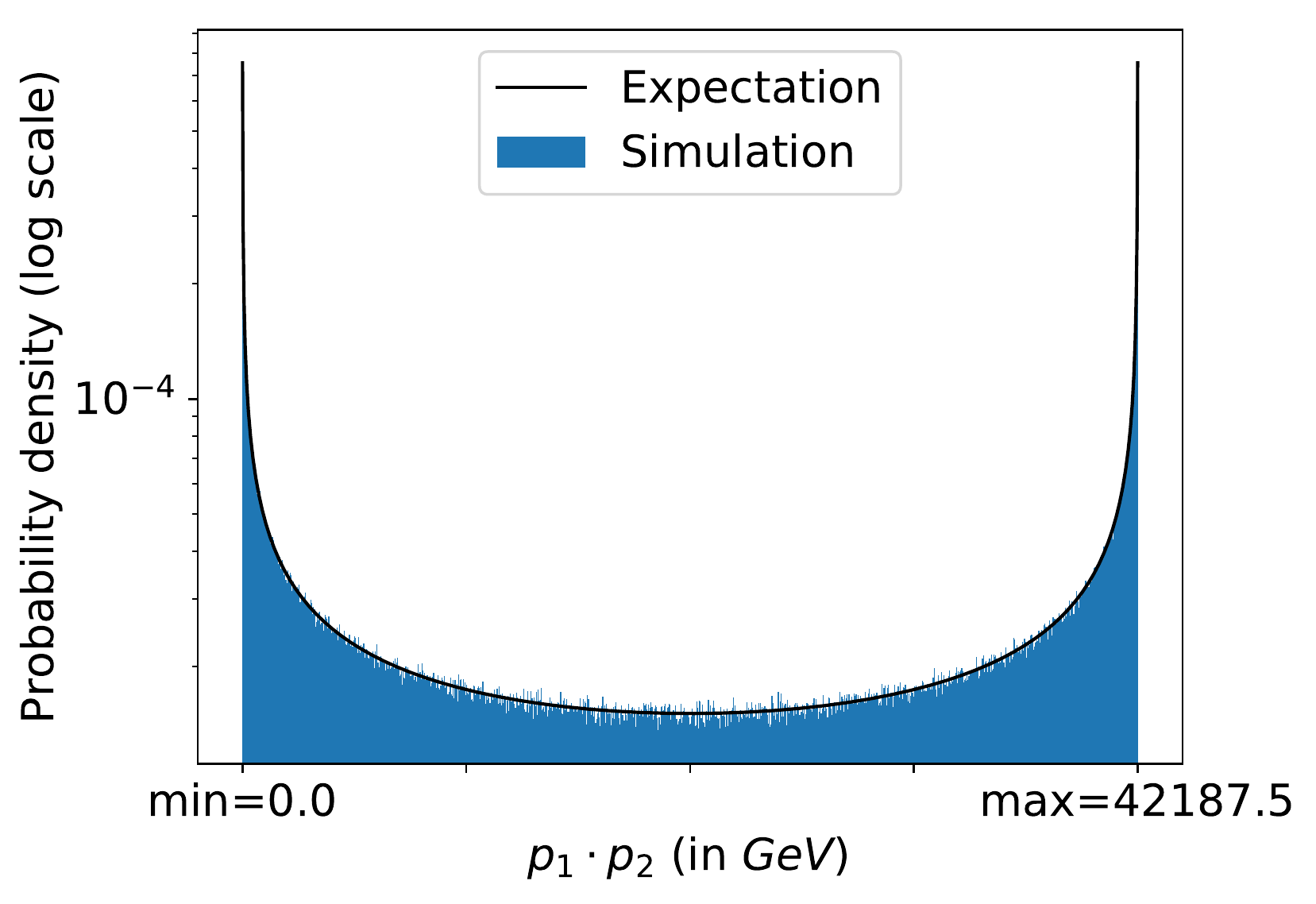}
 \caption{\label{fig:2step} Distribution of the singularity variable $p_1\cdot p_2$ on a linear scale (left panel) and on a log scale (right panel).
 With the mass spectrum from Table~\ref{tab:mass}, eq.~(\ref{p1p2singularities}) predicts the locations of the two singularities to be at
 $(p_1\cdot p_2)_{min}=0$ and $(p_1\cdot p_2)_{max}=42,187.5\ {\rm GeV}^2$. 
 The black solid line shows the theoretical prediction (\ref{dNdp1p2}) for the shape of the distribution 
 while the histogram shows the result from the numerical simulation. }
\end{figure}

The distribution of the singularity variable $p_1\cdot p_2$ is illustrated in Fig.~\ref{fig:2step}. In each panel, we 
plot the theoretical prediction (\ref{dNdp1p2}) superimposed on the result from our numerical simulations.
The spectrum was chosen as in Table~\ref{tab:mass}: $M_0=700$ GeV, $M_1=800$ GeV and $M_2=1000$ GeV,
which according to (\ref{p1p2singularitiesmax}) gives $(p_1\cdot p_2)_{max}=42,187.5\ {\rm GeV}^2$. We see that, as expected,
the distribution develops a sharp singularity at each end. These singularities are quite striking when viewed on a linear scale (as in the left panel). 
In order to better see the shape of the distribution in the intermediate $p_1\cdot p_2$ range, in the right panel of Fig.~\ref{fig:2step} 
we replotted the same data using a log scale for the y-axis.

Having derived the singularity coordinate for this case as either $p_1\cdot p_2$ or $m_{12}$ 
and identified the locations (\ref{p1p2singularities}) of the two singularities, 
we have accomplished two of the three stated goals at the end of Sec.~\ref{sec:prelim}.
The last goal, showcasing the focus point method of Ref.~\cite{Kim:2019prx} for mass measurements,
is not applicable in this case, since the singularity variable is constructed out of visible momenta only, 
with no reference to any hypothesized mass parameters. The situation is analogous to the one
already encountered in Sec.~\ref{sec:noPTISR} --- there we saw that when $P_T^{ISR}=0$, 
the singularity variable for the event topology in Fig.~\ref{fig:feynmandiag}(a) can be taken to be simply the $p_T$ of the visible particle, 
and a singularity occurs for {\em any} choice of test mass as shown in the upper left panels of Figs.~\ref{fig:1stepFP}-\ref{fig:1stepHM2}.
The best one can do in our case here, therefore, is to obtain one constraint on the three masses $M_0$, $M_1$ and $M_2$ 
from the measurement of the upper kinematic endpoint (\ref{p1p2singularitiesmax}).

\section{Single decay chain, three successive two-body decays} 
\label{sec:13}

In this section we shall work out the classic SUSY decay chain of three successive two-body decays depicted in Fig.~\ref{fig:feynmandiag}(c).
This cascade involves four new particles with masses $M_0$, $M_1$, $M_2$ and $M_3$, which in SUSY are typically identified with the 
lightest neutralino, a charged slepton, the second-to-lightest neutralino, and a squark, respectively. The same decay chain also pops up in 
other phenomenological models of new physics such as Universal Extra Dimensions with KK-parity \cite{Appelquist:2000nn,Rizzo:2001sd,Cheng:2002ab},
Little Higgs with $T$-parity \cite{ArkaniHamed:2002qy,Cheng:2004yc,Schmaltz:2005ky,Perelstein:2005ka}, etc. 
In any given such new physics scenario, there is a definite assignment for the spins of the new particles, however in order to stay as general as possible, 
we shall ignore any spin correlations (which are typically rather small anyway \cite{Athanasiou:2006ef,Wang:2006hk,Burns:2008cp}) 
and let the particles decay according to pure phase space. Unlike the toy example of the previous section, 
here and for the rest of the paper we shall work in $3+1$ dimensions, thus a single invisible particle with 4-momentum $q$ contributes $N_q=4$ 
unknown degrees of freedom.

We begin by listing the four on-shell conditions for the event topology of Fig.~\ref{fig:feynmandiag}(c):
\begin{subequations}
\begin{eqnarray}
q^2 &=& M_0^2, \\
(q+p_1)^2 &=& M_1^2, \\
(q+p_1+p_2)^2 &=& M_2^2, \\
(q+p_1+p_2+p_3)^2 &=& M_3^2,
\end{eqnarray}
\end{subequations}
which can be rewritten in analogy to (\ref{systemFigb}) as
\begin{subequations}
\begin{eqnarray}
q^2 &=& M_0^2, \label{cqq}\\
2 p_1\cdot q &=& M_1^2-M_0^2-m_1^2, \label{cp1q}\\
2 p_2\cdot q &=& M_2^2-M_1^2-m_2^2-2p_1\cdot p_2, \label{cp2q} \\
2 p_3\cdot q &=& M_3^2-M_2^2-m_3^2-2p_1\cdot p_3 -2p_2\cdot p_3. \label{cp3q}
\end{eqnarray}
\label{systemFigc}%
\end{subequations}
Since we already have $N_C=4$ constraints for $N_q=4$ unknowns, the condition (\ref{NceqNq}) 
is already met and we can proceed with the derivation of the Jacobian matrix (\ref{defJac}) 
\beq
D =
\left(
\begin{array}{cccc}
2\varepsilon & -2 q_{x} & -2 q_{y} & -2 q_{z} \\
2e_1 & -2 p_{1x} & -2 p_{1y} & -2 p_{1z}   \\
2e_2 & -2 p_{2x} & -2 p_{2y}  & -2 p_{2z}  \\
2e_3 & -2 p_{3x} & -2 p_{3y}  & -2 p_{3z}  
\end{array}
\right)
\eeq
and the singularity condition (\ref{det0toprow})\footnote{Note the alternative compact notation in analogy to (\ref{epsilonabg}):
\beq
\epsilon_{\mu\nu\rho\sigma}q^\mu p_1^\nu p_2^\rho p_3^\sigma = 0, \qquad \mu,\nu,\rho,\sigma=\{0,1,2,3\}.
\eeq
}
\beq
{\rm Det}\, D
= 16 
\left|
\begin{array}{cccc}
\varepsilon & - q_{x} & - q_{y} & - q_{z}  \\
e_1 & - p_{1x} & - p_{1y} & - p_{1z}   \\
e_2 & - p_{2x} & - p_{2y}  & - p_{2z} \\  
e_3 & - p_{3x} & - p_{3y}  & - p_{3z} 
\end{array}
\right|
=0.
\label{detDFigc}
\eeq
In order to obtain a singularity variable in terms of the visible momenta only, we need to eliminate the invisible momentum components 
with the help of (\ref{systemFigc}). However, the same task is more easily accomplished with the determinant trick used in the previous section:
after noting that
\beq
\mathrm{Det}~D = 0 \quad \Longleftrightarrow \quad
{\rm Det}\, D^T
= 16
\left|
\begin{array}{cccc}
\varepsilon & e_1 & e_2 & e_3   \\
-q_x & - p_{1x} & - p_{2x} & - p_{3x}   \\
-q_y & - p_{1y} & - p_{2y}  & - p_{3y} \\  
-q_z & - p_{1z} & - p_{2z}  & - p_{3z} 
\end{array}
\right|=0,
\eeq
we can combine the last two equations into
\beq
{\rm Det}\, D =0
\Longleftrightarrow
{\rm Det} \left( D \eta D^T\right) = 256
\left|
\begin{array}{cccc}
q^2 & ~p_1\cdot q~ & ~p_2\cdot q~ & ~p_3\cdot q~  \\ [1mm]
p_1\cdot q & p_1^2 & p_1\cdot p_2 & p_1\cdot p_3 \\[1mm]
p_2\cdot q & p_1\cdot p_2 & p_2^2  & p_2\cdot p_3 \\[1mm]
p_3\cdot q & p_1\cdot p_3 & p_2\cdot p_3  & p_3^2 
\end{array}
\right|=0,
\label{detDsquaredFigc}
\eeq
with $\eta$ being the Minkowski $4\times 4$ metric
\beq
\eta = {\rm diag}(1,-1,-1,-1).
\eeq
With the use of (\ref{systemFigc}), the dot products of momenta appearing in (\ref{detDsquaredFigc}) can now easily be traded for the relevant masses.
Once again, the result simplifies significantly in the case of massless visible particles, i.e., $p_1^2=p_2^2=p_3^2=0$, when (\ref{detDsquaredFigc}) reduces to
\beq
-256\,
\left|
\begin{array}{cccc}
M_0^2 & ~\frac{M_1^2-M_0^2}{2}~ & ~\frac{M_2^2-M_1^2-m_{12}^2}{2}~  & ~\frac{M_3^2-M_2^2-m_{23}^2-m_{13}^2}{2}~\\ [2mm]
\frac{M_1^2-M_0^2}{2}  & 0 & \frac{m_{12}^2}{2} & \frac{m_{13}^2}{2}\\ [2mm]
\frac{M_2^2-M_1^2-m_{12}^2}{2} & \frac{m_{12}^2}{2} & 0 & \frac{m_{23}^2}{2}\\ [2mm] 
\frac{M_3^2-M_2^2-m_{23}^2-m_{13}^2}{2} & \frac{m_{13}^2}{2} & \frac{m_{23}^2}{2} & 0
\end{array}
\right|
=0,
\label{eq:cDeteq0}
\eeq
where $m_{ij}$ are the pair-wise invariant masses of the three (massless) visible particles:
\beq
m_{ij}^2\equiv (p_i+p_j)^2 = 2 p_i\cdot p_j.
\eeq
Up to the numerical prefactor of 256, the left-hand side of (\ref{eq:cDeteq0}) is precisely the $\Delta_4$ variable introduced in \cite{Byckling:1971vca,Agrawal:2013uka}.
We have thus rederived from first principles\footnote{Alternative derivations leading to $\Delta_4=0$ as the defining condition for the {\em boundary} 
of the allowed phase space can be found in \cite{Costanzo:2009mq,Lester:2013aaa}. The fact that the event number density is in addition {\em singular} 
at that boundary was later emphasized in \cite{Agrawal:2013uka}.}
the well-known fact that $\Delta_4$ is the relevant singularity variable for the event topology of Fig.~\ref{fig:feynmandiag}(c),
and that the locations of singularities are those where $\Delta_4=0$.
Eq.~(\ref{eq:cDeteq0}) also confirms that the relevant observable phase space is only three-dimensional\footnote{The visible momenta $p_1$, $p_2$ and $p_3$ 
parametrize a 9-dimensional phase space, but the constraints (\ref{systemFigc}) are invariant under the 6-parameter Lorentz group,
leaving only 3 relevant visible degrees of freedom.}, and can be conveniently 
parametrized with the pair-wise invariant masses as
\beq
\left\{ m_{12}^2, m_{23}^2, m_{13}^2 \right\}.
\label{cparspace}
\eeq
Given the ubiquity of the decay chain of Fig.~\ref{fig:feynmandiag}(c) in SUSY and elsewhere, it is not surprising that the properties of the allowed region within this
invariant mass phase space have been extensively studied in the literature. Therefore, rather than reproducing previously published work, here we shall only state the
results most relevant to the current discussion, and for further details we refer the reader to the corresponding literature.

\begin{itemize}
\item {\em The shape of the allowed region in phase space.} The allowed region in the visible invariant mass space (\ref{cparspace}) is compact,
and is bounded by the (closed) two-dimensional surface defined by eq.~(\ref{eq:cDeteq0}). Three-dimensional plots of the allowed region
can be found in Fig.~1 of \cite{Costanzo:2009mq}, Fig.~9 of \cite{Kim:2015bnd} and on pages 568-572 of the TASI lectures \cite{Lester:2013aaa}. 
\item {\em Density enhancement on the boundary.} The phase space singularities occur on the two-dimensional boundary of the allowed region, i.e.,
the condition $\Delta_4=0$ (which is equivalent to (\ref{eq:cDeteq0})) defines both the boundary of the allowed region as well as the singularity locations
\cite{Agrawal:2013uka}. Since the allowed region is three-dimensional, it is difficult to visualize this enhancement unless one looks at two-dimensional slices through
the allowed region --- such plots can be found in Fig.~8 of \cite{Debnath:2016mwb} and Fig.~12 of \cite{Debnath:2016gwz}.
\item {\em The shape of the one-dimensional distribution of the singularity variable.} The differential distribution of $\Delta_4$ is known analytically. In terms of the unit-normalized variable 
$q\equiv \Delta_4/\Delta_4^{max}$ it is given by \cite{Debnath:2016mwb}
\beq
\frac{dN}{d q} =\frac{\arcsin(\sqrt{1-q})}{2\sqrt{q}}.
\eeq
The sharp peak at $\Delta_4=0$ can be used for discovering such new physics signal over the smooth SM background, as discussed in \cite{Debnath:2018azt}.
\item {\em Mass measurements and the focus point method.} Since computing the singularity variable $\Delta_4$ requires an ansatz for the mass spectrum, 
the focus point method for mass measurements \cite{Kim:2019prx} is in principle applicable, and one would be looking for a peak in the 4-dimensional parameter space of $\{M_0,M_1,M_2,M_3\}$.
However, the relevant observable parameter space (\ref{cparspace}), being only three-dimensional, is already simple enough so that in practice it may be easier to just perform a four-parameter fit to the 
boundary of the allowed region, as demonstrated in \cite{Debnath:2016gwz}.
\end{itemize}

\section{Two decay chains, each with one two-body decay} 
\label{sec:21}

In this section we shall simultaneously address the two event topologies shown in Figs.~\ref{fig:feynmandiag}(d) and \ref{fig:feynmandiag}(e).
The latter is known as the ``antler" topology \cite{Han:2009ss} and has been previously discussed in the context of both 
hadron colliders \cite{Baumgart:2006pa,Barr:2009mx,Han:2009ss,Choi:2009hn,Barr:2011he,Park:2011uz,Barr:2011ux,Edelhauser:2012xb,DeRujula:2012ns,Han:2012nm,Konar:2015hea}
and 
lepton colliders \cite{Christensen:2014yya,Choi:2015afa}.
At the same time, the diagram of Fig.~\ref{fig:feynmandiag}(d) is extremely common, and may represent many processes, 
including but not limited to $W$-pair production in the SM, squark or slepton production in SUSY, etc.
This diagram also has been extensively discussed in the literature ---  
at lepton colliders \cite{Feng:1993sd,HarlandLang:2012gn,Xiang:2016jni},
and especially at hadron colliders, where it offers a formidable challenge, despite its apparent simplicity  (for reviews, see \cite{Barr:2010zj,Matchev:2019sqa}).
In fact, it was precisely the diagram of Fig.~\ref{fig:feynmandiag}(d) which initially motivated a large number of 
now popular kinematic variables and techniques, including
the Cambridge $M_{T2}$ variable \cite{Lester:1999tx,Barr:2003rg}
and its variants \cite{Burns:2008va,Barr:2009jv,Konar:2009wn,Konar:2009qr}, 
the contransverse mass $M_{CT}$ \cite{Tovey:2008ui},
the $M_{CT2}$ variable \cite{Cho:2009ve,Cho:2010vz}, 
the MAOS method \cite{Cho:2008tj,Guadagnoli:2013xia}, and many others.

Given the similarities in the diagrams of Figs.~\ref{fig:feynmandiag}(d) and \ref{fig:feynmandiag}(e),
in this section we shall discuss them in one go by assuming the 4-momentum vector of the initial state ${\cal P}=({\cal P}_0,\vec{\cal P})$ to be fixed.
This is certainly true at lepton colliders, where the kinematics of the initial state is completely known: 
${\cal P}=(E_{CM},0,0,0)$, where $E_{CM}$ is the beam CM energy, which in the case of the antler diagram of Fig.~\ref{fig:feynmandiag}(e) 
can be tuned to be equal to the mass $M_2$ of the intermediate resonance, so that ${\cal P}$ becomes ${\cal P}=(M_2,0,0,0)$.
At hadron colliders, we will primarily focus on the antler diagram of Fig.~\ref{fig:feynmandiag}(e), 
for which the 4-momentum of the initial state can be written as ${\cal P}=(\sqrt{M_2^2+{\cal P}_z^2},0,0,{\cal P}_z)$, since
in general the resonance of mass $M_2$ will be produced with some non-zero
longitudinal momentum ${\cal P}_z$, whose size will be governed by the value of $M_2$ and the
parton distribution functions (pdfs) of the initial state partons. For completeness, we shall initially retain ${\cal P}_z$ 
in our formulas, but in the end, following \cite{DeRujula:2012ns},
we shall take the ``gluon collider" approximation ${\cal P}_z\approx 0$, 
which can be justified in cases where the pdfs of the initial state partons are the same (or similar) and are fast-falling functions, e.g., as in the SM process $H\to W^+W^-$ \cite{DeRujula:2012ns}.

\subsection{Derivation of a singularity coordinate}
\label{sec:devariable}

The event topologies of Figs.~\ref{fig:feynmandiag}(d) and \ref{fig:feynmandiag}(e) have the following constraints in common:
\begin{subequations}
\begin{eqnarray}
q^2 &=& M_0^2, \label{eq:delittleq}\\[1mm]
Q^2 &=& M_0^2, \label{eq:deQ} \\[1mm]
(q+p_1)^2 &=& M_1^2,  \label{eq:delittleqp1}\\[1mm]
(Q+P_1)^2 &=& M_1^2,  \label{eq:deQP1} \\[1mm]
q_x + Q_x&=& \mptx = -p_{1x} - P_{1x}, \label{eq:deqQx} \\ [1mm]
q_y + Q_y&=& \mpty = -p_{1y} - P_{1y}, \label{eq:deqQy} 
\end{eqnarray}
\end{subequations}
where in the last two equations we have assumed that there is no $P_T^{ISR}$ accompanying our event topology,
hence $\mpt= -\vec{p}_{1T}-\vec{P}_{1T}$.

Since we are treating the initial state kinematics as fixed, we can add two more relations representing energy conservation 
and longitudinal momentum conservation, respectively
\begin{subequations}
\begin{eqnarray}
\epsilon + {\cal E}&=& {\cal P}_0-e_1-E_1, \label{eq:deqQe} \\ 
q_z + Q_z&=& {\cal P}_z-p_{1z}-P_{1z}, \label{eq:deqQz} 
\end{eqnarray}
\end{subequations}
with some fixed ${\cal P}_0$ and ${\cal P}_z$ as discussed above.
Eqs.~(\ref{eq:deqQx}-\ref{eq:deqQz}) can now be used to eliminate the 4-momentum $Q$ as
\beq
Q = {\cal P} - p_1 - P_1 - q.
\eeq
Substituting this into eqs.~(\ref{eq:deQ}) and (\ref{eq:deQP1}), and again limiting ourselves for simplicity 
to the case of massless visible particles, $p_1^2=P_1^2=0$,
we can rewrite the four remaining constraints (\ref{eq:delittleq}-\ref{eq:deQP1}) as
\begin{subequations}
\begin{eqnarray}
q^2 &=& M_0^2, \label{deqq}\\
2\, p_1\cdot q &=& M_1^2-M_0^2, \label{dep1q}\\
2\, P_1\cdot q &=& -M_1^2+M_0^2+2({\cal P}-p_1)\cdot P_1, \label{decapPq} \\
2\, {\cal P}\cdot q &=& {\cal P}^2-2{\cal P}\cdot p_1. \label{dep3q}
\end{eqnarray}
\label{systemFigde}%
\end{subequations}

\begin{figure}[t]
 \centering
 \includegraphics[height=.3\textwidth]{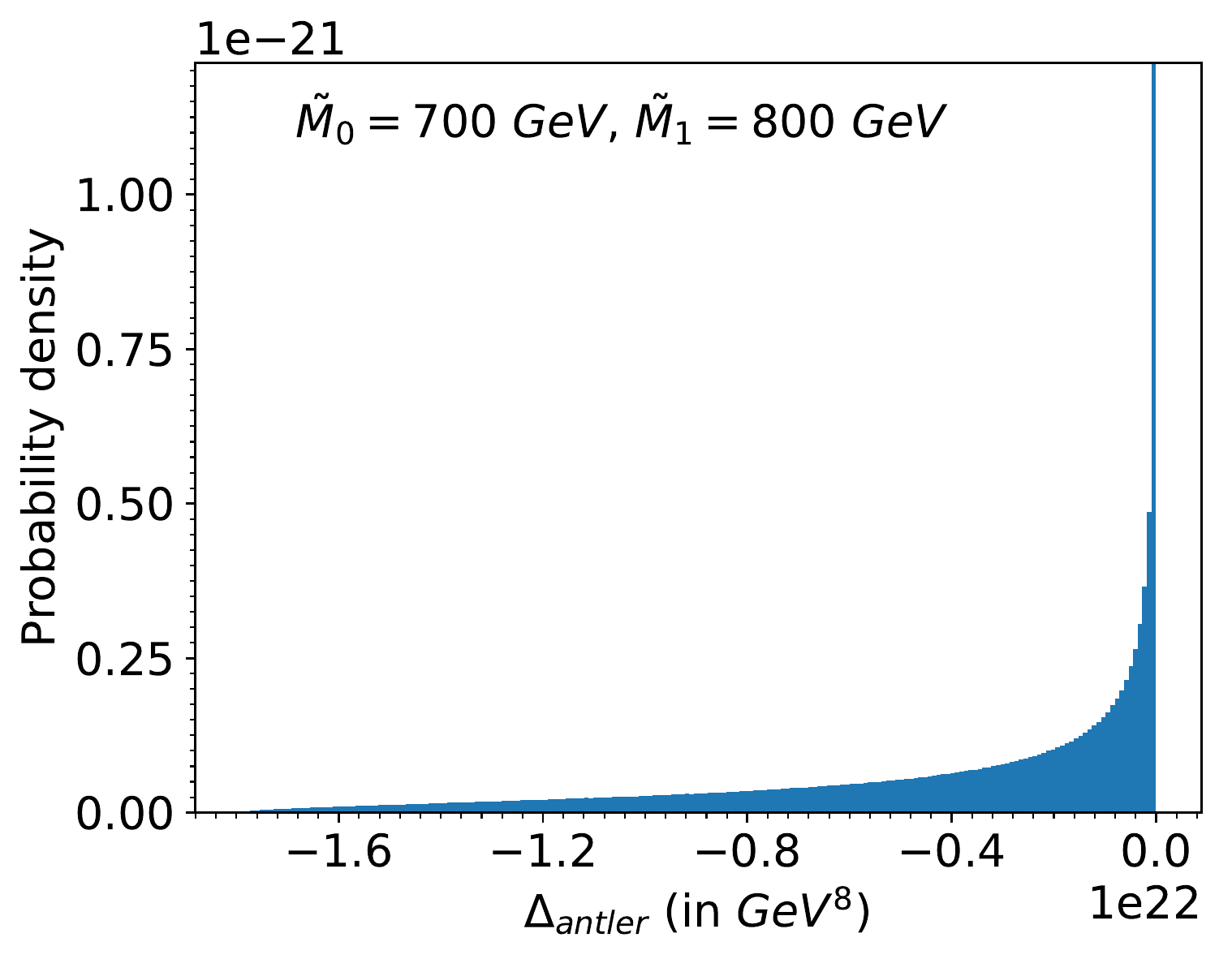}
 \hskip 5mm
 \includegraphics[height=.3\textwidth]{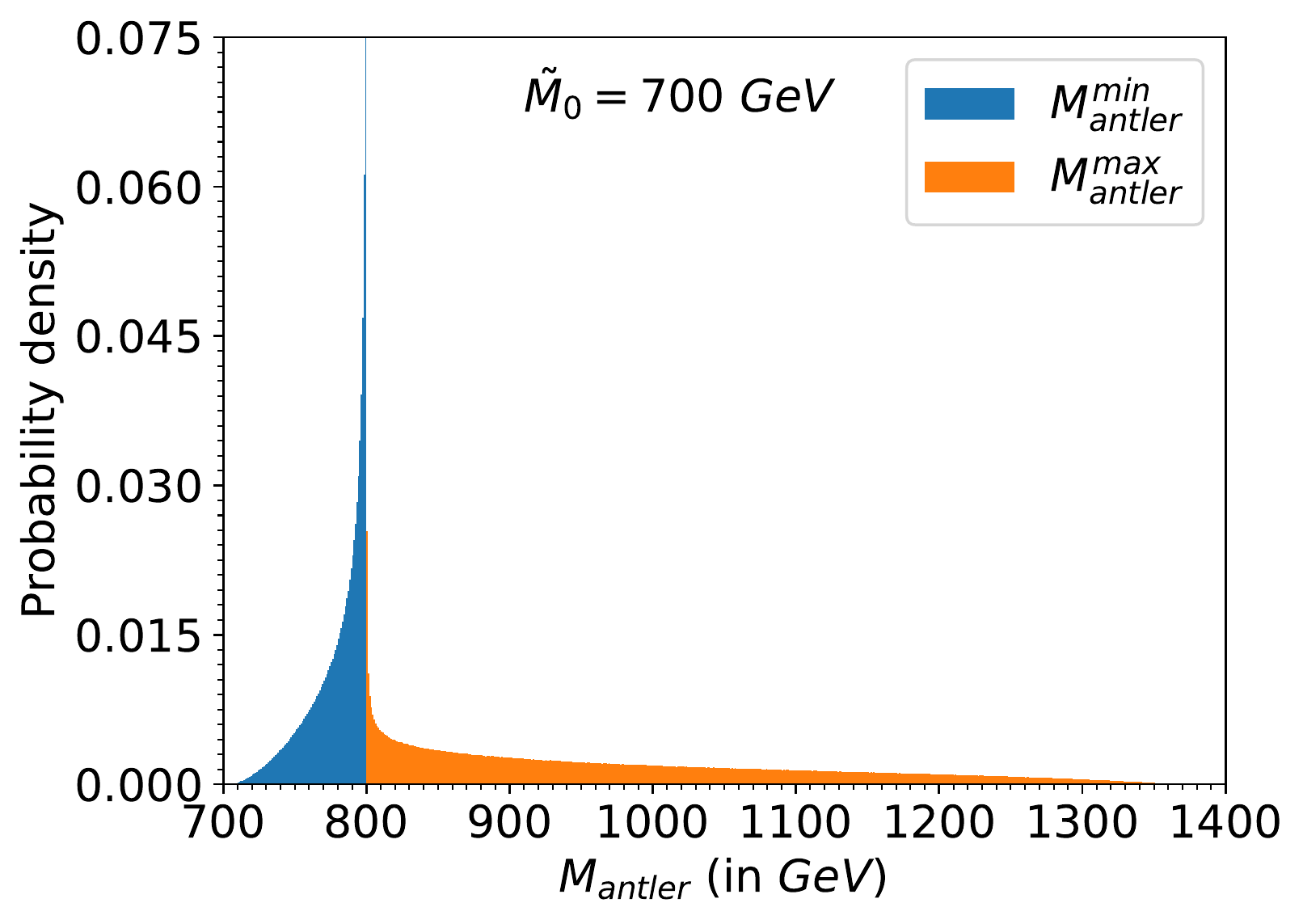}
 \caption{\label{fig:antler1d} One-dimensional distributions of the singularity variables $\Delta_{antler}(M_1,M_0)$ (left panel) and $M_{antler}(M_0)$ (right panel).
 In each plot, we use the true mass(es) for the respective mass parameter ansatz.
 Events were generated with the mass spectrum from Fig.~\ref{tab:mass} at a lepton collider with ${\cal P}=(M_2,0,0,0)$ and $M_2=3000$ GeV.
 In the right panel, the orange (blue) histogram corresponds to the larger solution $M_{antler}^{max}$ (smaller solution $M_{antler}^{min}$) of eq.~(\ref{eq:Mantler}).
  }
\end{figure}

From here we can compute the Jacobian matrix (\ref{defJac}) 
\beq
D =
\left(
\begin{array}{cccc}
2\varepsilon & -2 q_{x} & -2 q_{y} & -2 q_{z} \\
2e_1 & -2 p_{1x} & -2 p_{1y} & -2 p_{1z}   \\
2E_1 & -2 P_{1x} & -2 P_{1y}  & -2 P_{1z}  \\
2{\cal P}_0 & 0 & 0  & -2 {\cal P}_{z}  
\end{array}
\right)
\eeq
and the singularity condition then reads
\beq
{\rm Det}\, D = 0 \Longleftrightarrow
{\rm Det}\, (D\eta D^T) = 0,
\label{eq:DetaDTeq0}
\eeq
which implies
\beq
\left|
\begin{array}{cccc}
~2q^2~ & ~~2p_1\cdot q~~ & ~~2P_1\cdot q~~ & ~~2{\cal P}\cdot q~~  \\ [1mm]
2p_1\cdot q & 0 & 2p_1\cdot P_1 & 2{\cal P}\cdot p_1 \\[1mm]
2P_1\cdot q & 2p_1\cdot P_1 & 0  & 2{\cal P}\cdot P_1 \\[1mm]
2{\cal P}\cdot q & 2{\cal P}\cdot p_1 & 2{\cal P}\cdot P_1  & 2{\cal P}^2 
\end{array}
\right|
=0.
\eeq
Using (\ref{systemFigde}) to eliminate $q$, after a couple of row and column manipulations, this simplifies to
\beq
\Delta_{antler}\equiv
\left|
\begin{array}{cccc}
2M_1^2 & ~M_1^2-M_0^2~ & ~-M_1^2+M_0^2+2{\cal P}\cdot P_1~  & {\cal P}^2\\ [2mm]
M_1^2-M_0^2  & 0 & 2p_1\cdot P_1 & 2{\cal P}\cdot p_1\\ [2mm]
-M_1^2+M_0^2+2{\cal P}\cdot P_1 & 2p_1\cdot P_1 & 0 & 2{\cal P}\cdot P_1\\ [2mm] 
{\cal P}^2 & 2{\cal P}\cdot p_1 & 2{\cal P}\cdot P_1 & 2{\cal P}^2
\end{array}
\right|
=0,
\label{eq:deDeteq0}
\eeq
where the left-hand side is precisely the desired singularity variable $\Delta_{antler}$ for the case of Figs.~\ref{fig:feynmandiag}(d) and \ref{fig:feynmandiag}(e).
Eq.~(\ref{eq:deDeteq0}) indicates that the relevant phase space of kinematic observables is 
three-dimensional\footnote{This can be understood as follows. The observable momenta $p_1$ and $P_1$ are parametrized by 6 degrees of freedom,
however, three of those correspond to rotations in the CM frame (to which we can boost knowing ${\cal P}$), which will leave the 
set of constraints (\ref{systemFigde}) for $q$ invariant.} and can be parametrized, e.g., as
\beq
\left\{ {\cal P}\cdot p_1, ~{\cal P}\cdot P_1, ~p_1\cdot P_1 \right\}.
\label{3dotproducts}
\eeq
Just like the singularity variable $\Delta_4$ from Sec.~\ref{sec:13},
the singularity variable $\Delta_{antler}$ depends not only on the phase space (\ref{3dotproducts}), but also requires an ansatz $\tilde M_1$ and $\tilde M_0$ 
for the two mass parameters, i.e., $\Delta_{antler}(\tilde M_1, \tilde M_0 )$. If the ansatz is correct ($\tilde M_1=M_1$ and $\tilde M_0=M_0$), then
the singularity condition (\ref{eq:deDeteq0}) guarantees that the distribution of $\Delta_{antler}$
exhibits a singularity peak at 
\beq
\Delta_{antler} (M_1, M_0 ) =0.
\label{eq:Dantler}
\eeq
This is demonstrated in the left panel of Fig.~\ref{fig:antler1d}, which shows the one-dimensional unit-normalized distribution of 
$\Delta_{antler} (M_1, M_0 )$ at a lepton collider with $E_{CM}(=M_2)=3000$ GeV, and with the mass spectrum from Table~\ref{tab:mass},
$M_1=800$ GeV and $M_0=700$ GeV. As expected, there is a sharp peak at $\Delta_{antler}=0$. We also note that
with our conventions the values of $\Delta_{antler}$ are negative --- this can be traced back to eq.~(\ref{eq:DetaDTeq0}) and 
the fact that with the correct choice of mass parameters $D$ is guaranteed to be real and thus ${\rm Det}\, DD^T>0$, while ${\rm Det}\, \eta=-1$.

\begin{figure}[t]
 \centering
 \includegraphics[width=.5\textwidth]{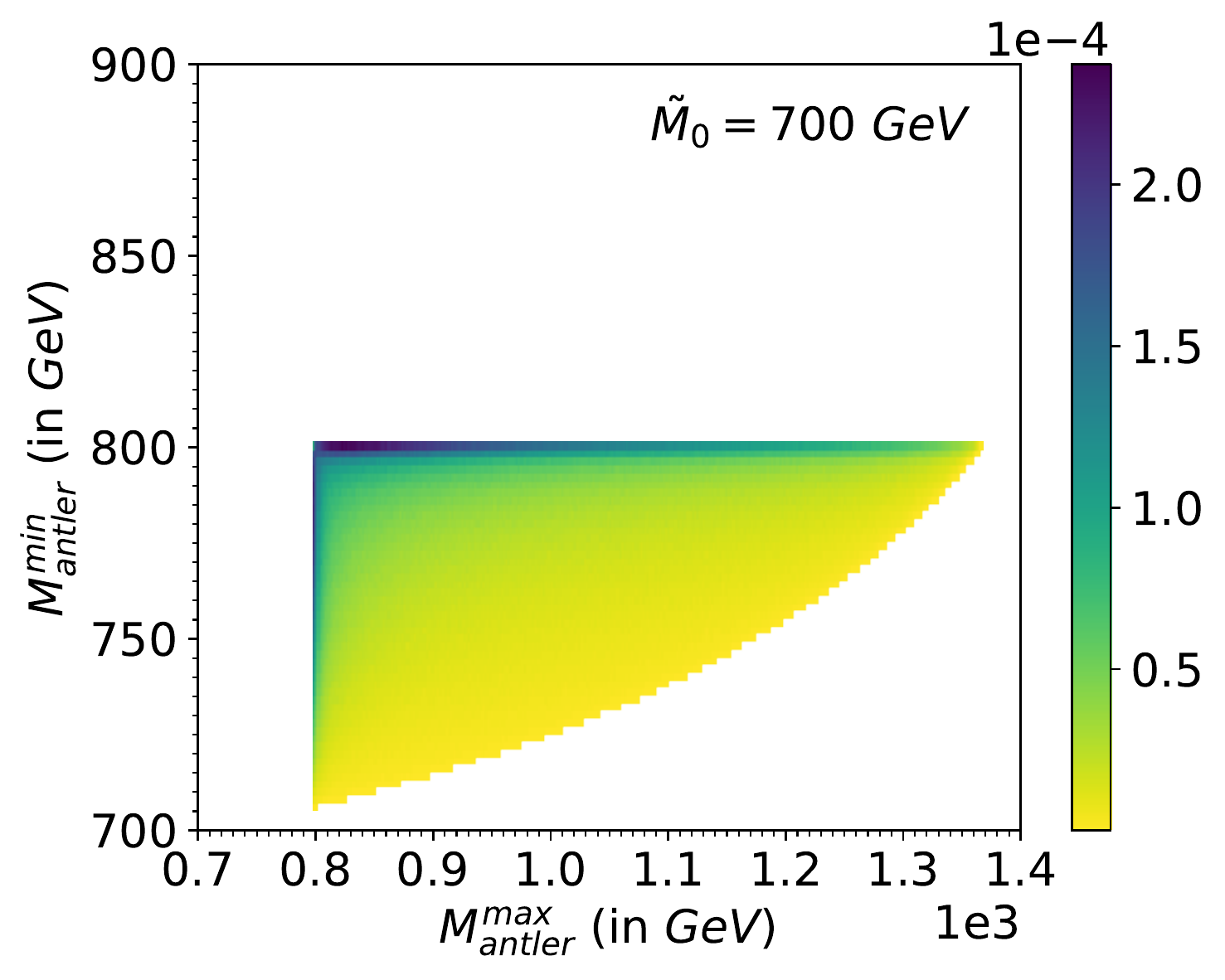}
 \caption{\label{fig:antler2d} A heatmap of the two-dimensional distribution of events versus the two solutions for the singularity variable $M_{antler}$, namely
 $M_{antler}^{max}$ (plotted on the $x$-axis) and $M_{antler}^{min}$ (plotted on the $y$-axis). }
\end{figure}

However, the mass spectrum may not always be known {\em a priori}, thus one would like to reduce the number of mass ansatze as much as possible.
To this end, we can follow the idea of the transverse mass $m_T$ from Sec.~\ref{sec:11}, 
which takes only $\tilde M_0$ as an input and then uses the corresponding singularity condition~(\ref{aMTcond}) to define the
singularity variable. In our case, we can use (\ref{eq:Dantler}) to define implicitly an alternative singularity variable $M_{antler}(\tilde M_0)$ 
as the solution to the equation
\beq
\Delta_{antler} (M_{antler}, \tilde M_0 ) =0.
\label{eq:Mantler}
\eeq
The distribution of $M_{antler}$ for $\tilde M_0=M_0$ is shown in the right panel of Fig.~\ref{fig:antler1d}.
Notice that the defining equation (\ref{eq:Mantler}) leads to a quadratic equation for $M_{antler}$ and therefore to two possible solutions,
both of which are entered in the plot. The larger of the two solutions, which we label $M_{antler}^{max}$, comprises the 
orange histogram, while the smaller one, labelled $M_{antler}^{min}$, makes up the blue histogram, and the
two histograms are individually normalized to $1$. As expected, the right panel of Fig.~\ref{fig:antler1d}
exhibits the presence of a sharp peak at the correct mass of the parent particle, $M_1=800$ GeV.
It is interesting to note that both solutions $M_{antler}^{min}$ and $M_{antler}^{max}$ are contributing to the singularity --- from below and from above, respectively.
Another noteworthy feature of the plot is that the orange and blue histograms do not overlap at all --- in fact, they meet at the true value of the mass $M_1=800$ GeV
(which is also the location of the singularity peak). This can be seen even more clearly in Fig.~\ref{fig:antler2d}, which 
shows the two-dimensional distribution of events as a heatmap in the plane of $(M_{antler}^{max},M_{antler}^{min})$.

The heatmap in Fig.~\ref{fig:antler2d} reveals an overdensity of events at both $M_{antler}^{min}=M_1$ 
(the horizontal blue-shaded band) and $M_{antler}^{max}=M_1$ (the vertical blue-shaded band).
This confirms that both solutions for $M_{antler}$ play a role in forming the singularity peak observed in the right panel of Fig.~\ref{fig:antler1d}.
Note also the absence of any events with $M_{antler}^{min}>M_1$ and $M_{antler}^{max}<M_1$,
which implies that the following hierarchy is always true:
\beq
M_{antler}^{min} \le M_1 \le M_{antler}^{max}.
\eeq
In other words, the true parent mass $M_1$ is always located between the two found solutions for $M_{antler}$, and furthermore,
$M_1$ is the upper kinematic endpoint of $M_{antler}^{min}$ (the blue histogram in the right panel of Fig.~\ref{fig:antler1d})
and at the same time it is also the lower kinematic endpoint of $M_{antler}^{max}$ (the orange histogram in the right panel of Fig.~\ref{fig:antler1d}).
This can be understood as follows. The two solutions for $M_{antler}$ obtained from the singularity condition (\ref{eq:Mantler}) 
represent the two values for the trial mass $\tilde M_1$ where the number of solutions for the invisible momentum $q$ changes \cite{Kim:2019prx}, in this case between 0 and 2.
Since the true mass $M_1$ will always give valid solutions for $q$, it belongs to the allowed interval for $\tilde M_1$ with 2 solutions, 
which is sandwiched between the two disallowed regions with 0 solutions.

In summary, the discussion in this subsection (and Fig.~\ref{fig:antler1d} in particular) 
shows that both $\Delta_{antler}$ and $M_{antler}$ are valid singularity variables,
albeit the latter has the added advantages of having a clear physical meaning and being of mass dimension 1 only.

\subsection{The phase space geometry of the singularity condition} 
\label{sec:antlersingularity}

Having derived the singularity variables for the event topologies of Figs.~\ref{fig:feynmandiag}(d) and \ref{fig:feynmandiag}(e),
we can now discuss the geometry of the singularity surface in the relevant observable phase space. The latter
can be parametrized as in (\ref{3dotproducts}), but the equation of the singularity surface (\ref{eq:deDeteq0}) 
can be written more compactly if we use an alternative set of observables
\begin{subequations}
\bea
{\cal X}&=& 2({\cal P}\cdot p_1+ {\cal P}\cdot P_1), \\ [1mm]
{\cal Y}&=& 2({\cal P}\cdot p_1- {\cal P}\cdot P_1), \\[1mm]
{\cal Z}&=& \frac{4({\cal P}\cdot p_1)({\cal P}\cdot P_1)}{{\cal P}^2} - 2(p_1\cdot P_1)
\eea
\label{eq:XYZdef}%
\end{subequations}
which reduces (\ref{eq:deDeteq0}) to the constraint
\beq
\frac{\left[{\cal X} - {\cal P}^2\left(1-\frac{M_0^2}{M_1^2}\right)\right]^2}{\left(1-\frac{M_0^2}{M_1^2}\right)^2-\frac{\cal Z}{M_1^2}} + \frac{{\cal Y}^2}{\frac{\cal Z}{M_1^2}}
= {\cal P}^2\left({\cal P}^2 - 4M_1^2\right).
\label{eq:antlersamosa}
\eeq
This equation describes a closed surface whose cross-sections at fixed ${\cal Z}$ are ellipses in the $({\cal X},{\cal Y})$ plane.
As discussed earlier, for illustration purposes, we shall now fix the momentum of the initial state as ${\cal P}=(M_2,0,0,0)$,
which can be viewed as a lepton collider running at a CM energy $E_{CM}=M_2=3000$ GeV and producing either one
of the diagrams in Figs.~\ref{fig:feynmandiag}(d) and \ref{fig:feynmandiag}(e), or as a hadron collider producing the antler topology, 
where one neglects the longitudinal momentum of the heavy $s$-channel resonance \cite{DeRujula:2012ns}.
In that limit, the variables (\ref{eq:XYZdef}) become
\begin{subequations}
\bea
{\cal X}&=& 2M_2(e_1+ E_1), \label{Xdef}\\ 
{\cal Y}&=& 2M_2(e_1- E_1), \label{Ydef}\\
{\cal Z}&=& 2(e_1E_1+\vec{p}_1\cdot \vec{P}_1)\equiv M_C^2. \label{Zdef}
\eea
\label{XYZdef}%
\end{subequations}
In the last line we recognize the quantity $M_C$ introduced in \cite{Tovey:2008ui}, which is invariant under contra-linear (back-to-back) boosts.
We can use this boost invariance to bring the two intermediate particles of mass $M_1$ to their corresponding rest frames (along with their decay products), 
which then allows us to express the quantity $M_C$ as
\beq
M_C^2 \equiv \frac{M_1^2}{2}\left(1-\frac{M_0^2}{M_1^2}\right)^2(1+\cos\theta^\ast),
\label{eq:MCdef}
\eeq
where $\theta^\ast$ is the angle between $\vec{p}_1$ and $\vec{P}_1$ after the respective boosts.  With the help of (\ref{XYZdef}) and (\ref{eq:MCdef}),
the equation of the singularity surface (\ref{eq:antlersamosa}) becomes
\beq
\frac{\left[(e_1+E_1) - \frac{M_2}{2}\left(1-\frac{M_0^2}{M_1^2}\right)\right]^2}{\sin^2\frac{\theta^\ast}{2}} + \frac{(e_1-E_1)^2}{\cos^2\frac{\theta^\ast}{2}}
= \frac{M_2^2}{4}\left(1-\frac{M_0^2}{M_1^2}\right)^2\left(1- \frac{4M_1^2}{M_2^2}\right).
\label{eq:antlersamosa2}
\eeq

\begin{figure}[t]
 \centering
 \includegraphics[width=.45\textwidth]{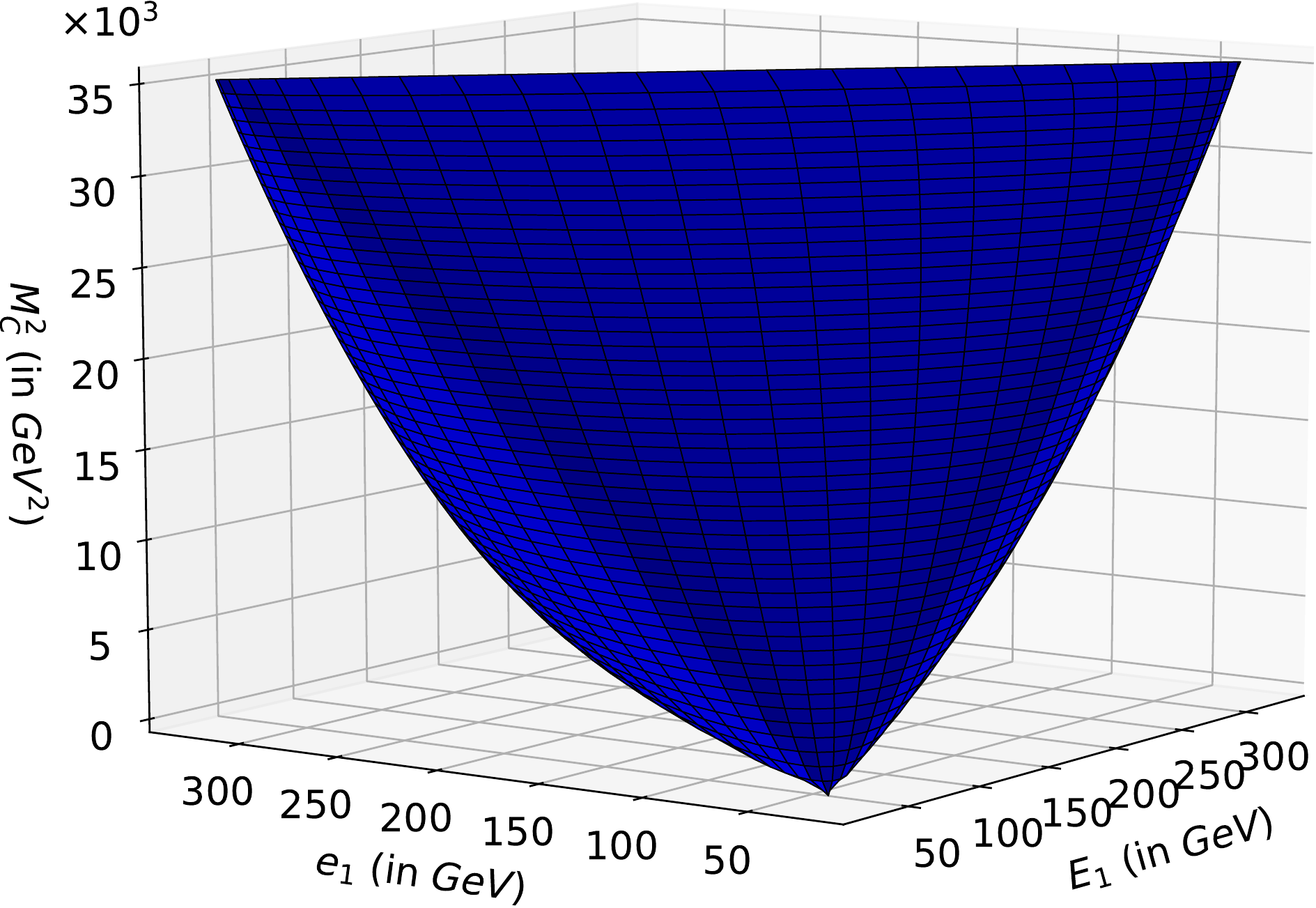}
 \hskip 5mm
 \includegraphics[width=.45\textwidth]{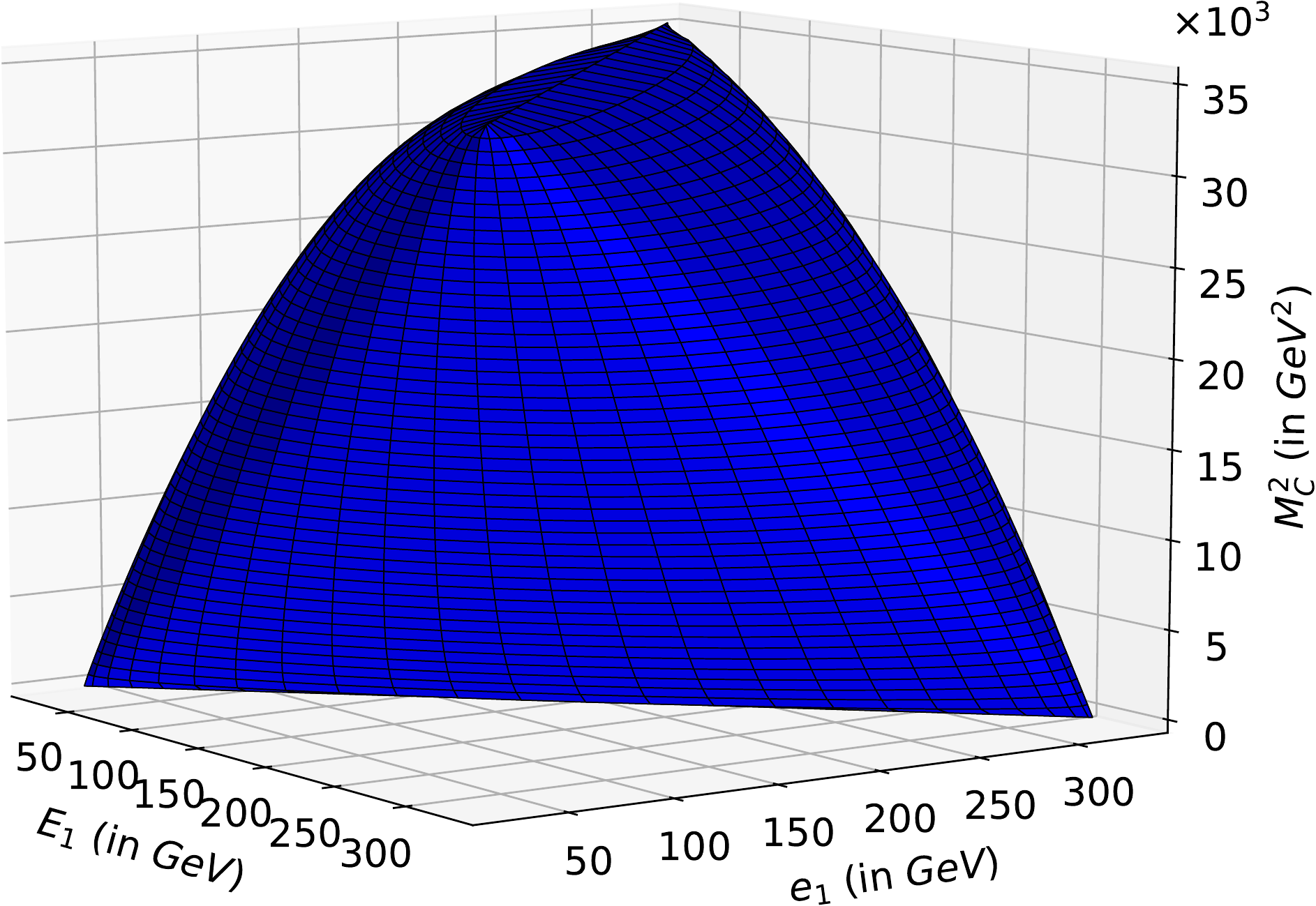}
 \caption{\label{fig:antler3D} Two different views of the singularity surface defined by eqs.~(\ref{eq:antlersamosa}) and (\ref{eq:antlersamosa2}). 
 The observable phase space is parametrized as $(e_1,E_1,M_C^2)$.}
\end{figure}

\begin{figure}[t]
 \centering
 \includegraphics[width=.3\textwidth]{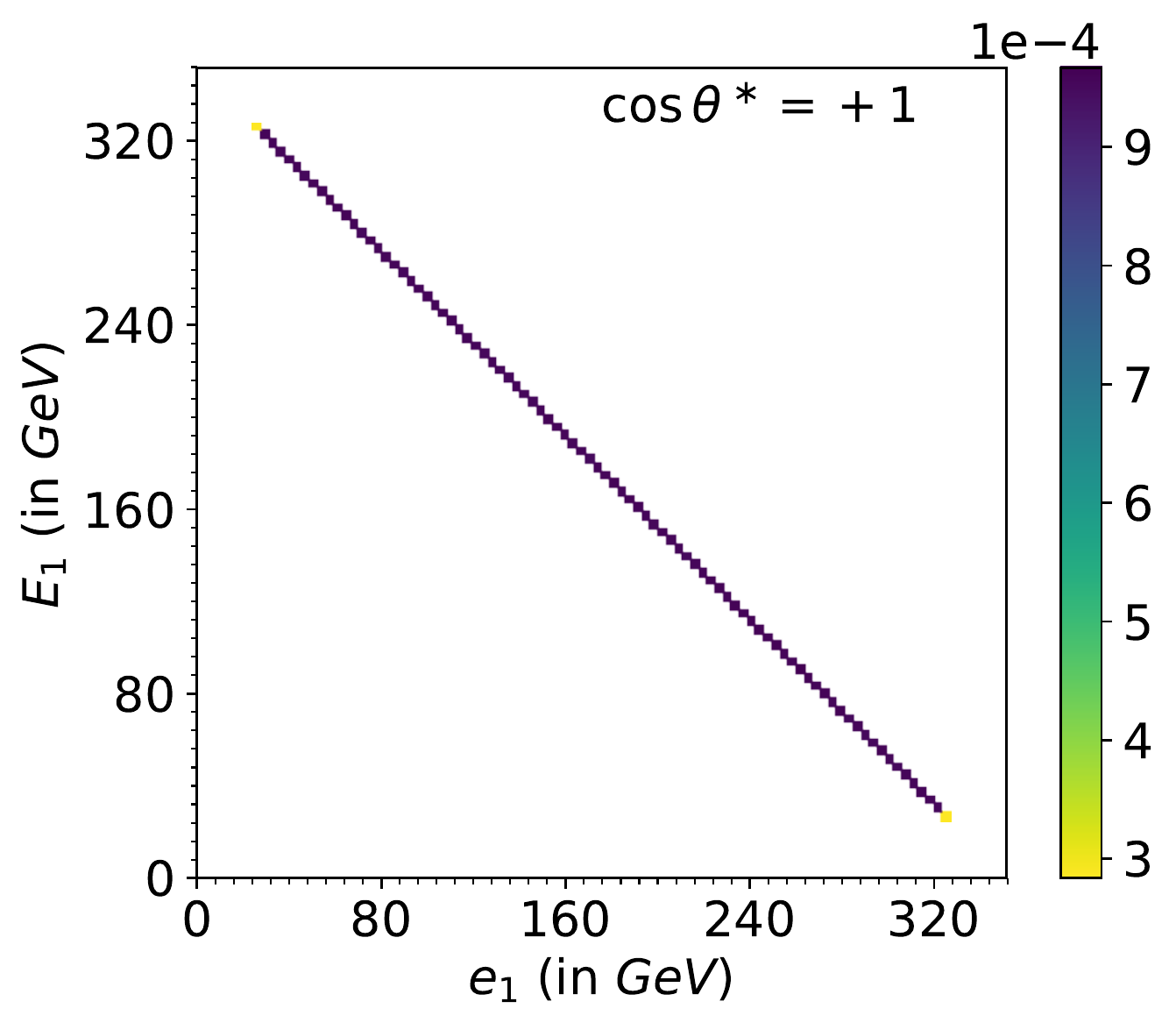}
 \hskip 5mm
 \includegraphics[width=.3\textwidth]{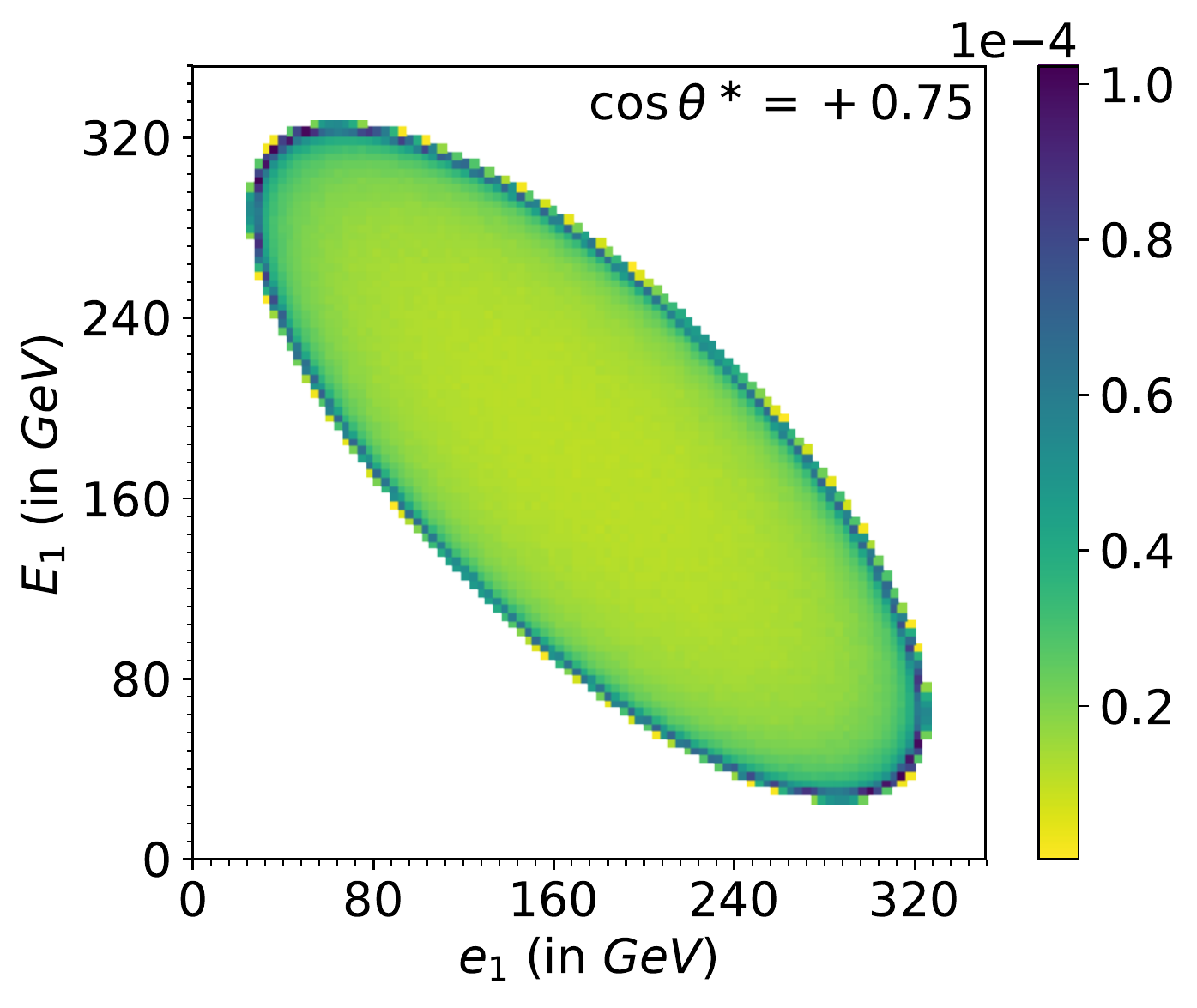}
 \hskip 5mm
 \includegraphics[width=.3\textwidth]{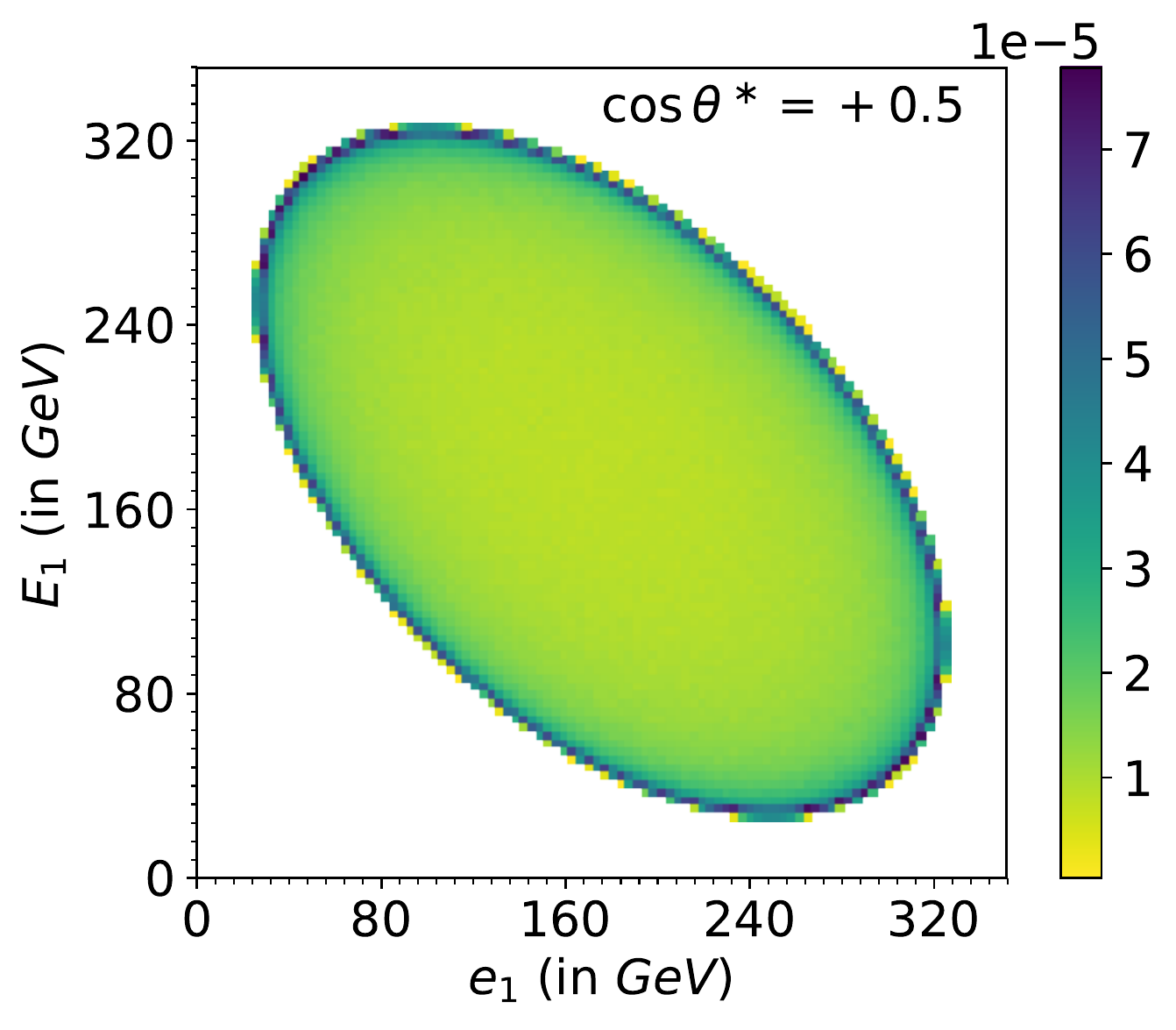}\\
 \includegraphics[width=.3\textwidth]{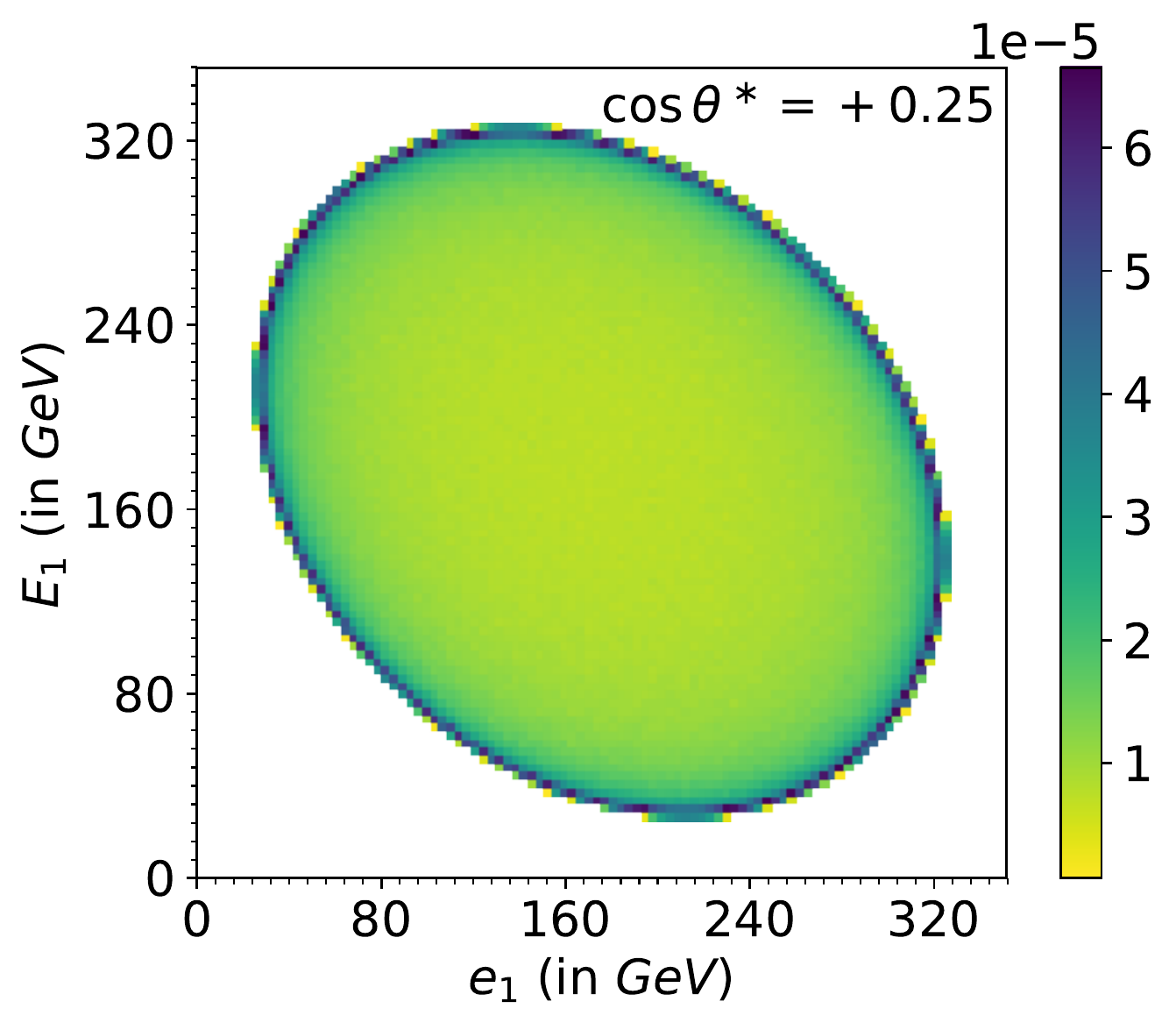}
 \hskip 5mm
 \includegraphics[width=.3\textwidth]{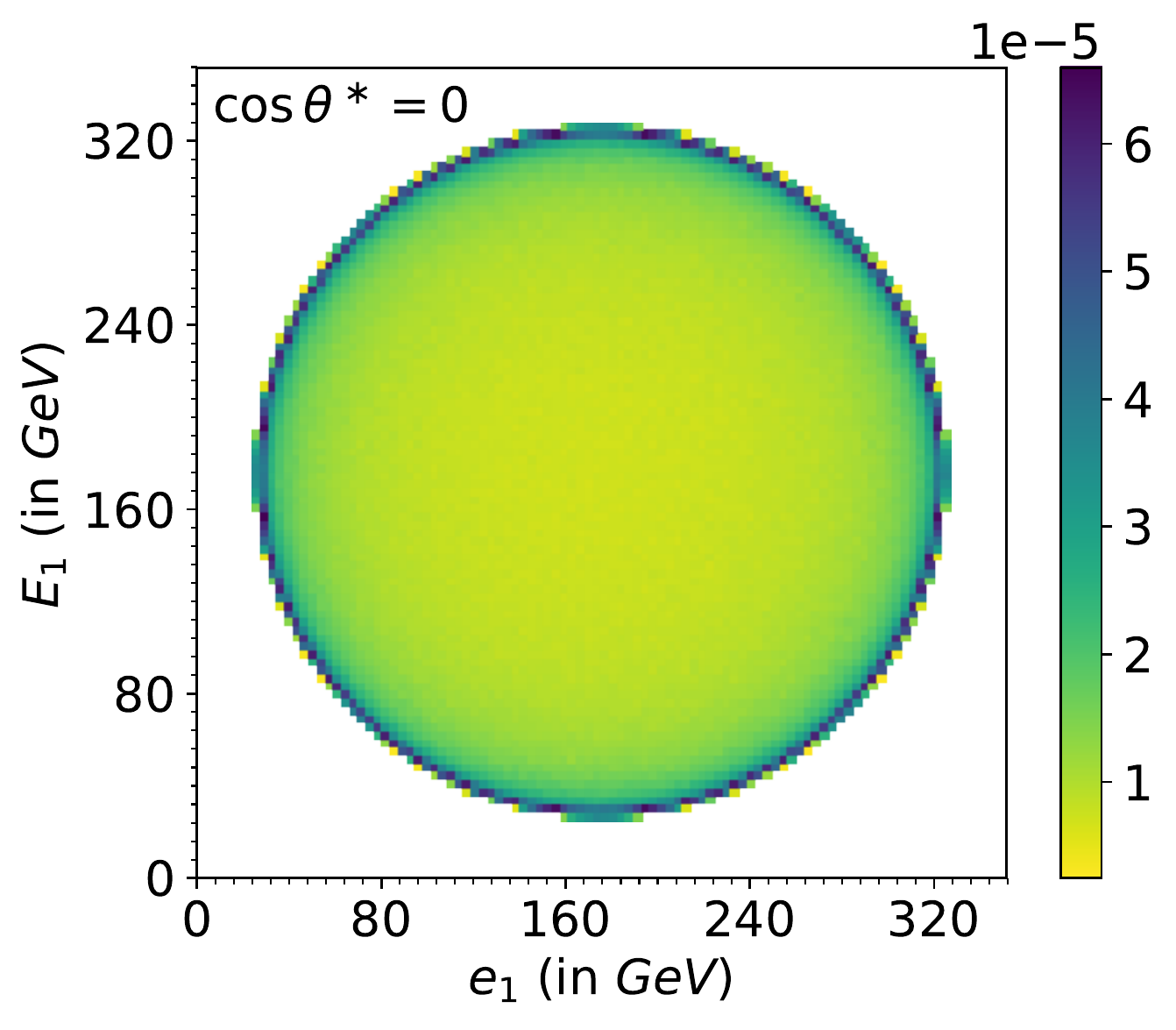}
 \hskip 5mm
 \includegraphics[width=.3\textwidth]{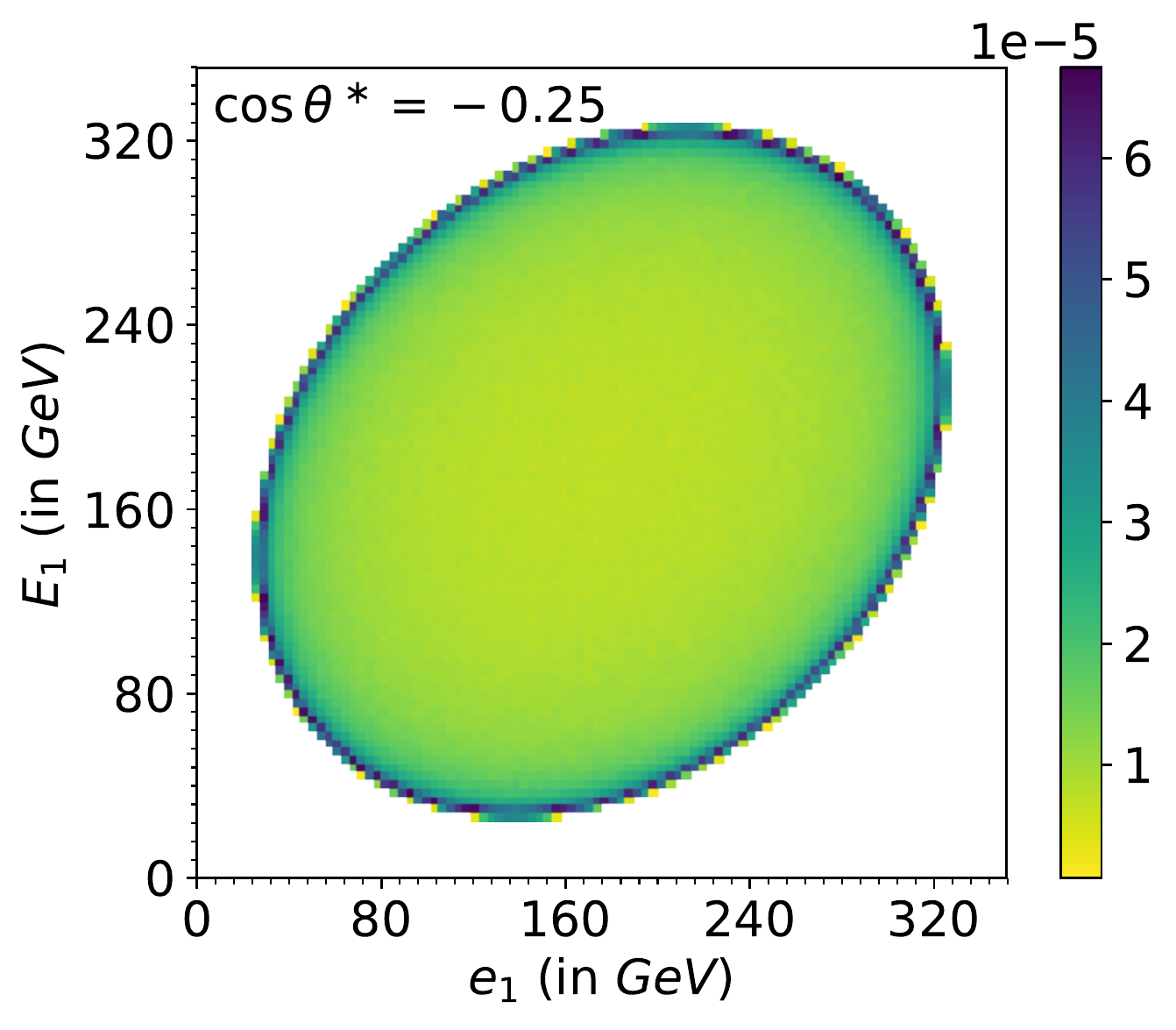}\\
 \includegraphics[width=.3\textwidth]{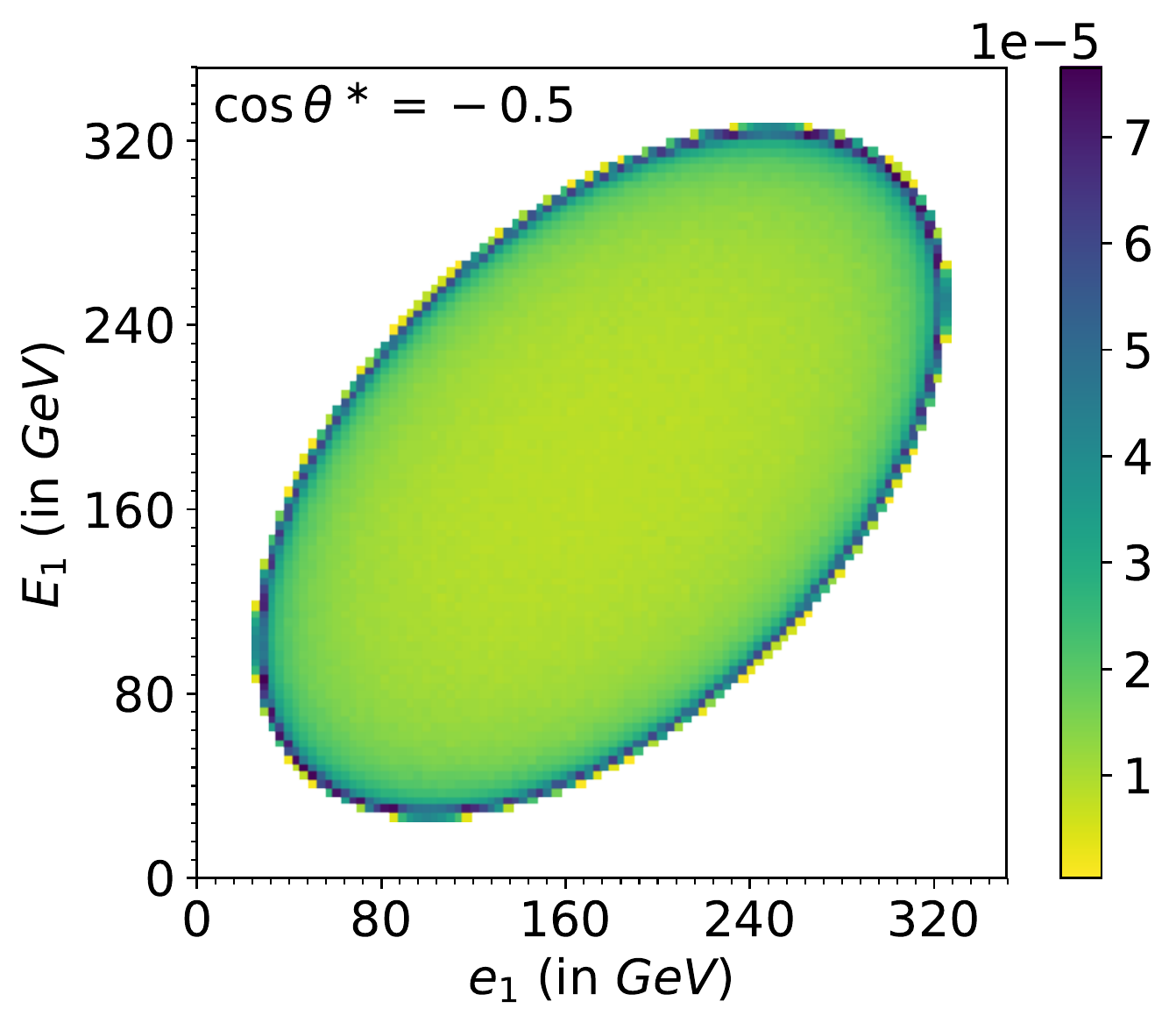}
 \hskip 5mm
 \includegraphics[width=.3\textwidth]{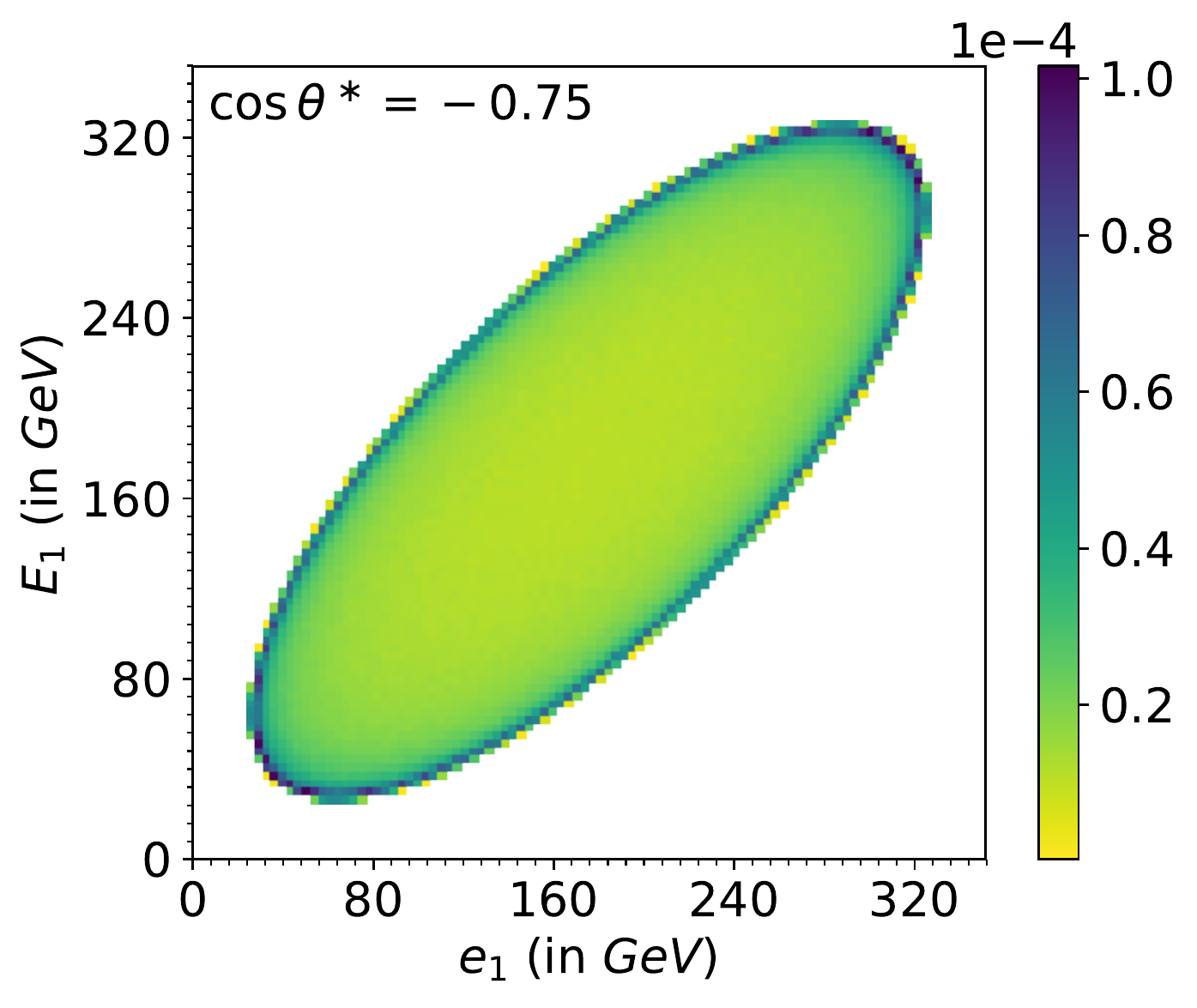}
 \hskip 5mm
 \includegraphics[width=.3\textwidth]{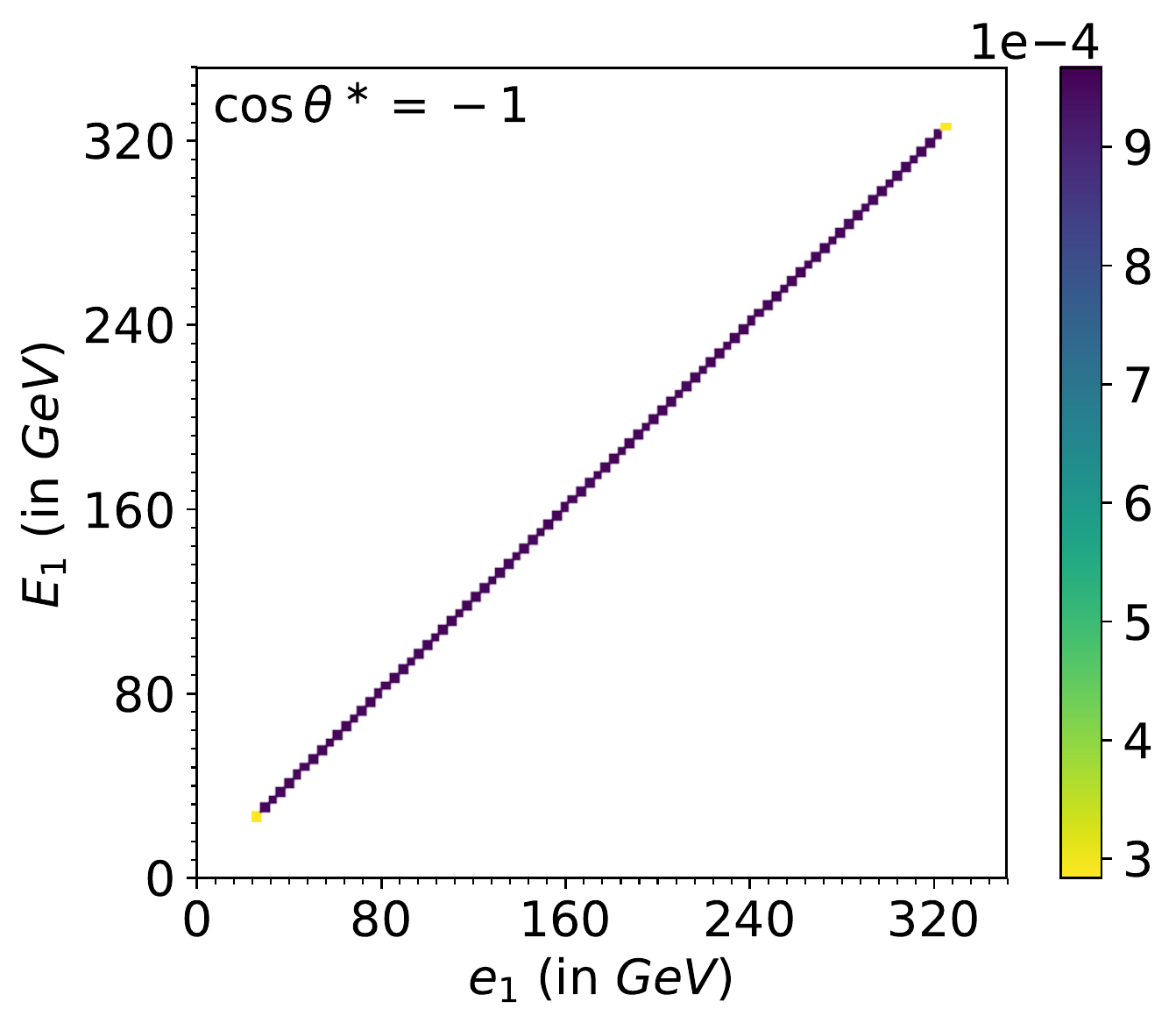}
 \caption{\label{fig:antlerphasespace} Allowed values for the energies $e_1$ and $E_1$ of the two visible particles 
  for different fixed values of $M_C$, or equivalently, $\cos\theta^\ast$.
  The plots from top to bottom are for fixed $\cos\theta^\ast=\{+1.0,+0.75,+0.50,+0.25,0,-0.25,-0.50,-0.75,-1.0\}$.
  }
\end{figure}

The singularity surface defined by this equation is pictorially illustrated in Figs.~\ref{fig:antler3D} and \ref{fig:antlerphasespace}.
Fig.~\ref{fig:antler3D} is analogous to the plot in the left panel of Fig.~\ref{fig:ellipses} from Sec.~\ref{sec:PTISR}, 
where the space of relevant observables was three-dimensional as well.
In Fig.~\ref{fig:antler3D}, we show two different views of the singularity surface when, as suggested by eq.~(\ref{XYZdef}), 
the observable phase space is parametrized as $(e_1,E_1,M_C^2)$.\footnote{The plots in Fig.~\ref{fig:antler3D} can be contrasted to the plots in Fig.~3 
of Ref.~\cite{DeRujula:2012ns}, which were done with the parametrization $\left(p_{1T},P_{1T}, \frac{ \vec{p}_{1T}\cdot \vec{P}_{1T}}{p_{1T}P_{1T}}\right)$. } 

Fig.~\ref{fig:antlerphasespace}, on the other hand, shows a series of plots in analogy to Fig.~\ref{fig:1stepISR}.
Each individual panel depicts the allowed range of values for the energies $e_1$ and $E_1$ of the two visible particles 
for a given fixed value of $M_C$, or equivalently, for a fixed value of $\cos\theta^\ast$, since the two are related by eq.~(\ref{eq:MCdef}).
For this figure we prefer to work with $\cos\theta^\ast$ (and equally spaced fixed values for it) since the distribution in $\cos\theta^\ast$ is uniform, 
thus each panel in Fig.~\ref{fig:antlerphasespace} has the same total number of events.
The event number density in the $(e_1,E_1)$ plane is indicated by the color bar. We see that, as expected, 
the events tend to cluster on the singularity boundary, whose shape is an ellipse of varying eccentricity depending on the value of $\cos\theta^\ast$.
The distortion in the shape of the elliptical boundary can be easily tracked and understood with the help of eq.~(\ref{eq:antlersamosa2}).
Consider, for example, the case of $\cos\theta^\ast=+1$ shown in the upper left panel of Fig.~\ref{fig:antlerphasespace}.
This implies that $\theta^\ast=0$ and the first term in the left hand side of (\ref{eq:antlersamosa2}) dominates, which in turn implies the linear relation
 $e_1+E_1=\frac{M_2}{2}\left(1-\frac{M_0^2}{M_1^2}\right)=351.6$ GeV seen in the plot.
As the value of $\cos\theta^\ast$ decreases, the elliptical boundary becomes less eccentric, and for $\cos\theta^\ast=0$, i.e.,
$\theta^\ast=\frac{\pi}{4}$, it eventually becomes a circle, as seen in the middle plot of the middle row. 
As the value of $\cos\theta^\ast$ decreases further, the ellipse begins to stretch along the orthogonal
$e_1=E_1$ direction, and for $\cos\theta^\ast=-1$ it simply becomes the line $e_1=E_1$.

\subsection{The focus point method} 
\label{sec:deFP}

\begin{figure}[t]
 \centering
 \includegraphics[height=.3\textwidth]{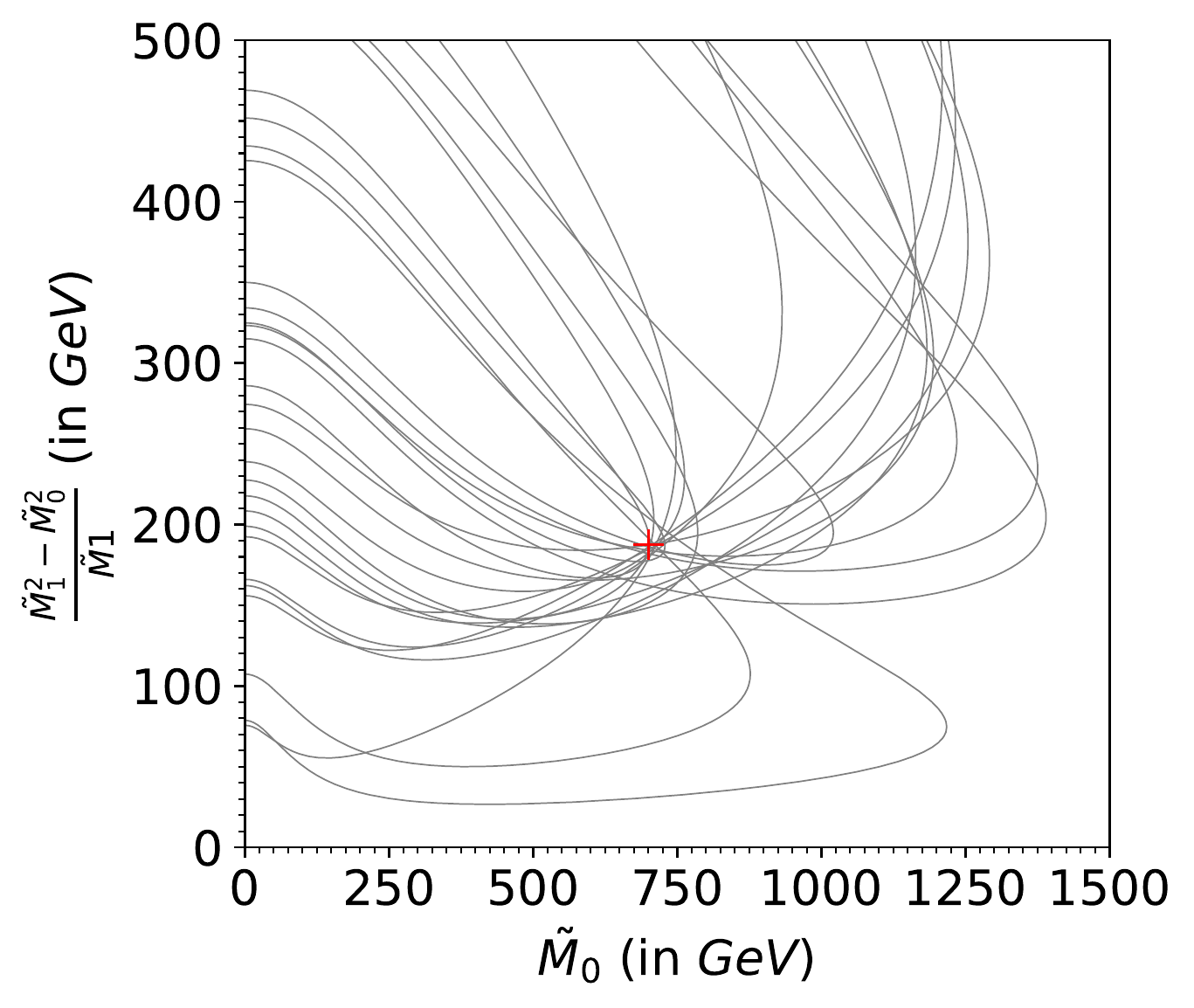}
 \hskip 5mm
 \includegraphics[height=.3\textwidth]{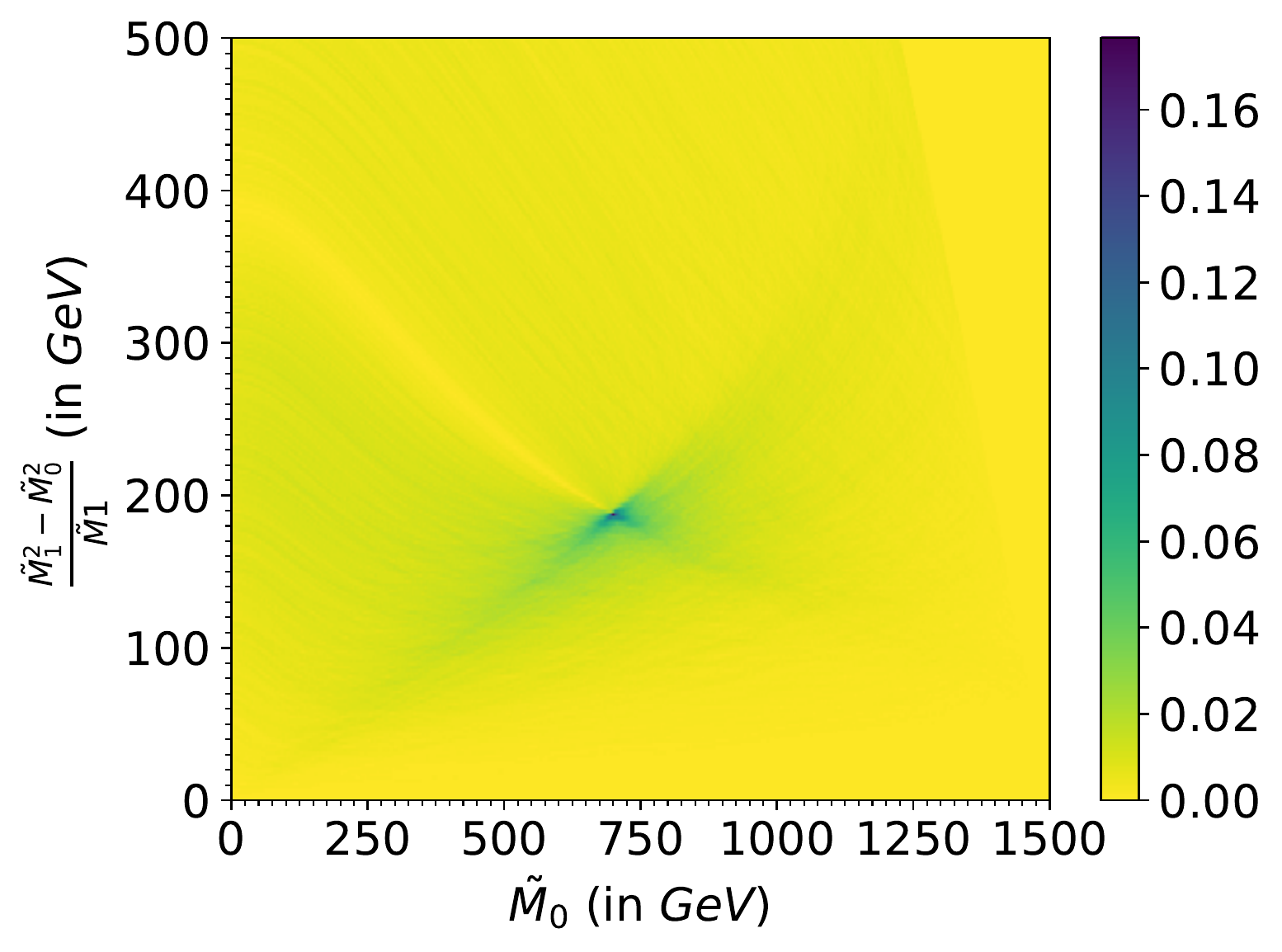}
 \caption{\label{fig:antlerFP} Focus point plots in the $(\tilde M_0, (\tilde M_1^2-\tilde M_0^2)/\tilde M_1)$ plane
 for the event topologies of Figs.~\ref{fig:feynmandiag}(d) and \ref{fig:feynmandiag}(e) with fixed ${\cal P}=(M_2,0,0,0)$ and $M_2=3000$ GeV.
 We use 20 randomly chosen events in the left panel, in which the red ``$+$" symbol marks the location of the true 
masses, and 10,000 events to produce the heatmap in the right panel, where the color indicates
 the fraction of events whose solvability boundaries pass through a given $2\times 2$ GeV bin. }
\end{figure}

Before concluding this section, we shall demonstrate that the focus point method of Ref.~\cite{Kim:2019prx}
applies to the event topologies of Figs.~\ref{fig:feynmandiag}(d) and \ref{fig:feynmandiag}(e) as well.
The point is that the singularity variable $\Delta_{antler}$ derived in Sec.~\ref{sec:devariable}
requires a mass ansatz $\tilde M_1$ and $\tilde M_0$ as input. The need for such mass ansatze was once considered undesirable,
but, as more recent studies have shown, it is precisely the dependence on the test masses that opens the door to 
new methods for extracting useful information about the mass spectrum, case in point being the kink method for measuring $M_0$
\cite{Cho:2007qv,Gripaios:2007is,Barr:2007hy,Cho:2007dh,Matchev:2009fh}.

In our case, the ansatz for $\tilde M_1$ and $\tilde M_0$ is needed to provide the needed number of 
kinematic constraints, which would allow us to solve for the invisible momenta. However, not all 
choices of $\tilde M_1$ and $\tilde M_0$ will lead to real solutions. Each event will thus delineate a viable region in the 
$(\tilde M_0,\tilde M_1)$ mass parameter space. The idea of the focus point method is to superimpose the boundaries 
of the allowed regions selected by different events, as illustrated in the left panel of Fig.~\ref{fig:antlerFP}.
The plot shows the solvability boundaries defined by
\beq
\Delta_{antler}(\tilde M_1, \tilde M_0)=0
\eeq
for 20 randomly chosen events, where for better visualization, we have rescaled the $y$-axis as was 
previously done in Figs.~\ref{fig:1stepFP2} and \ref{fig:1stepHM2}.
We see that even with just a handful of events, the solvability boundary curves tend to focus 
near the true mass point, marked with the red ``$+$" symbol.
With a lot more statistics, we obtain the heatmap shown in the right panel of Fig.~\ref{fig:antlerFP}
(contrast to the analogous heatmaps seen in Fig.~\ref{fig:1stepHM2}).
The heatmap clearly identifies the singularity peak which is situated at the true values of the masses,
thus establishing the viability of the method.

\section{Two decay chains, each with two successive two-body decays} 
\label{sec:22}

For completeness, in this section we shall review the final event topology from Fig.~\ref{fig:feynmandiag}, namely, the dilepton $t\bar{t}$ 
event topology in Fig.~\ref{fig:feynmandiag}(f), which was also the one used in Ref.~\cite{Kim:2019prx} to introduce
and illustrate the idea of the focus point method for mass measurements. Correspondingly, we shall not repeat the analysis of Ref.~\cite{Kim:2019prx} here, 
and simply refer the readers interested in mass measurements aspects to that paper. 
Here we shall focus more narrowly on the derivation of a singularity coordinate for that case, following the general method 
outlined in Sec.~\ref{subsec:idea} and illustrated with the examples from the previous sections.

Let us begin by listing the kinematic constraints for the event topology from Fig.~\ref{fig:feynmandiag}(f):
\begin{subequations}
\begin{eqnarray}
q^2 &=& M_0^2, \label{eq:flittleq}\\[1mm]
Q^2 &=& M_0^2, \label{eq:fQ} \\[1mm]
(q+p_1)^2 &=& M_1^2,  \label{eq:flittleqp1}\\[1mm]
(Q+P_1)^2 &=& M_1^2,  \label{eq:fQP1} \\[1mm]
(q+p_1+p_2)^2 &=& M_2^2,  \label{eq:flittleqp1p2}\\[1mm]
(Q+P_1+P_2)^2 &=& M_2^2,  \label{eq:fQP1P2} \\[1mm]
q_x + Q_x&=& \mptx, \label{eq:fqQx} \\ [1mm]
q_y + Q_y&=& \mpty, \label{eq:fqQy} 
\end{eqnarray}
\end{subequations}
which can be rewritten as
\begin{subequations}
\begin{eqnarray}
q^2 &=& M_0^2, \label{fqq}\\
2\, p_1\cdot q &=& M_1^2-M_0^2-m_1^2, \label{fp1q}\\
2\, p_2\cdot q &=& M_2^2-M_1^2-m_2^2-2p_1\cdot p_2, \label{fp2q} \\
Q^2 &=& M_0^2, \label{fcapQQ}\\
2\, P_1\cdot Q &=& M_1^2-M_0^2-m_1^2, \label{fcapP1q} \\
2\, P_2\cdot Q &=& M_2^2-M_1^2-m_2^2-2P_1\cdot P_2, \label{fcapP2q} \\
q_x + Q_x&=& \mptx, \label{fqQx} \\ 
q_y + Q_y&=& \mpty. \label{fqQy} 
\end{eqnarray}
\label{systemFigf}%
\end{subequations}
Within the dilepton $t\bar{t}$ example of the SM, the three masses are known:  
$M_2$ is the mass of the top quark, $M_1$ is the mass of the $W$ boson and $M_0$ is the neutrino mass.
However, the event topology of Fig.~\ref{fig:feynmandiag}(f) is relevant not only for top physics, but also for new physics searches,
where the masses $M_2$, $M_1$ and $M_0$ may correspond to new BSM particles and thus may not be known {\em a priori},
again forcing us to use an ansatz $\{\tilde M_2, \tilde M_1, \tilde M_0\}$ for the mass spectrum.
Either way, eqs.~(\ref{systemFigf}) provide 8 constraints for the 8 unknown components of the invisible momenta $q$ and $Q$,
and one can thus solve for $q$ and $Q$ in terms of the mass ansatz by standard means
\cite{Sonnenschein:2005ed,Sonnenschein:2006ud,Betchart:2013nba}, obtaining 
\begin{subequations}
\bea
\tilde q&=&q(\tilde M_2, \tilde M_1, \tilde M_0), \\
\tilde Q&=&Q(\tilde M_2, \tilde M_1, \tilde M_0).
\eea
\label{qQsolutions}%
\end{subequations}

\begin{figure}[t]
 \centering
 \includegraphics[height=.5\textwidth]{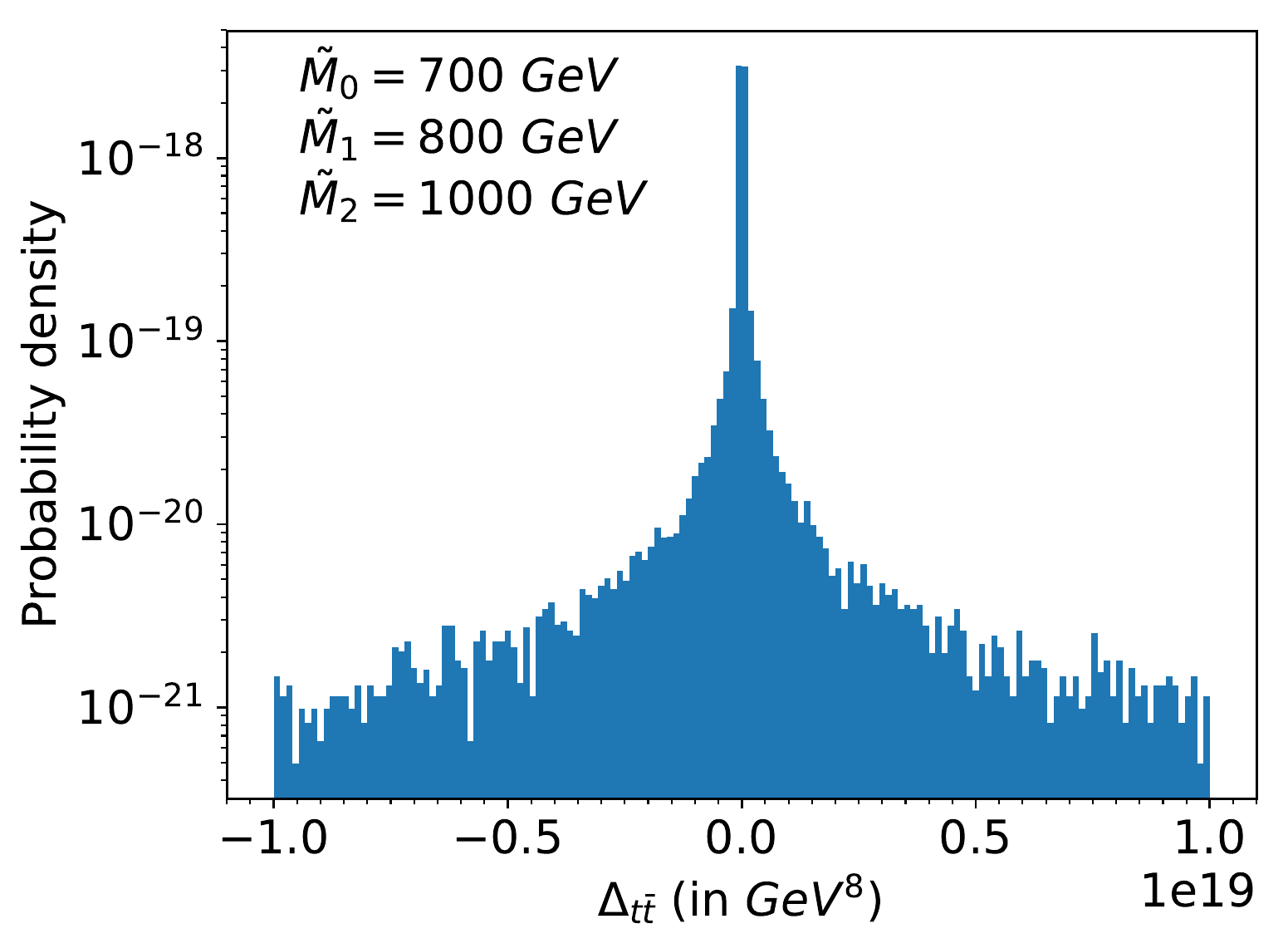}
 \caption{\label{fig:Deltatt} One-dimensional distribution of the singularity variable $\Delta_{t\bar{t}}(M_2,M_1,M_0)$,
computed with the true mass spectrum from Table~\ref{tab:mass}.  Note that the $y$-axis is plotted on a log scale.
  }
\end{figure}

The Jacobian matrix (\ref{defJac}) for the system (\ref{systemFigf}) is
\beq
D_{t\bar{t}} (\tilde M_2, \tilde M_1, \tilde M_0 )\equiv
\left(
\begin{array}{cccccccc}
2\tilde\varepsilon & -2 \tilde q_{x} & -2 \tilde q_{y} & -2 \tilde q_{z} & 0 & 0 & 0 & 0 \\
2e_1 & -2 p_{1x} & -2 p_{1y} & -2 p_{1z} & 0 & 0 & 0 & 0   \\
2e_2 & -2 p_{2x} & -2 p_{2y} & -2 p_{2z} & 0 & 0 & 0 & 0   \\
0 & 0 & 0 & 0 & 2\tilde {\cal E}  & -2\tilde Q_x & -2\tilde Q_y & -2\tilde Q_z \\
0 & 0 & 0 & 0 & 2E_1 & -2 P_{1x} & -2 P_{1y}  & -2 P_{1z}  \\
0 & 0 & 0 & 0 & 2E_2 & -2 P_{2x} & -2 P_{2y}  & -2 P_{2z}  \\
0 & 1 & 0 & 0 & 0 & 1 & 0 & 0 \\
0 & 0 & 1 & 0 & 0 & 0 & 1 & 0 
\end{array}
\right),
\eeq
where the dependence on the mass ansatz enters through the solutions for the invisible momenta $\tilde q$ and $\tilde Q$ from (\ref{qQsolutions}).
We can now take the singularity variable for the event topology of Fig.~\ref{fig:feynmandiag}(f) to be 
\beq
\Delta_{t\bar{t}}(\tilde M_2, \tilde M_1, \tilde M_0 )\equiv {\rm Det}\, D_{t\bar{t}} (\tilde M_2, \tilde M_1, \tilde M_0 ).
\label{eq:Deltattdef}
\eeq
As indicated, its computation again requires an ansatz for the masses, but
this is precisely the property which makes it relevant for mass measurements, since the singularity condition 
\beq
\Delta_{t\bar{t}}(M_2, M_1, M_0 )=0
\eeq
holds only if we use the true mass spectrum \cite{Kim:2019prx}.
The distribution of $\Delta_{t\bar{t}}(M_2, M_1, M_0 )$ is shown in Fig.~\ref{fig:Deltatt}.
One subtlety of the $\Delta_{t\bar{t}}$ computation is that there can be multiple (up to four) solutions (\ref{qQsolutions}) 
for the invisible momenta.
In making Fig.~\ref{fig:Deltatt}, we made sure that each event contributes equally to the plot, by entering the result for each solution with a weight $1/N_s$, 
where $N_s$ is the total number of solutions found in that event\footnote{When using the true spectrum as our ansatz, we are guaranteed at least one valid solution (\ref{qQsolutions}).}.

Having derived the singularity variable for the event topology of Fig.~\ref{fig:feynmandiag}(f), our next task would have been to 
illustrate the singularity surface in the relevant observable phase space, similarly to Fig.~\ref{fig:ellipses} (left panel) and Fig.~\ref{fig:antler3D}.
Unfortunately, the observable phase space in this case is nine-dimensional\footnote{The momenta of the four visible particles are 
parametrized by $4\times 3=12$ degrees of freedom, three of which can be removed by an azimuthal rotation and separate $z$-boosts 
for each of the two decay chains in Fig.~\ref{fig:feynmandiag}(f).}, and we shall not attempt to visualize it here.  The third and final task, the demonstration of the focus point method, 
was already accomplished in Ref.~\cite{Kim:2019prx}.

\section{Conclusions and Outlook}
\label{sec:conclusions}

In this paper we outlined the general prescription for deriving a singularity variable for a given event topology with missing energy, 
i.e., where some of the final state particles are invisible in the detector. We then illustrated the procedure with several common 
event topologies shown in Fig.~\ref{fig:feynmandiag}. We started with the case of a single two-body decay in Sec.~\ref{sec:11}
and re-derived the well known result that the distribution of the transverse mass $m_T$ has a Jacobian peak. In the subsequent sections, 
we demonstrated that similar Jacobian peak features are present in the distributions of the relevant kinematic variables 
for the remaining five event topologies. We also identified, parametrized and studied the shapes of the singularity surfaces in the 
appropriate visible phase spaces. In some special circumstances (see Secs.~\ref{sec:noPTISR} and \ref{sec:12}) 
the singularity variable can be computed directly in terms of the available kinematic information, without the need for any additional hypothesized inputs.
However, more often than not, the singularity variable depends on the masses of the intermediate resonances and/or the masses of the invisible final state particles,
and thus its computation requires a mass ansatz. If the event topology is applied to a SM process, the input masses will be known,
but when applied to a BSM process, the mass ansatz is {\em a priori} unknown. However, this can be used to our benefit --- it is precisely this 
dependence on the mass ansatz that makes the focus point method for mass measurements possible,
as explicitly demonstrated in Sec.~\ref{sec:aFP} and \ref{sec:deFP}.\footnote{For an application of the focus point method to the
event topologies of Figs.~\ref{fig:feynmandiag}(c) and \ref{fig:feynmandiag}(f), see Refs.~\cite{Debnath:2016gwz} and \cite{Kim:2019prx}, respectively.}
The main advantage of the focus point method is that it maximally utilizes the singularity structures in phase space.
As a consequence, the true masses are identified as a kinematic {\em peak} instead of a kinematic endpoint ---
endpoints are more difficult to observe experimentally, once we include the finite widths and detector resolution effects \cite{Chatrchyan:2013boa}.
It is also worthwhile to contrast the focus point method to the polynomial method~\cite{Cheng:2007xv,Cheng:2008mg,Cheng:2009fw,Webber:2009vm}.
While both methods use the same type of kinematic constraints, the latter requires a larger set of constraints in order to avoid the need for a mass ansatz.

The ideas presented in this paper may find immediate application in a large number of LHC analyses targeting the event topologies of Fig.~\ref{fig:feynmandiag}.
\begin{itemize}
\item {\em Standard Model measurements.} Precision studies of SM processes typically require the identification of a specific event topology, e.g., 
Fig.~\ref{fig:feynmandiag}(a) for $W$ production, 
Fig.~\ref{fig:feynmandiag}(b) for top quark decay, 
Fig.~\ref{fig:feynmandiag}(d) for $W$ pair-production, and
Fig.~\ref{fig:feynmandiag}(f) for dilepton $t\bar{t}$ production.
The corresponding singularity variables are ideal for tagging such events, and may also be used as input features for event selectors based on machine learning.
\item {\em New physics searches.} The Jacobian peaks in the distributions of the singularity variables can be used to discover new physics processes over the SM background \cite{Debnath:2018azt}.
Likewise, the peak structures in the heatmaps constructed in the focus point method can also be used for discovery of new physics, with the added advantage that 
background processes will not develop fake peaks\footnote{Even though the background event distribution may not have a singularity,
the phase space integration involved in reducing the singularity surface to a one-dimensional singularity coordinate may 
introduce fake peaks due to volume enhancement \cite{Debnath:2018azt}. } in the signal regions.
\end{itemize}

In this paper we mostly focused on the case (\ref{NceqNq}) when the number of unknowns $N_q$ matches the number of kinematic constraints $N_C$.
However, the under-constrained case $N_C < N_q$ is also worth investigating, e.g., following the analysis of Ref.~\cite{DeRujula:2012ns}.
All of these topics are being pursued in a future publication \cite{MS}.

\acknowledgments
We thank D.~Kim, K.~Kong, F.~Moortgat, L.~Pape and M.~Park for useful discussions. PS is grateful to the LHC Physics Center at Fermilab for 
hospitality and financial support as part of the Guests and Visitors Program in the summer of 2019.
This work was supported in part by the United States Department of Energy under Grant No. DE-SC0010296.

\end{document}